# Longitudinal Antigenic Sequences and Sites from Intra-Host Evolution (LASSIE) Identifies Immune-Selected HIV Variants

**Peter Hraber[1*], Bette Korber[1*], Kshitij Wagh[1], Elena E. Giorgi[1], Tanmoy Bhattacharya[1,2], S. Gnanakaran[1], Alan S. Lapedes[1], Gerald H. Learn[3], Edward F. Kreider[3], Yingying Li[3], George M. Shaw[3], Beatrice H. Hahn[3], David C. Montefiori[4], S. Munir Alam[4], Mattia Bonsignori[4], M. Anthony Moody[4], Hua-Xin Liao[4], Feng Gao[4], Barton F. Haynes[4]**

[1] Los Alamos National Laboratory, Los Alamos, NM 87545 USA
[2] Santa Fe Institute, Santa Fe, NM 87501 USA
[3] Perelman School of Medicine, University of Pennsylvania, Philadelphia, PA 19104 USA
[4] Duke University Medical Center, Durham, NC 27710 USA

* Authors to whom correspondence should be addressed; E-Mail: phraber and btk@lanl.gov; Tel.: +1-505-665-7491; Fax: +1-505-665-3493.

---

**Abstract:** Within-host genetic sequencing from samples collected over time provides a dynamic view of how viruses evade host immunity. Immune-driven mutations might stimulate neutralization breadth by selecting antibodies adapted to cycles of immune escape that generate within-subject epitope diversity. Comprehensive identification of immune-escape mutations is experimentally and computationally challenging. With current technology, many more viral sequences can readily be obtained than can be tested for binding and neutralization, making down-selection necessary. Typically, this is done manually, by picking variants that represent different time-points and branches on a phylogenetic tree. Such strategies are likely to miss many relevant mutations and combinations of mutations, and to be redundant for other mutations. Longitudinal Antigenic Sequences and Sites from Intrahost Evolution (LASSIE) uses transmitted-founder loss to identify virus "hot-spots" under putative immune selection and chooses sequences that represent recurrent mutations in selected sites. LASSIE favors earliest sequences in which mutations arise. With well-characterized longitudinal Env sequences, we confirmed selected sites were concentrated in antibody contacts and selected sequences represented diverse antigenic phenotypes. Practical applications include rapidly identifying immune targets under selective pressure within a subject, selecting minimal sets of reagents for immunological assays that characterize evolving antibody responses, and for immunogens in polyvalent "cocktail" vaccines.

---

## 1. Introduction

It is not yet known how to elicit protective immunity against HIV-1, and neutralizing antibody induction remains a central focus of the HIV vaccine field. Neutralizing antibodies are immune correlates of protection in all antiviral vaccines licensed to date [1,2]. Administration of neutralizing antibodies can confer protection in SHIV challenge models with rhesus macaques [3,4]. During the natural course of HIV infection, typically a single transmitted-founder (TF) virus establishes infection [5,6]. The virus population grows exponentially, with random mutations that initially accumulate following a Poisson distribution of inter-sequence distances [5,6]. The viral load eventually declines and resolves to a quasistationary set-point [7], influenced by both host and viral factors [8]. HIV-1 is maintained as a continuously evolving quasispecies population throughout chronic infection [9], with diversification driven by adaptive immune responses, including antibody [10-18] and T cell responses [19-21]. Mutations that facilitate immune evasion are positively selected and become more frequent, while mutations that result in a relative fitness disadvantage do not persist. Though neutral mutations may also drift to higher frequency with rates that depend on the effective population size, positive selection exceeds drift at driving envelope evolution within hosts [22-24].

During the chronic phase of infection, fifty percent of chronically HIV-1 infected individuals' antibody responses cross-neutralize 50% of HIV-1 primary isolates and breadth of neutralization responses varies uniformly across individuals, from neutralization of only a few heterologous viruses, to sera with great breadth and potency [17]. Plasma samples from individuals with the most potent and broad antibody neutralization are frequently singled out for detailed study [25-28]. Such studies include investigations of both viral and B cell lineages to understand the immunological processes that elicit effective neutralization responses and to inform strategies for vaccine design [15,16,18,29-31]. In general, autologous strain-specific neutralizing antibodies (nAbs) begin to develop within the initial months after infection, and rapidly select for viral escape variants [11,14]. High titers of more broadly neutralizing antibodies (bnAbs) develop in a subset of cases, but only after years of infection, and bnAb development is associated with persistently high levels of viral replication [32,33]. Previous longitudinal studies of HIV-1 infected individuals have shown that viral diversification precedes the acquisition of breadth, which suggests that antigenic diversity may be necessary for bnAb induction *in vivo* [15,18]. Several antibody lineages can place selective pressure on the same epitope at the same time, and escape from one antibody lineage can enhance recognition by another lineage, in a delicate evolutionary balance [16].

Viruses in individuals with bnAbs characterized to date have escaped from otherwise broadly-reactive neutralizing antibody responses [34]. Antibodies that recapitulate much of the potency and breadth of polyclonal sera have been isolated from subjects with high bnAb titers [35,36]. The developmental pathway of B cell immunoglobulin genes from the initial triggering of an HIV-1 specific clonal lineage through the acquisition of potency and breadth later in infection is an active research frontier. Properties of evolving viral Env proteins that stimulate or facilitate the important transition from autologous to heterologous reactivity are only beginning to be understood [15,18]. Understanding the events in natural HIV-1 infection that result in broader humoral responses should ultimately enable new vaccine strategies to elicit potent, broadly cross-reactive nAbs. Thus, a continuing research priority has been to characterize virus coevolution with antibodies in individuals who develop the greatest potency and breadth of neutralization [15,16,18,31,37,38]. Working back from mature bnAb clonal lineages, through ancestral intermediates, to the unmutated germline precursor, has begun to help define the process of bnAb development [15,18,29,30,35,37-40]. To study antibody/viral co-evolution, analysis of longitudinally (i.e. serially) sampled sequences, which represent both the anti-body population as it undergoes affinity maturation and the virus population as it evolves to evade the ongoing immune responses, explores mutational patterns that are selected over time [15,18].

Here we describe a new computational approach, called LASSIE, a two-step process to conduct longitudinal sequence analyses and inform reagent design. The first part of the LASSIE approach allows one to define and visualize sites that are potentially under positive selective pressure in the viral population. Identifying sites under putative immune selection can immediately help guide inference of antibody specificities that are active in the plasma, enable tracking of their appearance through time, and identify key mutations to be characterized during experimental follow-up studies. The second part of the approach is an algorithm that objectively down-selects sequences from a much larger sequence sample, yielding a subset of viral variants that represent mutations among the selected sites. We call the resulting subset of sequences an "antigenic swarm," which captures mutations at sites that are under the most potent selective pressure as they first emerge in the evolving HIV-1 quasispecies [15,18,31]. The size of the sequence subset involves a trade-off between the experimental cost of including more variants and the degree of selection to be represented. LASSIE involves two parameters that can be adjusted to balance these factors, explore the data, and choose the most representative set given experimental feasibility and sample-size limitations. To validate LASSIE, we retrospectively analyzed the Env quasispecies complexity in an African HIV-infected individual (CH505), for which we already have extensive information regarding antibody interactions and targeted Env epitopes [15,16]. This approach allowed us to determine how well the relevant diversity was captured by our computational method.

LASSIE can be used to select Envs (or similarly diversifying variants) for expression and binding studies, or to generate pseudoviruses for use in neutralization assays. In turn, the resulting reagents can be used to study relationships between viral phenotype and genotype, and to investigate in detail how neutralizing antibody responses develop by affinity maturation. Moreover, the chosen Envs can be used as experimental polyvalent "swarm" HIV vaccine immunogens.

## 2. Materials and Methods

### *2.1. Overview*

LASSIE analyzes a multiple sequence alignment in two phases. The first phase identifies protein sites most likely to be under positive (diversifying) selection, by considering the extent to which the TF amino acid state is "lost," i.e. replaced by mutations or deletions, at any one time-point during longitudinal sampling. This yields a list of sites of interest, from which we tabulate and track amino acid mutations that appear over time. TF loss is a simple and useful strategy to identify candidate sites undergoing positive selection in a scenario such as HIV evolution *in vivo* [41]. Profound immunological pressure can result in a selective sweep, where a mutation in a site is fixed and persists after its initial introduction, and such a mutational event may occur only once in a reconstructed phylogenetic history.

Such sites may be critically important to the evolving immunological phenotype of the virus, and would be good candidates for attention in experimental work. However, they would be undetected in commonly used statistical methods to detect positively selected sites. This is because the common methods require recurrence of mutational patterns in a tree for statistical validation, to identify ratios of non-synonymous to synonymous mutation rates significantly greater than one [42-44].

In contrast, our goal is to identify inclusively the sites that are *candidates* for being under positive selection pressure, as quantified by the extent of transmitted-founder loss [41]. To motivate use of the TF loss criterion, we performed a detailed comparison using both MEME [44] and LASSIE with longitudinal sequences from CH505, described in Results. LASSIE also provides plotting functions to review variation in selected sites over time. Comparing the list of selected sites with previously described antibody contacts from the growing set of anti-HIV bnAbs can facilitate the search to characterize immune responses that drive sequence diversification in newly obtained samples. The second phase of analysis uses the list of selected sites to choose sequences that represent the natural accrual of mutational variants that occur more than once among selected sites. The two phases of analysis, and parameters that influenced the number of sites and sequences thereby obtained, are detailed below.

An important feature of the single-genome amplification (SGA) sequences analyzed below is that they were obtained by limiting-dilution PCR, which provides genetic linkage across all of the *env* gp160, without *in vitro* recombination artifacts, and limited nucleotide substitution errors in cDNA synthesis [45,46]. Unlike Sanger sequencing from bulk PCR or large numbers of fragmentary high-throughput reads from unlinked templates, SGA sequences can provide high-quality sequence data spanning intact genes, ideally suited to understand how viruses adapt to host immune responses over time [5,14,21,45-47].

### *2.2. Site Selection*

We analyzed *env* cDNA amplicon sequences from plasma viral RNA by single-genome amplification (SGA), also known as limiting-dilution PCR [45], sampled longitudinally, beginning early (4 weeks) after infection, with three years of clinical follow-up. The sequencing effort was intended to obtain about 20 sequences (median of 25, range 18-53) from each of 14 samples. SGA sequencing from homogeneous infections commonly yields multiple identical sequences, all of which we kept. (Because they produce “rakes” of identical sequences on phylogenetic trees, monotypic sequences contain information about evolution and allele frequencies in the population sampled, but are incompatible with methods that require phylogenies with strictly dichotomous branching. We revisit this issue in the Discussion.) A naming convention for Env sequences was used to ensure consistency and to enable parsing of sample time-point labels from sequence names. By default, time-point labels are assumed to be in the first dot-delimited field of each sequence name, though any character could separate fields and they need not be in the first position. To study variant dynamics, we computed the number of days elapsed after the earliest sample from sample dates, and added the number of days post-infection estimated from the earliest sample. For homogeneous infections sampled before the onset of immune selection, such as subject CH505, a simple Poisson model of random sequence evolution provides the estimate [5,6].

We added the HXB2 reference sequence to standardize site numbering, codon-aligned the sequences, translated them, and inferred a phylogeny. Accurate alignment of the hypervariable loops, which evolve by insertion and deletion, is difficult for automated algorithms. Because no algorithm aligns the HIV envelope perfectly, particularly when a translation is needed, we manually edited the preliminary alignment, which was based on an HIV-specific hidden Markov model [48,49], for consistent and compact loop regions. PhyML v3 inferred maximum-likelihood trees from translated amino-acid sequences with the HIVw (HIV-specific, within-host) substitution model [50-52]. The phylogeny was used to order sequences for visualization and provides an organizing principle for sequence evolution from the ancestral TF virus. Reordering the alignment to follow the phylogeny was used to review alignment accuracy and consistency among closely related sequences. To identify potential N-linked glycosylation (PNG) sites readily, we annotated PNG sites, replacing asparagine sites that match the Nx[ST] motif to become Ox[ST]. In the PNG motif, x indicates any amino acid except proline, and the third position is either serine or threonine. To identify selected sites, the gaps inserted to maintain an alignment are considered a character state, because HIV frequently evolves by insertion and deletion, in addition to point mutations. For each aligned site, we computed TF loss per time-point sequenced, identified the maximum, and compared this peak TF loss with a threshold. We adjusted the threshold and considered the resulting number of sites. This produced a list of sites, which we considered as interesting evolutionary “hot spots”, to be represented by a swarm of Envs for experimental design.

We identify selected sites by transmitted-founder (TF) loss, defined as the proportion of sequences sampled per time-point that mutated away from the ancestral TF state. This is an efficient and inclusive way to identify rapidly evolving sites [41]. Because immune pressures can be transient, a variant may be relevant in only a particular phase of antibody lineage development. Thus, by considering TF loss per time-point, we seek to identify mutations that may be transiently important for resistance profiles of developing B-cell lineages. Here we considered no other information than TF loss per time-point, though other information could be used, so an option to specify additional sites for inclusion with the automatically selected sites is available. For example, signature sites associated with neutralization assay outcomes, antibody contact residues from structural data, key glycosylation sites associated with common bnAb epitopes, and additional sites identified using other methods (e.g. [42-44]) can be included by listing them for subsequent evaluation, even when these sites do not meet the minimum TF loss criterion. Similarly, if excessive diversity in hypervariable loops needlessly increases the number of sequences selected, such sites can be excluded simply by listing them.

Calculations of cumulative TF loss weighed the TF loss for two consecutive samples by the amount of time elapsed between when the two samples were drawn. Use of cumulative TF loss distinguished between sites that reverted to the TF form and sites that quickly mutated away from the TF and never reverted, and between sites that changed at different rates. For an informative representation of the accumulation of mutations among selected sites during site selection, we sorted sites by two criteria: time to initial TF loss and cumulative TF loss. The ordered list of selected sites identified automatically by LASSIE, or added specifically by choice, forms a “concatamer” by stringing together the listed sites extracted from a full-length sequence.

### *2.3. Sequence Selection*

After having used the TF-loss criterion to select sites from the alignment, we identified a set of Envs to represent the mutations that occur at these sites. By simple combinatorics, there are at least $10^{100}$ distinct ways to choose *k* representatives from *n* individuals for *n* above 427 and *k* over 100. On the scale of the current example, choosing 50 representatives from 385 candidates gives over $10^{63}$ distinct alternatives. To search such a vast space of possible solutions is intractable for even the fastest computers. Worse, in the regime of interest, the number of alternative solutions grows exponentially with *k*, where $k << n/2$. To our knowledge, there is no established optimality criterion for choosing representative variants from a larger set of aligned sequences. The second phase of analysis provides a practical solution to the problem, which we consider as a working proxy for an optimal solution [53].

As outlined in **Figure 1**, the swarm selection algorithm identifies sets of sequences that recapitulate viral evolution in key residues from a table of the amino acid mutations that occur in each site selected in the first analysis phase. Mutations that only ever appear once, or less than some other minimum number of times if specified, are disregarded. This parameter, the minimum variant count, is set to two by default. The resulting table lists all mutations to be included in the swarm set of Envs. Candidates for Env selection must be functionally viable, by lacking long deletions (as specified by the operator of the algorithm) and premature stop codons or incomplete codons, which typically result from frame-shift mutations.

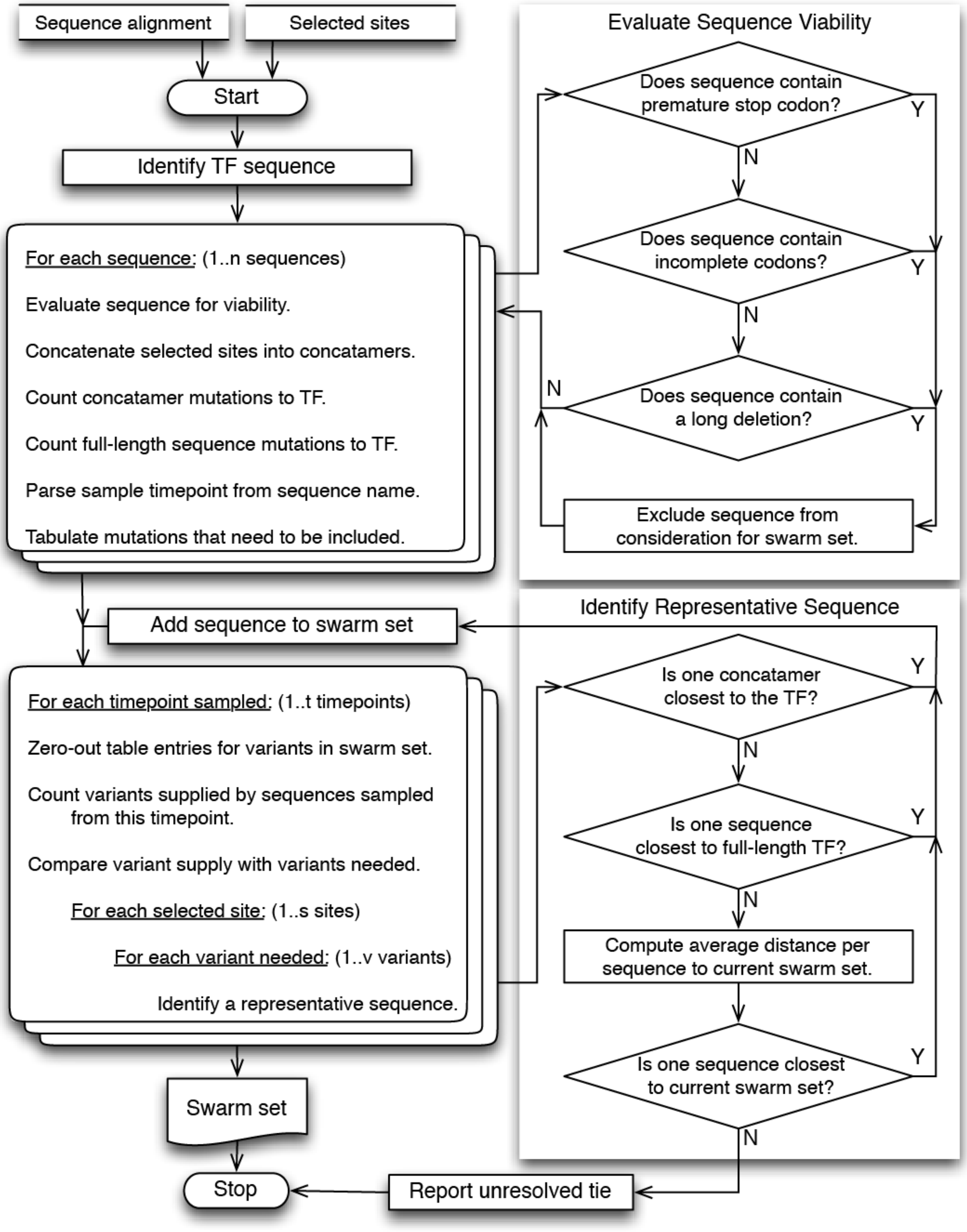

**Figure 1. Swarm selection algorithm.** From a sequence alignment and list of selected sites, this approach identifies viable Envs and tabulates mutations in selected sites. The table is used initially to define which mutations will be represented by the swarm, and is used subsequently to keep track of which mutations remain to be included. Rare mutations, i.e. mutations detected fewer times than the minimum variant count over the entire sampling

period, are disregarded. Selection among multiple sequences that carry a mutation is resolved by minimizing a series of distance criteria, first to minimize Hamming distance (number of mutations, gaps included) to the TF form among selected sites, then distance to the full-length TF sequence, and finally to minimize average distance to sequences in the current swarm set. The selected Env is included in the swarm set, counts in the table of needed mutations are set to zero, to indicate the particular mutation is now covered in the swarm, and iteration continues. This produces a "swarm" of Envs, which represents diversity in selected sites as it first developed within the subject, given sampling constraints. Stacked boxes signify iteration. Unresolved ties are reported, though we have not yet encountered them in several large experimental sequence sets we have tested; such an outcome would signal the need for an alternative distance metric or more selection criteria.

Building a swarm set starts with the TF sequence. If working with sequences that were first sampled during chronic infection, the natural sequence most similar to the consensus from the first available time-point is a good alternative. Sequences from the earliest time-point sampled are then considered, and scanned for the presence of sites with amino acids represented in the table of mutations. If a particular mutation of interest is present in one or more sequences from the first time-point, the choice among multiple Envs that carry a needed mutation is resolved by a series of criteria. The algorithm first tries to identify the sequence that uniquely minimizes the distance (number of mutations, including gaps) to the TF among selected sites. Then, in case of ties, a sequence is chosen that minimizes distance to the full-length TF. Finally, if ties remain, the sequence chosen minimizes the average distance to the current working set of sequences (**Figure 1**). An option exists to require that specific sequences be included, if desired. Such a sequence is added during iteration, when the time-point from which the sequence was sampled is being evaluated, rather than beforehand, to ensure inclusion of earlier (less divergent) sequences that carry mutations found on the specified sequence. After iterating over sample time-points, needed mutations, and selected sites, swarm selection is complete. Unresolved ties may exist among alternative sequences for some data. Such remaining ties would indicate a need for an alternative distance metric or additional selection criteria, though we have not yet encountered this outcome using Hamming distances from matching amino acid sequences and the three selection criteria described.

The algorithm is deterministic, which means it will always produce the same set of sequences from a given alignment, because it does not make random choices, and does not depend on the order in which sequences are provided in the input alignment. Overall, an advantage of this approach is that it selects no more sequences than are necessary to represent the mutational variants in selected sites, rather than some arbitrary number. By design, this greedy approach favors inclusion of early point mutations. This strategy produces larger sets of sequences than would result from favoring later, more divergent Envs that carry a greater number of needed mutations, but it better recapitulates the gradual increase of diversity in the evolving virus population. Among selected sites, each mutation observed more than once will ultimately be included in the antigenic swarm, generally multiple times if it remains common in subsequent time-points. The algorithm identifies the first appearance of mutations of interest in the least divergent sequence background possible, among available sequences sampled. It does this by progressively covering mutations that occurred in selected sites in the first time-point they appeared, and by representing them with the sequence most similar to the TF or, to resolve ties, the sequence most similar to those under consideration (lower-right quadrant in **Figure 1**).

The algorithm was made efficient through use of vector operations, and computes distance matrices only when they are needed to choose between otherwise ambiguous alternatives. Its duration of execution is expected to require no worse than a linear increase with the number of sequences and sample time-points in the input alignment. That is, doubling the number of input sequences or sample time-points should no more than double the run time. This was consistent with our experiences applying LASSIE to other large longitudinal data sets.

The methods to select sequences and sites and to plot the results graphically were written as an open-source (GPL v2) `R` package called `lassie` (`github.com/phraber/lassie`). Development activities are underway to provide lassie as a web-enabled service via the LANL HIV database. The package includes aligned sequences from CH505 and a tutorial vignette. The methods used in this paper to plot variant frequencies are included with the `lassie` package. We have included the GenBank accession numbers (KC247375-KC247667 and KM284696-KM284799) with those data. The *env* sequences from chronic infection in CH0457 are also in GenBank (KT220796-KT221004).

To visualize variation among descendants sampled serially from within a host, we have paired phylogenetic trees (rooted on the TF virus, ladderized, then rendered as phylograms) together with pixel plots [54,55], which illustrate polymorphisms as either mutations or indels relative to the TF sequence. We find these to be informative representations for understanding evolution of the virus population in an acutely infected host, given the limited genetic diversity that occurs in early infections [15,56]. Renderings such as given below emphasize sites with evolutionary changes that produce the branching patterns in the tree, and enable detection of recombinant clades or evolutionary associations with phenotypic assays. The code we use to make such renderings was written as an open-source `R` package called `pixgramr` (`github.com/phraber/pixgram`), and uses `ape` to draw trees [57].

### *2.4. Positively Selected Sites by MEME and FEL Analyses*

We analyzed the CH505 *env* codon alignment using the DataMonkey server (datamonkey.org) [58]. Because the alignment was too large for GARD recombination analysis and used directly for MEME and FEL analyses [44,59], these results are limited by to caveat that recombination was not excluded. We used the GTR/REV model of base substitution [60,61], and all sites with p-values below 0.1 were considered. Nine sequences with premature stop codons were excluded. After removing these and the duplicated sequences, 333 sequences remained.

### *2.5. Clinical Sample Assays*

Procedures for binding and neutralization assays have been described elsewhere [15,16,56]. Clinical materials were obtained as a part of the CHAVI 001 observational study. All participants in that study gave their informed consent before they participated in the study, which was conducted in accordance with the Declaration of Helsinki. The Kilimanjaro Christian Medical Centre Research Ethics Committee, the Tanzania National Institute for Medical Research Ethics Coordinating Committee, and the Duke University Institutional Review Board approved studies involving human subjects from whom we obtained materials with results described herein.

## 3. Results

Building upon recent insights into antibody/Env coevolution in individual CH505 [15,16,56], we first applied LASSIE to this subject as a test case, to determine how well the method performed in a situation where key epitopes had already been defined and characterized. We aligned 385 sequences from 14 time-points sampled over three years across 953 Env sites and computed TF loss (**Figure 2**).

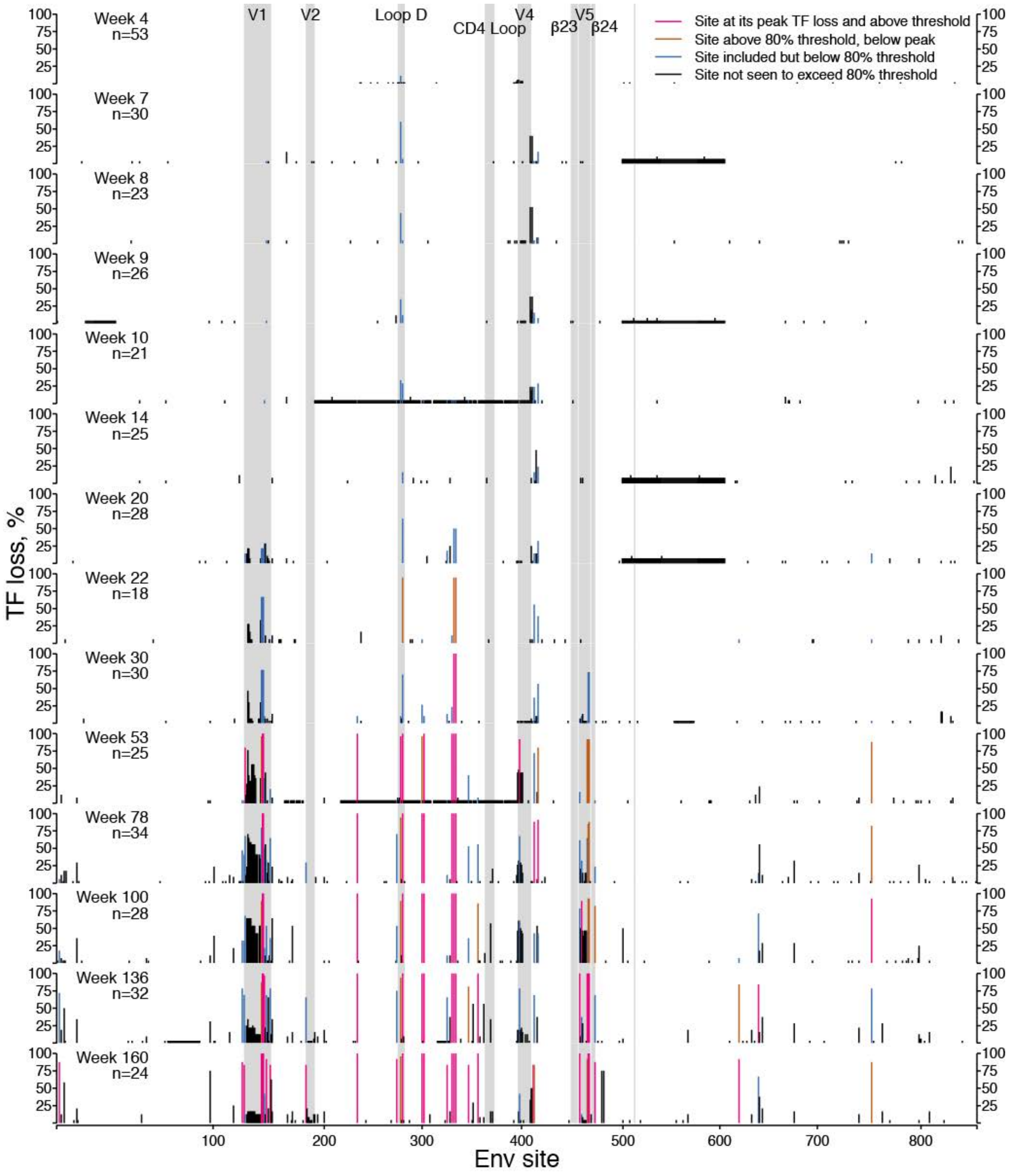

**Figure 2. Loss of ancestral transmitted-founder (TF) amino acids in Envs from CH505.** For 953 aligned Env sites spanning the full-length protein, TF loss is proportion of non-TF mutations per time-point sampled from the study participant CH505. TF loss is computed for each of 14 time-points sampled longitudinally, weeks 4 through 160, with the number of Envs sequenced (*n*) per time-point as shown. Bar colors vary over time to indicate 35 sites with at least 80% TF loss in any time-point, whether at peak TF loss (pink), below peak but above the 80% cutoff (brown), or below 80% (blue). Variable sites that were not selected for further consideration, because they never exceeded 80% TF loss in ant time-point during the study period, are also depicted (black bars). Grey boxes identify variable loops, which contain hypervariable regions that evolve by insertion and deletion and other gp120 landmarks. A thin grey line marks the boundary between gp120 and gp41. The time (*t*) of the sample is shown as days post infection (dpi), and the number of sequences (*n*) available from each time-point is shown. The gp160 site numbers indicate HXB2 positions in the CH505 protein alignment.

### *3.1. Site Selection*

**Figure 2** shows TF loss per site for each time-point sampled, from week 4 through week 160 post-infection. Clearly, most sites show little or no TF loss. Sites with high levels of TF loss are putative escape mutants due to immune selection. Because we counted an insertion or deletion relative to the TF virus as a change, the hypervariable V1, V2, V4, and V5 regions also showed TF loss, largely due to length variation. We used TF loss to list sites where frequency of the TF form fell below a fixed cutoff percentage. The cutoff is a parameter that can be adjusted as needed; here we used the value of 80% TF loss in samples from at least one time-point.

#### 3.1.1. TF Loss Varied across Sites

Initially dominated by the TF form, the quasispecies acquired mutational variants over time, displaying different dynamics among sites with high TF loss. **Figure 3** depicts variant frequencies in subject CH505 over time in the 35 sites shown, which had over 80% TF loss in at least one time-point. The rate of TF loss was slower in some sites than in others. Such slow transitions could reflect the evolving immune response and newly arising selective pressure. In qualitative terms, we saw four dynamic categories, designated i-iv. First (*i*), some sites showed complete replacement of the TF form with another single variant, whether fast or slow, e.g. the shift of a glycosylation site from position 334 to 332 (top-right panel "N332" in **Figure 3**, N → O, and O334 in the next row, from O → S, where "O" is an asparagine embedded in a glycosylation motif, as described in Section 2.2). Next (*ii*), in some sites, the initial TF replacement was followed by one or more additional mutations arising sequentially. For example, site 279, located in Loop D, was initially an asparagine, but a transient lysine mutation yielded to an aspartic acid after transient reversion to the TF asparagine (**Figure 3**, top-left, N → K → N → D). Third (*iii*), some sites reverted to the TF form after high TF loss. For example, site 417, initially histidine, was predominantly an arginine from about six months to nearly two years after infection, but then reverts to the ancestral histidine (**Figure 3**, top row, second panel, H → R → H). Finally (*iv*), some sites exhibited sustained polymorphisms. These were particularly common in hy-pervariable loops, in which insertions and deletions are the predominant evolutionary processes. In such cases, distinct subpopulations may carry forward divergent forms. For example, several distinct insertions arise in the V1 loop, which follow HXB2 position 144, and some of these forms are maintained at different frequencies (**Figure 3**, middle of the second row, 144g, 144h, and 144i). When such patterns are located in hypervariable regions, they may be alignment dependent, and should be reviewed in the context of the full alignment to ensure that they represent distinct and common forms of the hypervariable loops of the viral quasispecies, and not simply alignment artifacts.

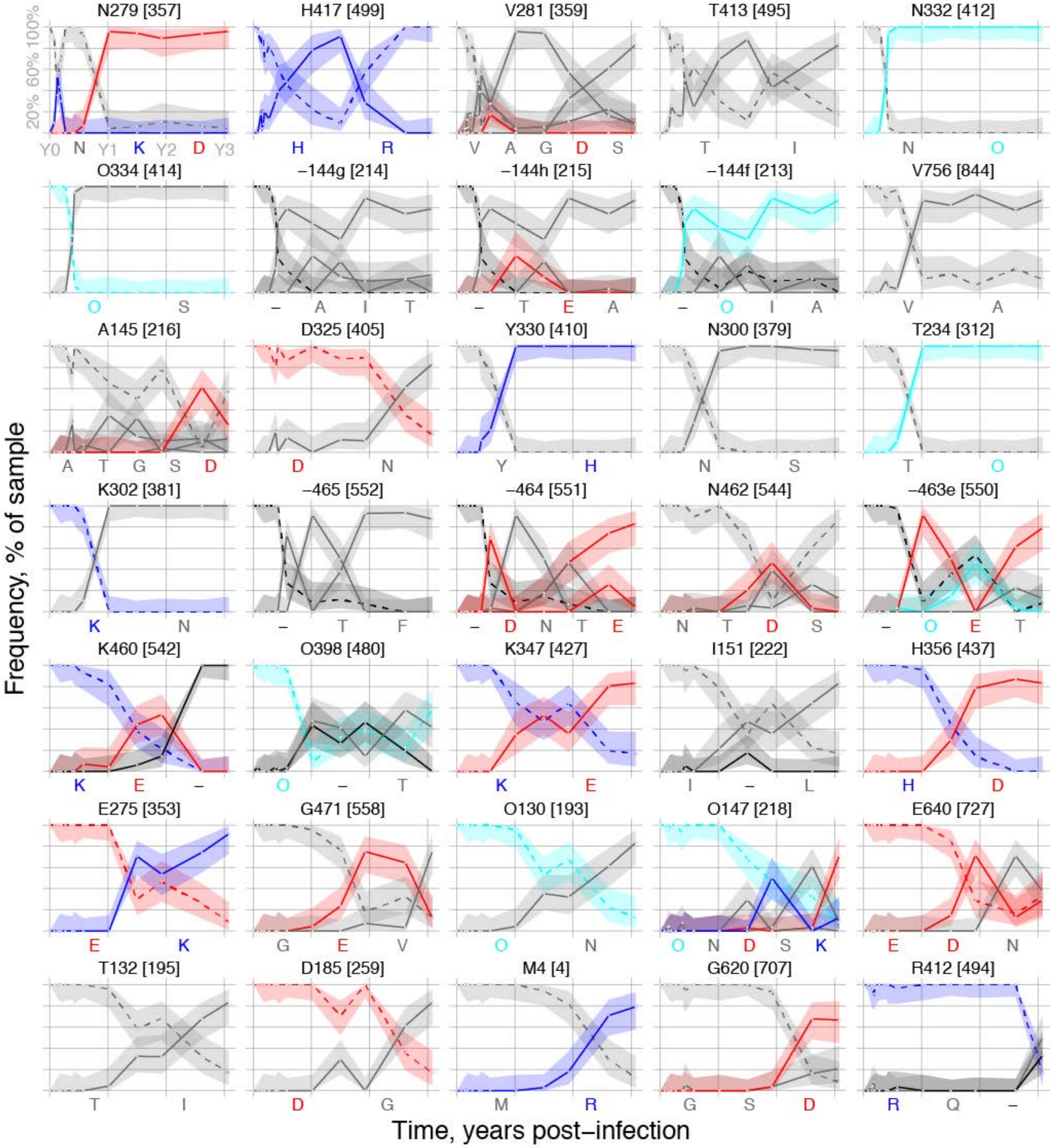

**Figure 3. Variant frequency dynamics within sites.** The single TF virus amino acid (dashed lines) yields to putative escape mutations over the sampling period. Letters below

each plot list mutations in order of appearance, and panels are ordered according to the timing of TF loss. Numbers above each plot denote TF form, HXB2 position, and alignment column, e.g. "N279 [357]" indicates HXB2 position 279 (alignment column 357) and depicts loss of the transmitted asparagine. Lower-case letters denote insertions at the C-terminal end of the HXB2 site given. Colors indicate positive (blue) and negative (red) charges and an "O" is used instead of "N" to indicate a potentially gylcosylated asparagine (cyan), i.e. an N that is embedded in a glycosylation motif of Nx[ST], where x can be any amino acid except Pro, followed by either Ser or Thr. The vertical bars indicate the sampling year, abbreviated in the upper left panel as Y0, Y1 and Y2, with a single TF virus starting at Y0, then followed by the first time-point sampled, estimated to be 28 days post infection for CH505 [15]. Shaded regions show 95% confidence intervals for variant frequencies, computed from the binomial probability distribution, given the number of sequences sampled per time-point. Several distinct insertions arise after HXB2 position 144 in the V1 loop, shown in panels in the middle of the second row as 144f, 144g, and 144h. Lower-case letters (f through h) specify the relative positions of the inserted region in the alignment [62]. The TF lacks an amino acid in this position, which is characterized as a gapped state and represented by a dash (–) to maintain the alignment. Over time, new and distinct insertions arise that span this position, with major and minor variants carried along with distinctive insertions occurring at positions 144f, 144g, and 144h (e.g. in 144g, three different insertions are maintained, which include A, I and T).

We sampled a median of 25 (range 18-53) sequences across 14 time-points, and sampling 18 sequences, our minimum, has 85% probability to detect variants with prevalence 10% or more. In **Figure 3** we show the dynamics with 95% confidence intervals estimated for the frequencies. For clarity, the mutations that never attain 15% frequency in any sample are not shown. Given our sampling constraints, we could estimate relative frequencies of the different mutational classifications described in the paragraph above. Simple shifts, like (*i*), were the most common form of TF loss, evident in 16 of the 35 selected sites (46%). Serial mutations, like (*ii*), were also common and could be the direct result of serial escape, due to new pressures imposed by adaptation of the evolving antibody response to an initial escape mutation, driving continued selection. Alternatively, serial replacements could result from complex interactions with multiple antibodies in a polyclonal response [16], or pressures resulting from balancing fitness costs and/or compensatory mutations in a changing evolutionary milieu. Transient losses (*iii*) reverting to the TF form were rare, and clear restoration of the TF as the dominant form after a loss of at least 80% occurred in only 2 of 35 positions (6%), positions 417 and 462 (**Figure 3**). Position 279 had a transient reversion to the TF form, and position 145 may have been reverting at the last time-point sampled. Different underlying reasons for this pattern could be at play, such as a fitness cost for a mutation that was carried along with a neighboring mutation, or a changing immunological environment in the host, which could transiently favor a mutation with a modest fitness cost [14,63-65].

#### 3.1.2. Peak TF Loss Identified Selected Sites

We defined “peak” TF loss per site as the highest TF loss in that site over all time-points sampled (often the “peak” was maintained over many samples), and used it to select candidates for sites under immune selection. Together, 15 sites completely lost the TF form during the three-year sampling period, while the other 938 aligned sites never reached 100% TF loss. The cumulative distribution of peak TF loss per site (**Figure 4**) indicated that 588 of 953 sites (62%) were strictly invariant, and 64 (6.7%) lost over 50% TF. We selected the 35 sites with at least 80% peak TF loss for further study and Env selection. The 80% cutoff is the threshold we chose to use for this presentation of these data. Increasing the TF loss cutoff decreases the number of sites selected, and working with other cutoff values can adjust the number of selected sites for subsequent investigation. This value can be chosen in light of available resources.

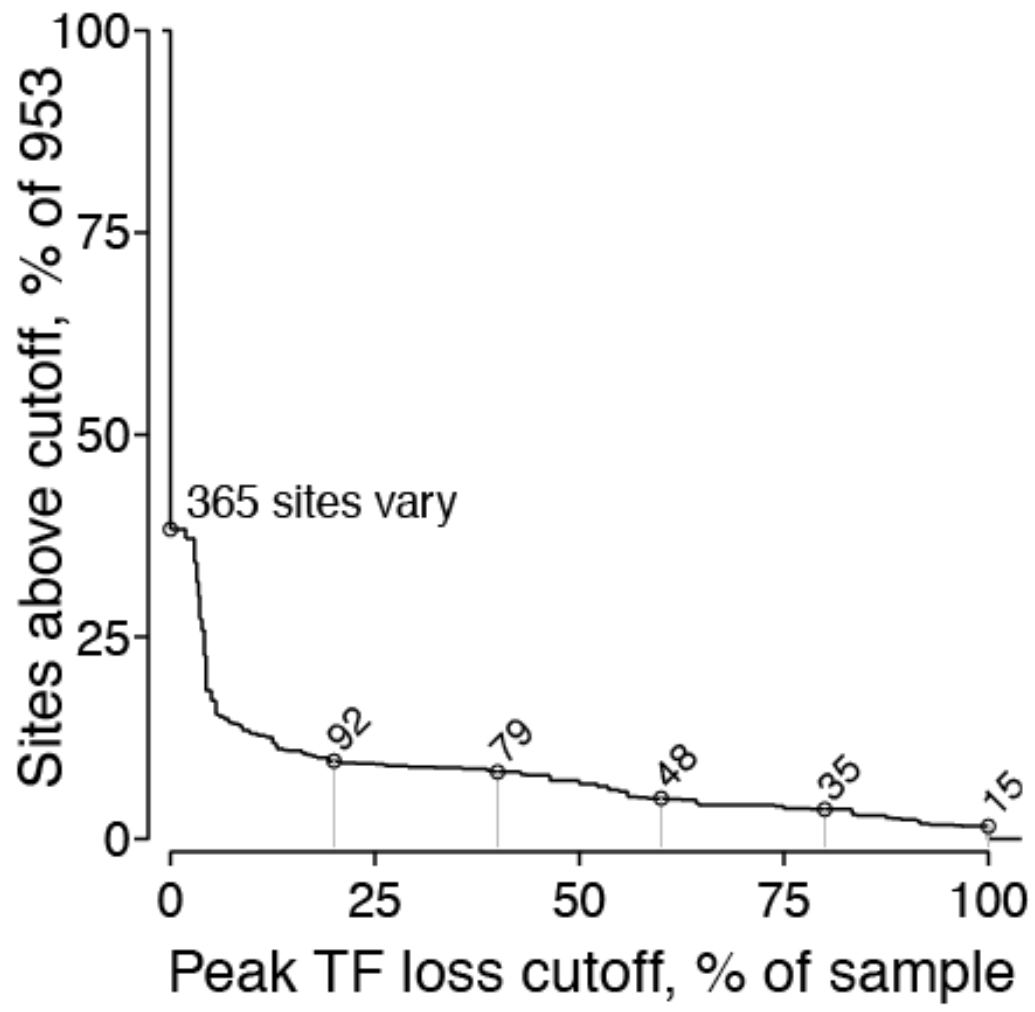

**Figure 4. Cumulative distribution of peak TF loss over 953 aligned Env sites.** Peak TF loss is the greatest proportion of non-TF variants in any time-point sampled, which corresponds to the minimum for each dashed line in **Figure 3**. Of 953 aligned sites, 365 (38.3%) varied and the others did not vary among sequences sampled throughout the period studied. We selected 35 sites with at least 80% peak TF loss for further study. Other cutoff values would yield more (e.g. 48 at 60% TF loss) or fewer sites (e.g. 15 at 100% loss) for consideration.

### 3.1.3. Selected Sites were Consistent with Antibody-Driven Selection

The time at which TF loss started to emerge at each selected site in the sampled virus population naturally varied from one site to another (**Figures 2** and **3**). The cumulative amount of TF loss also varied. Cumulative TF loss over time had a simple geometric interpretation as the area above the dashed TF line in the plots of frequency over time that appeared in **Figure 3**.

**Table 1** lists the 35 selected sites with at least 80% peak TF loss, ranked by the earliest time at which any non-TF variant exceeded 10%, with ties resolved by cumulative TF loss sorted in descending order. Most (91%) of the selected sites occurred in gp120.

**Table 1. Selected sites.** Summary of the extent of TF loss and timing of the mutation in the subject. These CH505 Env sites showed at least 80% TF loss in at least one time-point. Symbol color in left column indicates known immune pressures in CH505 that likely drove selection (**Figures 5c** and **5f**).

| HXB2 Site | Peak loss | When up | Rank | MEME p-value[1] | Immune pressure | Notes |
|---|---|---|---|---|---|---|
| 4 | 87.5 | d701 | 33 | NA | NA | Signal peptide |
| 130 | 87.5 | d547 | 28 | 0.086 | CD4bs | PNG site at base of V1, near VRC01 contact [66] |
| 132 | 83.3 | d547 | 31 | * | CD4bs | V1 indels cause CH103 resistance [16] |
| 144f | 100 | d141 | 9 | * | CD4bs | V1 indels cause CH103 resistance [16] |
| 144g | 100 | d141 | 7 | * | CD4bs | V1 indels cause CH103 resistance [16] |
| 144h | 100 | d141 | 8 | * | CD4bs | V1 indels cause CH103 resistance [16] |
| 145 | 96.8 | d141 | 11 | * | CD4bs | V1 indels cause CH103 resistance [16] |
| 147 | 91.7 | d547 | 29 | * | CD4bs | V1 indels cause CH103 resistance [16] |
| 151 | 83.3 | d371 | 24 | * | CD4bs | V1 indels cause CH103 resistance [16] |
| 185 | 83.3 | d547 | 32 | 0.013 | CD4bs | Signature site for CD4bs bnAb b12 [67] |
| 234 | 100 | d211 | 15 | NA | CD4bs | Signature site for CD4bs VRC01 & NIH45-56 [67] |
| 275 | 91.7 | d547 | 26 | 0.044 | CD4bs | Loop D, CH103 contact, CH235 resistant [15,16] |
| 279 | 95.8 | d28 | 1 | 0.090 | CD4bs | Loop D, CH235 resistant, CH103 sensitive [15,16] |
| 281 | 100 | d64 | 3 | **0.00007** | CD4bs | Loop D, CH235 resistant, CH103 sensitive [15,16] |
| 300 | 100 | d211 | 14 | NA | V3 loop | V3 autologous nAb in CH505 [56] |
| 302 | 100 | d211 | 16 | NA | V3 loop | V3 autologous nAb in CH505 [56] |
| 325 | 83.3 | d141 | 12 | NA | V3 loop | V3 autologous nAb in CH505 [56] |
| 330 | 100 | d157 | 13 | NA | V3 loop | V3 autologous nAb in CH505 [56] |
| 332 | 100 | d141 | 5 | | V3 loop | V3 autologous nAb in CH505 [56] |
| 334 | 100 | d141 | 6 | 0.029 | V3 loop | V3 autologous nAb in CH505 [56] |
| 347 | 83.3 | d371 | 23 | 0.0074 | CD4bs | 15-17 Angstroms from CH103 contacts |
| 356 | 100 | d547 | 25 | 0.018 | CD4bs | Adjacent to CD4bs bnAb 12A12 signature [67] |
| 398 | 91.3 | d371 | 22 | 0.0088 | CD4bs | 15-17 Angstroms from CH103 contacts |
| 412 | 83.3 | d1121 | 35 | NA | CTL | CTL epitope V4 loop [16] |
| 413 | 88.2 | d64 | 4 | **0.00004** | CTL | CTL epitope V4 loop [16] |
| 417 | 91.2 | d51 | 2 | **0.00073** | CTL/CD4bs | CTL epitope V4 loop; CD4bs b12 contact [16] |
| 460 | 100 | d371 | 21 | * | CD4bs | V5, CH103 contact region, resistance [15,16] |
| 462 | 89.3 | d211 | 19 | * | CD4bs | V5, CH103 contact region, resistance [15,16] |
| 463e | 100 | d371 | 20 | * | CD4bs | V5, CH103 contact region, resistance [15,16] |
| 464 | 100 | d211 | 18 | * | CD4bs | V5, CH103 contact region, resistance [15,16] |
| 465 | 100 | d211 | 17 | * | CD4bs | V5, CH103 contact region, resistance [15,16] |
| 471 | 87.5 | d547 | 27 | 0.0057 | CD4bs | CH103 contact [16] |
| 620 | 91.7 | d953 | 34 | **0.0026** | NA | gp41 |
| 640 | 83.9 | d547 | 30 | 0.0054 | NA | gp41 |
| 756 | 92.9 | d141 | 10 | 0.0035 | NA | gp41 cytoplasmic tail |

[1] In the MEME p-value column, an asterisk (*) indicates a site in hypervariable regions of V1 or V5, where evolution is enhanced by insertions and deletions. Because this mode of evolution fails to satisfy assumptions of the statistical model, the p-values (false-positive rates) are not readily interpreted. Regardless, MEME found support for positive selection in these regions. "NA" indicates no association with positive selection was found with p below 0.1. Bold text indicates MEME q-values (false discovery rates) below 0.2.

In the context of the Env trimer structure, the selected sites formed three localized clusters on the outer domain of gp120 (**Figure 5**). The clustered patches of selected sites on gp120 corresponded to the three known regions of immunological pressure in individual CH505. The first cluster of three selected mutations (HXB2 positions 412, 413, 417) was in a CTL epitope, which was targeted early in infection in CH505 and conferred early CTL escape [16]. The second cluster of six selected sites (300, 302, 325, 330, 332, 334) was located within the V3 loop, or in the glycosylation site at its base. Two autologous neutralizing anti-V3 antibodies, DH151 and DH228, were isolated from CH505. These antibodies could only neutralize heterologous Tier 1 viruses at the heterologous population level, but potently neutralized a subset of autologous CH505 Tier 2 viruses, shown by linear peptide array binding to target a linear epitope in the V3 loop, which encompass the V3 loop sites in this second cluster [56]. Thus, this second cluster of V3 sites may be relevant to escape from members of this nAb lineage.

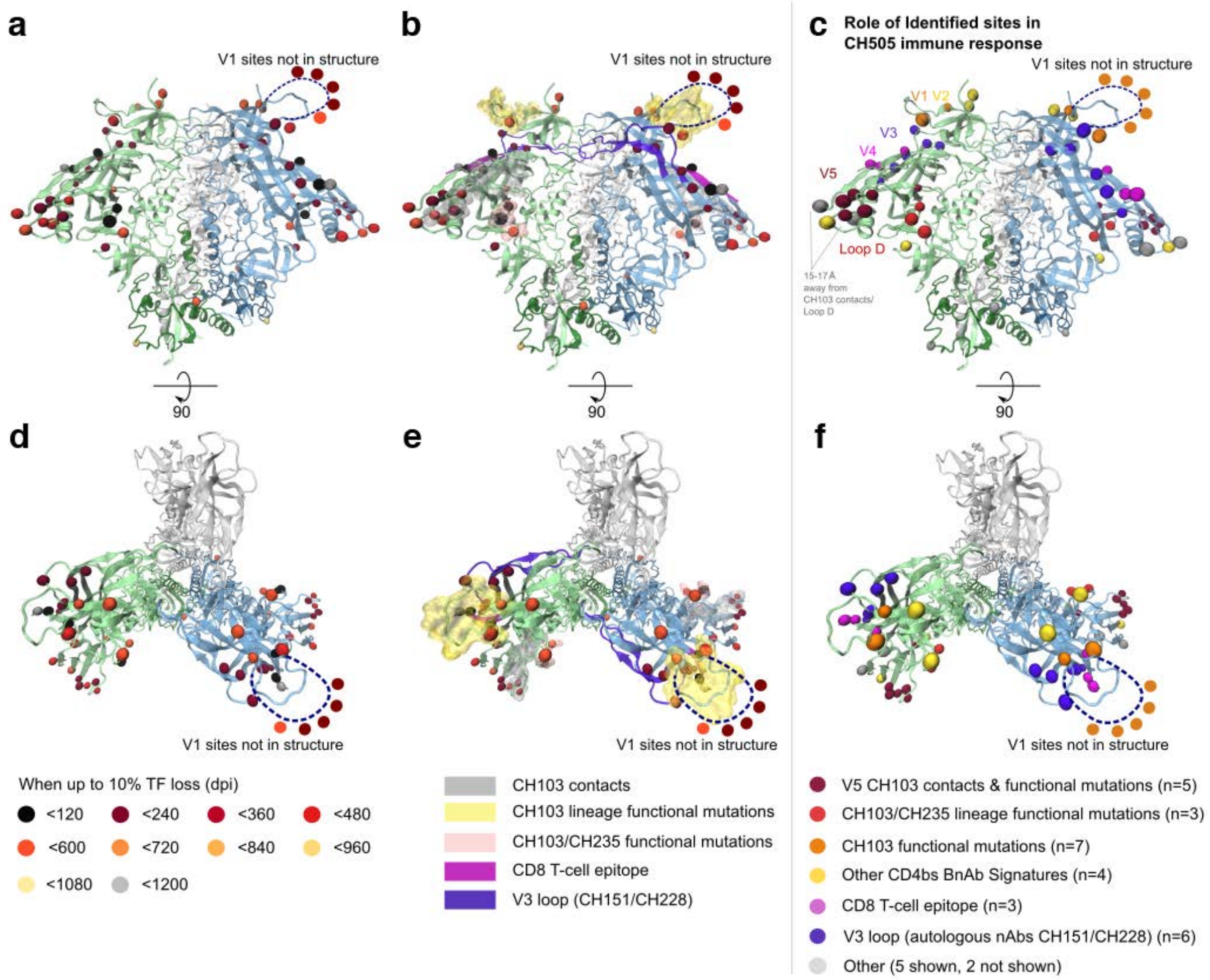

**Figure 5. Selected sites are localized to the known immunogenic regions in CH505**, as visualized by mapping on the BG505 SOSIP trimer structure (PDB ID 4TVP [68]). Selected sites are depicted as spheres, colored to indicate the timing of their emergence. (**a**) Side view, oriented with viral membrane towards bottom. (**b**) Additional colored highlights indicate known immunogenic regions. (**c**) Selected sites are colored to show which immune pressures are known to have induced TF loss. (**d-f**) The corresponding details from top view, as seen from host cell membrane. **Table 1** lists symbol colors for each selected site.

The third cluster, in the CD4 binding site (CD4bs), is the most complex. The CD4bs is the target of both the CH103 bnAb lineage [15] and the CH235 nAb cooperative lineage [16] in subject CH505. Many of the 32 selected gp120 sites included structurally defined contacts for CD4 [66,69], and several previously studied CD4bs bnAbs, including VRC01 [66,69], NIH45-46 [70], and b12 [71] (**Table 1**). Although the current study is retrospective, this pattern of mutations indicates the presence of CD4bs antibodies in the subject, even prior to isolation of CH235 and related nAbs at week 41 [16], as this region was apparently under selective pressure very early, particularly in the loop D region [72]. As expected, CH103 contacts and previously characterized resistance mutations were well represented among the selected sites [16]. Three selected sites (279, 281, 275) were localized to CH103 light-chain contacts near loop D (**Figure 5**), a region that rapidly accumulated mutations as a result of escape from the autologous CD4bs neutralizing antibody CH235; these mutations rendered the virus more susceptible to the CH103 early lineage members. Six CH103 heavy-chain contacts in and near V5 (460, 462-465, 471) were also among the selected sites, and mutations in this region conferred CH103 resistance [73]. V1 loop mutations also conferred CH103 resistance, and seven sites in V1 were among the 35 selected sites (132, 144f, 144g, 144h, 145, 147, and 151). Three of these were inserted together in V1 after position 144. Finally, 5 additional selected sites are known to be important for other CD4bs bnAb interactions, providing indirect evidence that they may be important for either CH103 or CH235, or both. These are: 417, a contact for the CD4bs bnAb b12 [71], and 185, a V2 region signature site for b12 [67]; 234, a signature site for CD4bs bnAbs VRC01 and NIH45-46 that is near Loop D [67]; the glycosylation site N130, adjacent to a VRC01 contact [66]; and position 356, adjacent to a 12A12 signature [67]. We identified selected sites that were relevant to the other antibodies noted in this section using the Los Alamos HIV-database genome browser and CATNAP tool (`hiv.lanl.gov`).

Together, 29 of the 35 selected sites (83%) are related to the three epitopes that were functionally defined in this subject during the time period under study, despite these sites being simply and objectively identified based solely on the TF loss criterion (**Table 1**). Of the six sites that were not directly related, three were gp41 sites (620, 640, 756) and one was in the signal peptide (position 4). The other two (398 and 347) were clustered near position 356 in gp120, and both were near but not in the CH103 contact region (indicated in **Figure 5c** as 15-17 Angstroms away from CH103 contacts). These six mutations in sites selected by TF loss are indirectly implicated as having biological or immunological significance, and may suggest strong leads for follow-up experimental investigation.

### 3.1.4. Comparison of Selected Sites Identified by LASSIE and Phylogenetic Methods

MEME is an analysis method that identifies sites under episodic diversifying selection within a phylogenetic reconstruction [44], and represents a class of tools that identifies positive selection by comparing relative levels of synonymous and non-synonymous mutations in the dataset. While such tools are very useful, their application comes with several caveats. Here we compare the sites identified by the two methods, MEME and LASSIE, which correspond to several dynamic categories. Using LASSIE, we identified 35 sites with the TF loss criterion, while MEME identified 14 sites of highest interest, with a q-value (false-discovery rate) below 0.2.

Both methods identified multiple sites in the V1 and V5 hypervariable regions, 8 sites with LASSIE and 9 sites with MEME. Identifying these regions as foci of positive selection is appropriate, because changes in both V1 and V5 have been shown to confer distinct immunological phenotypes with respect to the evolving nAb lineages in subject CH505 [15,16,73]. However, a concern for interpreting the MEME results is that the model used to compute positive selection probabilities assumes that codons evolve by base substitutions. In hypervariable regions, this assumption is violated, and frequent insertions and deletions dominate the mutational framework. This is readily evident in the full alignment, but to illustrate it further, **Tables S1** and **S2** (in the Supplement) list each distinct variant form of the V1 and V5 hypervariable regions sampled from subject CH505. This shows clearly the effects of insertions and deletions on the evolution of these regions. MEME identified some of these positions, as it should, but the p- and q-values are not readily interpreted in the context of the evolutionary model. LASSIE also identifies V1 and V5 hypervariable regions as candidates for positive selection, simply by counting changes from the transmitted-founder state over time.

Outside of the V1 and V5 hypervariable regions, MEME identified an additional site with a q-value below 0.2, which LASSIE did not identify. This site, HXB2 position 411 (codon 493) was in the known CTL epitope, HXB2 positions 409-418 (codons 491-500) [16], which may confer CTL resistance. The TF frequency for this site never falls below 40% in a time-point. While this site is likely driven by immune selection, incomplete resistance or viral fitness costs associated with escape mutations at this site presumably make other local escape mutations more competitive. Regardless, this site may be of interest, and with LASSIE one could include it with the list of sites for consideration during the sequence-selection phase of analysis, if desired. An additional 14 sites identified by LASSIE were also found by MEME with a more inclusive p-value (false-positive rate) cutoff of 0.1, meaning they did not withstand the correction for multiple tests in MEME, but were identified as interesting by both methods.

A class of sites likely to be of biological interest is those captured by LASSIE and missed by MEME. These sites in CH505 are HXB2 positions 4, 234, 300, 302, 325, 330, and 412 (**Table 1** and **Figure 3)** and are illustrated in **Figure S1** to show the TF loss pattern clearly. These sites generally have either a single mutation (or very few) among internal branches of a phylogenetic reconstruction, and cannot be statistically validated using an approach like MEME. However, once such a mutation emerges, the TF form is essentially lost, and replaced by the mutation, a selective sweep. LASSIE identified 7 such sites, which MEME and another phylogeny-based method, FEL [59], did not. Mutations in such sites are good candidates for changes that confer a profound selective advantage, due to immune selection, viral fitness, or both. Once they are introduced, lineages that survive to later time-points all carry the mutation.

In addition to site 411 in the CTL epitope, MEME identified 27 more sites with p-values below 0.1 and q-values above 0.2 (**Figure S2**), which LASSIE did not find. Most of these mutations were transient, rare, or not obviously related to the epitopes we know were targeted in CH505. For example, many of these sites occurred in gp41 (**Figure S2**).

### 3.1.5. Threshold Considerations

To consider systematically what sites might be missed by the 80% peak TF-loss criterion we used here (**Figure 4**), we explored the locations of mutated sites that did not attain 80% TF loss at any time-point in the study period. The 365 sites that varied were dispersed over the entire protein, as expected. In particular, sites that had only one mutation among all available sequences are scattered throughout the structure. By requiring multiple mutations among all 385 available sequences, regional patterns appeared in the spatial distribution of mutations (**Figure S3**). Positions with as few as three or four mutations began to show a clear focus towards the immunologically targeted regions. This suggests that the relatively high threshold of 80% TF loss at any one time-point excludes from consideration some mutations that occur in immune-targeted regions, which may have phenotypic consequences for immunological sensitivity. Several more localized clusters of sites were apparent, which contained variable positions but did not attain the within-time-point 80% TF loss cutoff criterion (**Figure S3**). Such structural clustering suggests presence of other selected regions. However, these sites were not under the same high degree of selective pressure as sites in which the TF form was depleted. These sites may be targets for transient or less potent antibodies, or antibodies that are just beginning to impose selective pressure at the end of the study period. They might otherwise result from CTL escape, tolerance of neutral variation, or other mechanisms of molecular change not directly related to immune selection. LASSIE allows investigators to target the most highly selected sites for further study, and adjust the threshold as practical for reagent design.

### 3.1.6. Concatamers of Selected Sites

A concise representation of selected sites strings the sites together to form concatamers of 35 amino acids. The order of sites in concatamers can, but need not, follow the primary Env sequence; here we ordered them by when non-TF mutations first emerged. Modified sequence logos [74,75], in which symbol height indicates frequency in a sample, clearly show this progression over time (**Figure 6)**.

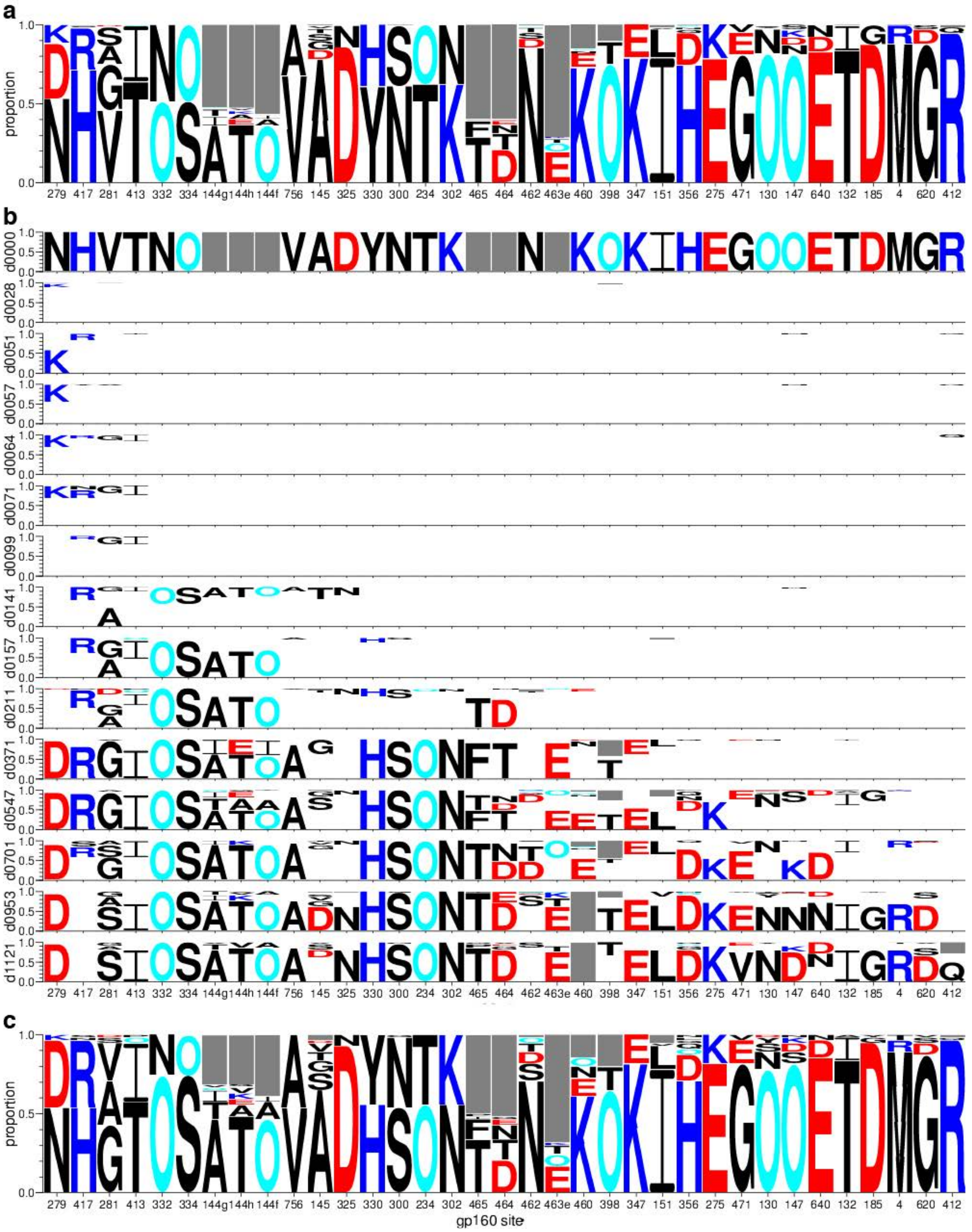

**Figure 6. Variant frequency across 35 sites selected from CH505 Env gp160.** (**a**) Population variant frequencies, computed from all 385 aligned sequences. (**b**) Temporal development of variant frequencies. To emphasize TF loss progression, frequency of the TF form below the first row is blank. Each row corresponds to one time-point sampled for the three-year study period, days 0-1121 (d0000 through d1121). (**c**) Variant frequencies in swarm set of 54 selected Envs. Symbol height is proportional to amino acid frequency per site. Colors correspond to **Figure 3**. The gaps inserted to maintain the alignment appear as grey boxes to represent indels. Site order follows ranks listed in **Table 1**.

The top row (**Figure 6a**) summarizes variant frequencies per site from all 385 Envs sequenced over the first three years of infection in this individual. Below that (**Figure 6b**), rows were stratified to summarize frequency in each sample, first for the TF virus (day 0), then for 14 plasma samples (day 28–day 1121, i.e. week 4–week 160, post-infection). Electrostatic charges of amino-acid side chains, depicted by symbol colors (cf. **Figure 3**), changed polarity in 25% of the gp120 sites (279, 144h, 463e, 460, 347, 356, 275, 147) but not in gp41 sites. Gain or loss of potentially glycosylated asparagines (O) appeared in 13 of the 32 (41%) gp120 sites, but none of the gp41 sites. The swarm of representative sequences selected by the next stage of analysis (Section 3.2) was also depicted in this manner (**Figure 6c**).

Comparing timing of TF loss in the CH505 virus population with neutralization titers assayed longitudinally from contemporaneous plasmas (**Figure S4**) suggests that neutralization breadth follows Env diversification in selected sites. Autologous neutralization is evident at week 14, and heterologous neutralization breadth continues to increase thereafter. TF loss at selected sites starts to emerge at week 4 and net TF loss at all selected sites combined continues to increase until the end of the time-period studied.

### *3.2. Swarm Selection*

We designed and implemented a simple, efficient algorithm to select sequences that represent variants at sites selected by TF loss. The approach is greedy, meaning it adds variants iteratively, rather than refine the entire set for potentially better solutions. Although such a greedy approach is unlikely to give the best possible overall solution, in situations where addition of variants leads monotonically to better solutions with diminishing returns, a greedy algorithm efficiently provides reasonably good approximations to the optimum solution [53], and can be refined to include other criteria as needed. The algorithm starts with the alignment used to select sites, and assumes that sample time-points can be identified from sequence names. A merit of the approach is that, by considering the time of sampling, it starts with sequences most like the form that established the infection. It then progressively builds diversity in a manner that follows the natural course of infection. In this way, common mutations and mutations that eventually reach fixation are sampled repeatedly in varied genetic contexts.

#### 3.2.1. Representative Variants Among Selected Sites

The algorithm identified 54 Envs that covered variant diversity at the 35 sites selected by TF loss. **Table 2** summarizes these as concatamers. The LASSIE algorithm selection criteria had at least two clear consequences. First, the gradual accumulation of mutations found in early infection was deliberately mimicked using this strategy. Second, the appearance of each new mutation of interest is, by design, relatively isolated from other accumulating mutations emerging in the within-host virus population. Therefore, to the extent possible with the given sampling, each mutation in each selected site was expressed in a context as close as possible to the form of the Env in which it appeared when it first began to emerge in the viral population at a level high enough to be sampled. Thus, if a particular mutation conferred a phenotypic change in either antigenicity or neutralization susceptibility of an isolate, then that change would be included for study in its natural context. Selected mutations that appear early, but are retained at later time-points, are resampled together with later variants (**Table 2**).

**Table 2. Selected Envs.** Concatamers (35 sites with at least 80% TF loss) in antigenic swarm of 54 Envs, selected to represent polymorphisms among 385 Env gp160s from CH505. Dots match TF state.

| Name | Accession | Concatamer |
|---|---|---|
| w000.TF | KC247556 | NHVTNO---VADYNTK--N-KOKIHEGOOETDMGR |
| w004.31 | KC247583 | .....................-............. |
| w004.54 | KC247604 | K.................................. |
| w007.8 | KM284749 | KR................................. |
| w007.21 | KM284732 | ..................................Q |
| w007.25 | KM284734 | ............................N...... |
| w007.34 | KM284744 | ...I............................... |
| w008.20 | KM284762 | ..A................................ |
| w009.19 | KM284781 | ..G................................ |
| w010.7 | KM284714 | .N................................. |
| w020.15 | KC247489 | ....OS....T........................ |
| w020.11 | KC47485 | ..........TN....................... |
| w020.24 | KC247495 | .RA......AT........................ |
| w020.25 | KC247496 | .R....ATO.......................... |
| w022.6 | KC247523 | ..AIOSATO...H...................... |
| w022.5 | KC247522 | .RA.OSATO....S..................... |
| w022.9 | KC247525 | ..GIOS.................-........... |
| w022.22 | KM284717 | ...O............................... |
| w030.20 | KC247541 | ..AIOSATOA......TNTO............... |
| w030.17 | KC247532 | D.GOOSATO.......TD................. |
| w030.21 | KC247535 | .RA.OSATO...........E.............. |
| w030.36 | KC247549 | ....OSATO.....O.TD................. |
| w030.26 | KC247539 | .RG.OS....-....NTD................. |
| w030.13 | KC247529 | ..DPOS............................. |
| w030.32 | KC247546 | .RG.OSATO.......TD-................ |
| w053.15 | KC247614 | D.G.OSATOA..HSONFT.E.-.L........... |
| w053.29 | KC247625 | .RAIOSATOA..HSONFT.E.-....E........ |
| w053.22 | KC247620 | DRGIOSIEIAG.HSONFT.E.TE............ |
| w053.8 | KC247632 | DRGIOSATOA..HSONFT.E.T..Q.......... |
| w053.31 | KC247628 | DRGIOSAT....HSON....N-............. |
| w053.9 | KC247633 | DRGIOSIEIAG.HSONFT.E.TE....N..I.... |
| w078.6 | KC247664 | DRGIOSOS.AS.HSONTN.OE-............. |
| w078.36 | KC247655 | DRGIOSTAAAS.HSON..S.O-.-....SD..... |
| w078.9 | KC247667 | DRG.OSTAA.S.HSONFT.E....QK..S...... |
| w078.26 | KC247645 | DRGIOSTAAAS.HSON..S.O.E.NK..S...... |
| w078.29 | KC247647 | DRGIOSTAAAS.HSONTN..-..-L..NS.A.... |
| w078.30 | KC247649 | ..A.OSATOA..HSON....N-......D...... |
| w078.33 | KC247652 | ..A.OSATOA..HSONT.O.N-.....N..I.T.. |
| w078.17 | KC247639 | DRG.OSATOA..HSONFT.EE..LDK.D..IG... |
| w078.15 | KC247637 | DRGIOSATOA..HSON..D..TEL.KES....... |
| w078.27 | KC247646 | DRGIOSATOA..HSONTDD..TEL.KES....R.. |
| w100.T3 | KC247401 | DRGIOSATO..NHSONTDD.ETEL.KEN..I.RS. |
| w100.B10 | KC247386 | DSG.OSATOA..HSONTDD....LDKEN..I.... |
| w100.B2 | KC247387 | .RAIOSIK.AG.HSON....N...D.V........ |
| w100.B4 | KC247389 | DRGIOSATO...HSON..S.O...DKE.K....D. |

| Name | Accession | Concatamer |
|---|---|---|
| w100.A11 | KC247376 | D.S.OSATOA..HSONTNTOE-..D.E.KD..... |
| w100.A13 | KC247378 | ..A.OSATOAV.HSONTNTOE-.......D..... |
| w136.B10 | KC247404 | D.GIOSATOADNHSONTD.E-TELDKES.DIY.S. |
| w136.B5 | KC247429 | ..A.OSATOAV.HSONTESK-.E.O..Y.DI.... |
| w136.B2 | KC247411 | D.G.OSTVAA-.HGONIDOT--E.O.......RD. |
| w136.B23 | KC247414 | D.A.OSIK..G.HSONTEST-..VD....N...D. |
| w160.C1 | KC247465 | ..A.OSVTOAV.HSONTGST-...D..Y.D..TV. |
| w160.T3 | KC247482 | D.SIOSATOA.NHSONTD.E-TELDKVND.IGRD- |
| w160.T4 | KC247483 | D.A.OSTVA.S.HSONPD..-...G...DN..... |

Swarm variant frequencies (**Figure 6c**) resembled variant frequencies sampled in the virus population (**Figure 6a**), with additions of under-represented mutations at selected sites, which were less readily apparent in the larger population. Mutations seen only once among all of the sequences obtained were not required for inclusion, but all mutations in selected sites seen in two or more of all the sequences were represented by the 54 selected Envs. Mutations that occurred only once were not considered, as they are more likely to represent random mutations or possible sequencing error than recurring mutations. Increasing this setting to include each mutation only if it occurs more often will decrease the number of Envs selected.

### 3.2.2. LASSIE Compared with Randomly Selected Sequences

We performed a resampling experiment to evaluate the swarm-selection algorithm against a null distribution, which might be sampled naively by less informed methods. The null distribution was sampled randomly from full-length Envs that had been normalized to eliminate multiple copies of the same Env sequence. Removing duplicates and excluding Envs with premature stop or incomplete codons gave 260 distinct Envs, from which we repeatedly resampled the same number of sequences as in the swarm set (54 Envs) without replacement. **Figure 7** compares the null distribution from resampled results with the algorithmically chosen swarm. By design, in our set of 54 selected Envs, no concatamers were duplicated, i.e. each Env carried a distinct combination of amino acids in the 35 positions of interest, and all recurrent mutations in selected sites were represented. Because the sites represented progressive adaptation of the virus in CH505, we expect each concatamer to have distinct antigenic and/or phenotypic properties, including sensitivity to the coevolving antibody response, which could be identified by assaying each variant against longitudinally obtained plasmas or mAbs isolated to represent a developing clonal lineage (Section 3.2.4). In contrast, among the randomly chosen sets of 54 Envs, redundant concatamers of selected sites were common. Resampling 1,000 replicates gave a median of 40 distinct concatamers with 95% CI from 34 to 45 (**Figure 7a**).

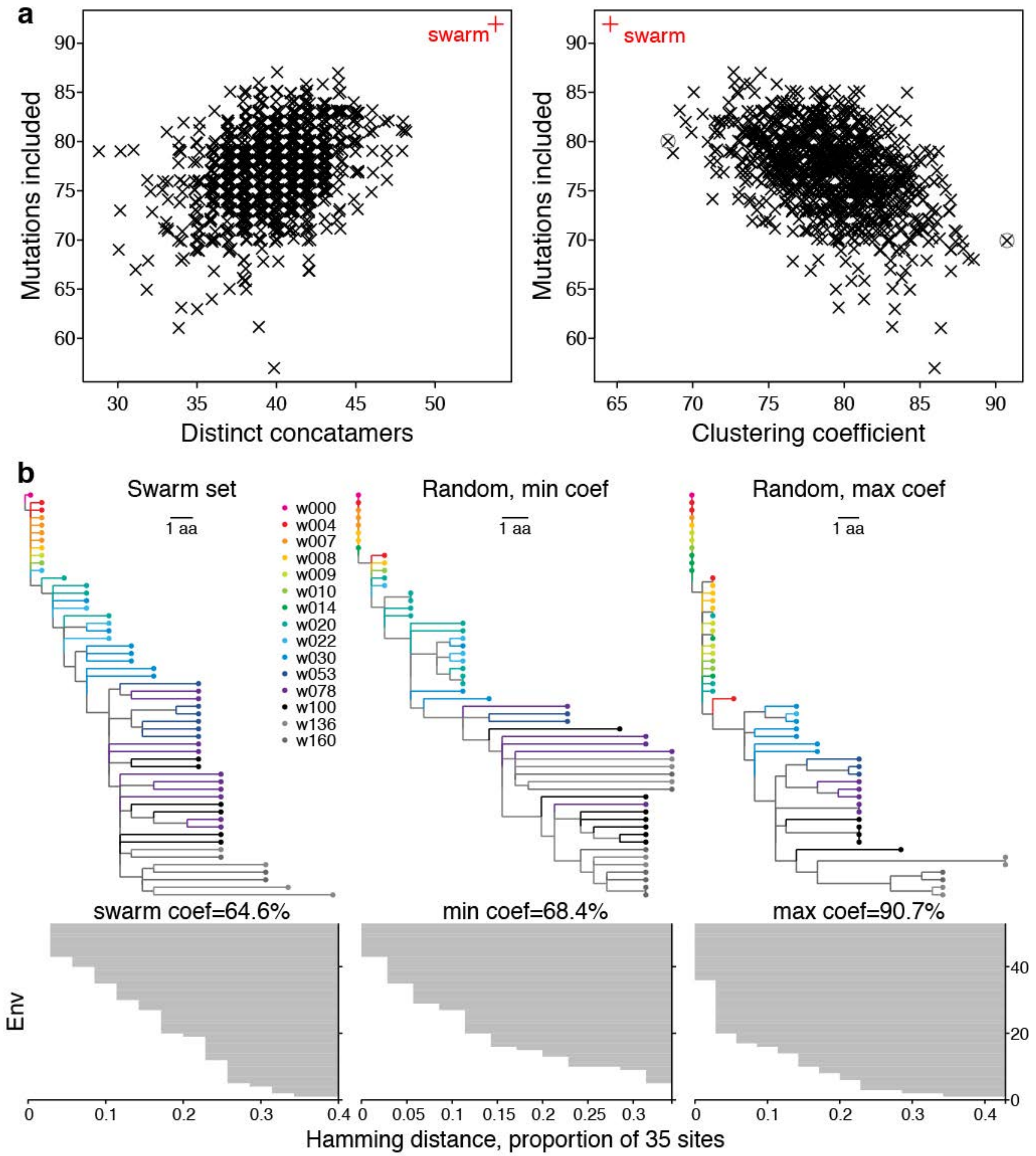

**Figure 7. The selected swarm set is distinct from randomly selected sets.** (**a**) Number of distinct concatamers, mutations included, and clustering coefficients from dendrograms of concatamer distances differ for the selected swarm of 54 Envs (red) and the null distribution from 1,000 sets of 54 Envs, randomly selected without replacement from the non-redundant set of 260 viable full-length Envs, with the TF form always included. Values have jitter added for less overplotting. (**b**) Clustering coefficient quantifies sequence differences among as the average normalized distance at which each sequence is merged into a cluster (horizontal grey bars in bottom row), compared for the selected and swarm set

two extreme randomly sampled sets (min and max, circled points in the right-hand plot in panel **a**, above).

We also compared how many of the non-TF mutations tabulated in the first pass of the algorithm through all 385 sequences were covered. The antigen swarm set selected using LASSIE was designed to cover all 92 distinct mutations that arose in the 35 selected sites. As expected, random sampling of Envs gave consistently lower coverage of the mutations of interest (median 77; 95% CI: 69 to 84) than the 92 mutations that were included by the swarm-selection algorithm (**Figure 7a**). This indicates that random sets of the same size do not capture all of the mutations we consider to have the most potential relevance to immune selection in general, and antibody sensitivity in particular.

Further, we computed hierarchical dendrograms from Hamming distance matrices for swarm and random sets, and summarized the outcomes as clustering coefficients. The results shown here were obtained using the single-linkage method, which is related to the minimum spanning tree [76]. The strength of the clustering can be measured as a dimensionless number between zero and one called the agglomerative coefficient [77]. It is the mean normalized distance at which each sequence clusters with others, and characterizes how well the data are clustered together. To provide an intuition for how this coefficient works, **Figure 7** also shows the dendrogram from the swarm set and compares it with the resampled sets that gave lowest (“min”) and highest (“max”) coefficients. The LASSIE selected Envs had a lower clustering coefficient (65%) than sets of randomly selected sequences, which had a median of 79% and 95% CI: 72-80% (**Figure 7**). The lower clustering coefficient indicates less hierarchical grouping structure, i.e. a more uniform distribution over the available sequence space or lower overall relatedness among subsets of concatamers from the selected Envs, than exhibited by the random sequence sets [77].

These metrics compared sequence sets from the swarm selection algorithm with null distributions that were obtained by random selection. Because the three metrics are only loosely correlated, they measure different aspects of selected sets of sequences. To our knowledge, this is the first attempt to establish criteria to quantify how well any subset of sequences from a larger related set represents diversity (number of distinct concatamers), polymorphisms (number of recurrent mutations included), and progressive divergence (clustering coefficient) of the larger set. We suggest these objective criteria may be useful to choose representative sets of sequences, regardless of how they may be chosen; the smallest set of sequences that maximizes these three criteria, given an ordered list of selected sites, would then be considered the optimal sequence subset when choosing representative reagents.

### 3.2.3. Phylogenetic Context

A consideration of the phylogenetic context of Envs included in the swarm set shows the persistence of selected sites against a scattered background of ephemeral mutations (**Figure 8**).

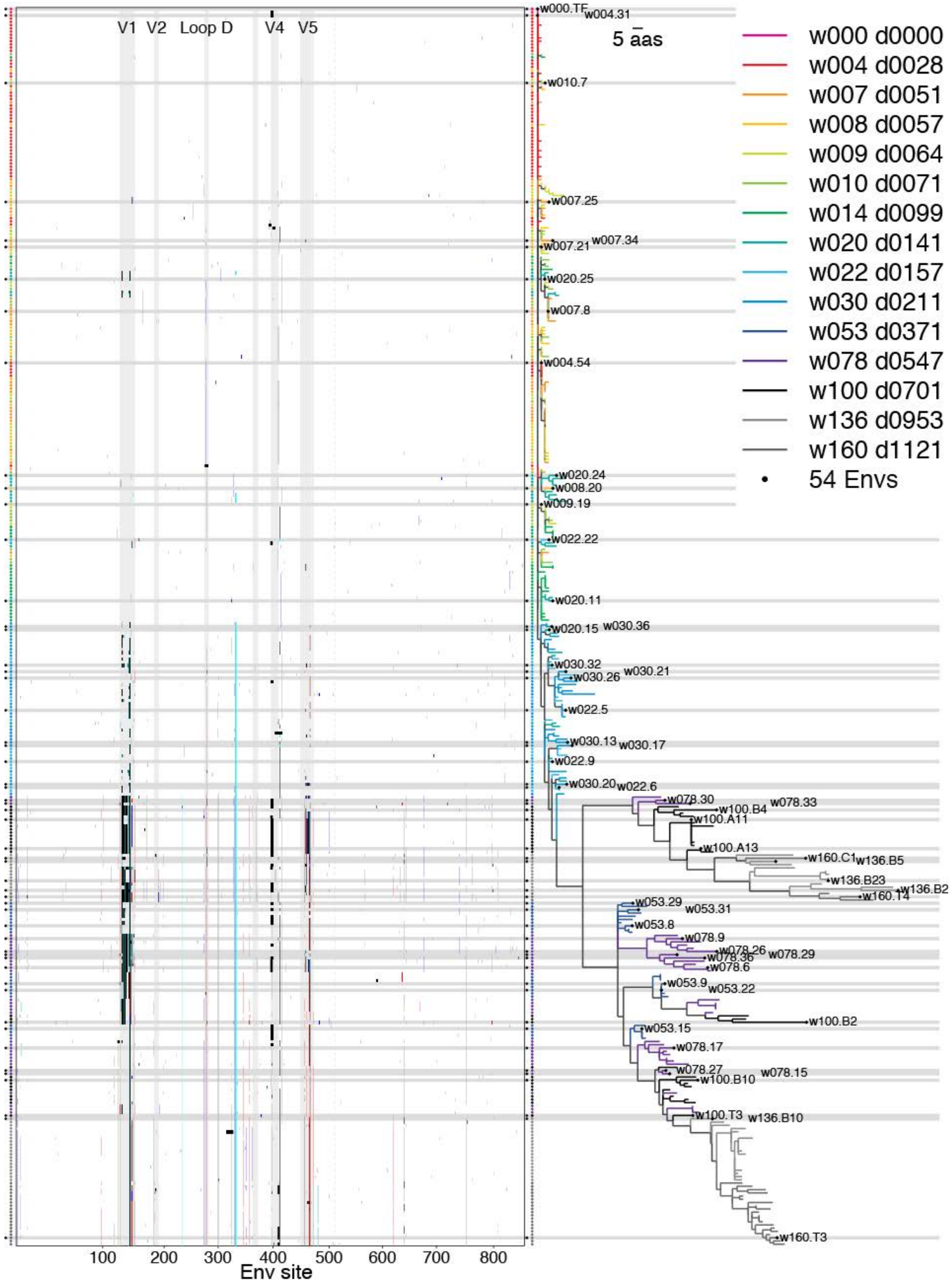

**Figure 8. Env variants in phylogenetic context.** A pixel plot is paired with the maximum-likelihood phylogeny, such that each row depicts one of 385 Envs sequenced by limiting-dilution PCR. The top row corresponds to the TF virus. In the pixel plot (left), sites that match the TF are blank and mutations are shaded indicate gain of negatively (red) or positively charged amino acids (blue), addition of an N-linked glycosylation motif (cyan), indels (black), or other mutations (grey). The colored vertical stripes that emerge with time correspond roughly to TF loss. Env landmarks appear as vertical bands throughout the pixel plot (light grey), and dashed lines delineate the signal peptide and gp41. Tree branches and symbols are color-coded to indicate sample time-point, and the 54 selected Envs are marked by a black circle and horizontal bar.

Selected Envs were widely distributed over the phylogeny. The earliest selected Envs (weeks 4-10) tended to carry single mutations, which are often carried forward and represented again by Envs at later time-points. Some later Envs represented large clades sampled only in one or two time-points, such as the sequence w160.T3 (KC247482), which appears almost at the bottom of the tree (**Figure 8**).

### 3.2.4. Antigenic Diversity

We opted to work with data from subject CH505 for initial LASSIE development and testing not only because a large set of longitudinal sequences was already published and available, but also because roughly one-fourth of the Env sequences from CH505 had been hand-selected and previously expressed as pseudovirus and proteins, then subjected to immunological evaluations [15,16]. These available data provided an opportunity to explore the antigenic diversity and neutralization sensitivity represented by a large subset of the antigenic swarm that LASSIE selected. Of note, the sequences we had originally used to analyze the CH505 immune response had been chosen manually from the phylogenetic tree, a practice typical in this field. Combined uncertainty about how well the manual selection covered all immunologically relevant mutations in CH505, how effectively it recapitulated the gradually emerging resistance to evolving antibody clonal lineages, and a desire to minimize the expense of building very large subject-specific Env reagent sets in future studies, all motivated development of LASSIE. LASSIE provided a principled solution for the first two of these biological problems, coverage and mimicking *in vivo* antigen evolution, while defining a minimal set of Envs to achieve these ends, and minimize experimental costs.

ELISA binding assays with mAbs from the CH505-derived CH103 CD4 binding site bnAb B cell lineage were available for 94 gp120s; a subset of 27 LASSIE-selected Envs were included in this set (**Figure 9**). Binding assay results confirmed that selected viruses exhibited diverse antibody sensitivities, which increased with maturation of the bnAb lineage and generally followed the progression of mutations away from the TF virus (**Figure 9**).

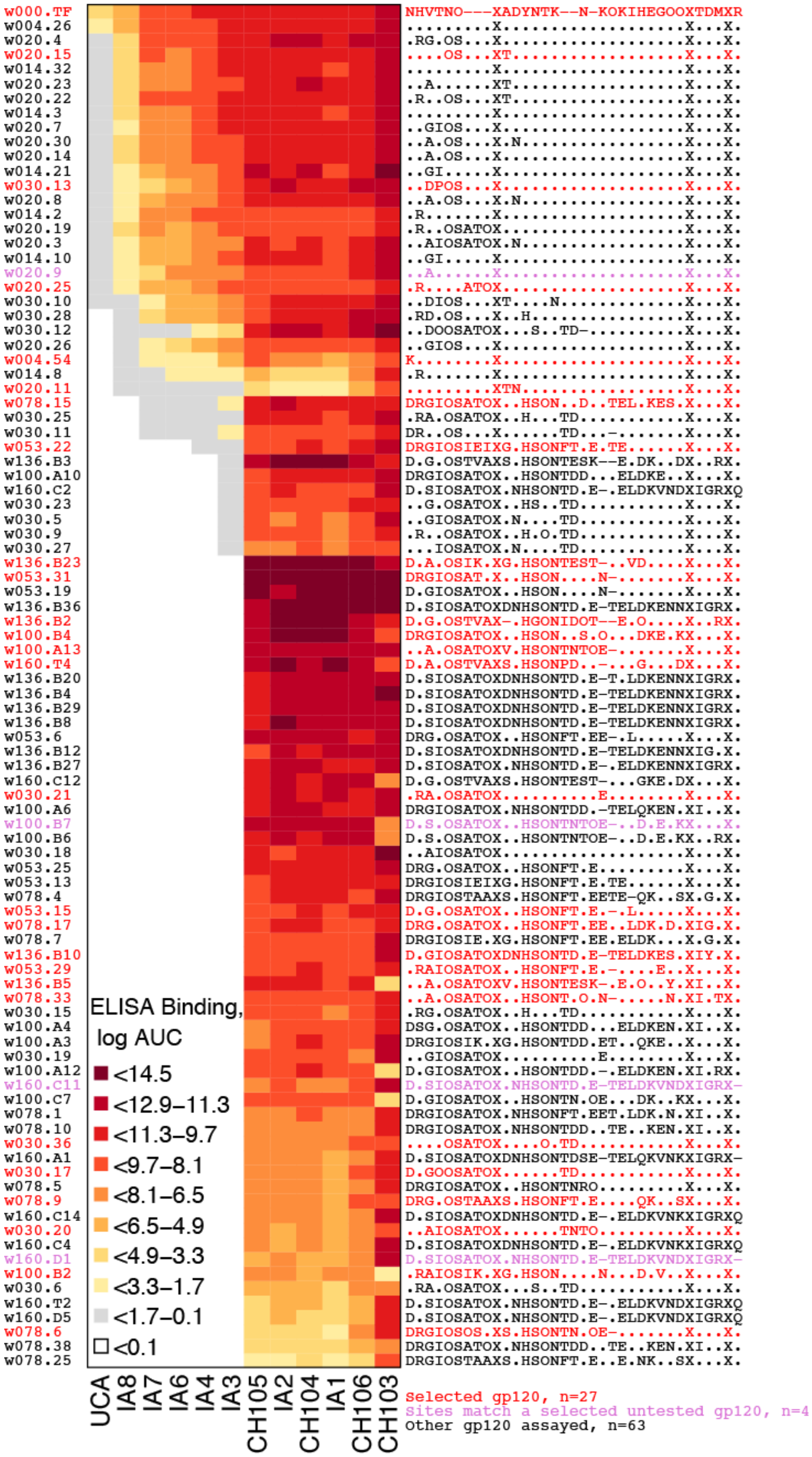

**Figure 9. Selected Envs represent diverse binding phenotypes.** Among the swarm of 54 Envs selected, 27 were synthesized as gp120s for ELISA binding assays (red text). Another four of the antigens tested contained selected sites that matched with those in selected Envs (purple text). Binding data are shown as colors to indicate log-transformed area un-

der the curve (AUC) from dilution series, which summarized experimental results better than EC50s. Both assays tested Env constructs against monoclonal antibodies of the CH103 lineage, from mAb isolates (e.g. CH103) to the unmutated ancestor (UCA) via intermediate ancestors IA1-IA8 [15]. Blank entries indicate no binding was detected. Selected Env sites correspond to concatamers in **Table 2**. An “X” appears for gp41 sites, which were not in the gp120 antigens tested. Data are listed in **Table S3**.

In a similar manner for neutralization sensitivity, 26 LASSIE-selected Envs were among 121 Env-pseudotyped viruses tested for neutralization sensitivity by CH103 lineage mAbs [16] (**Figure 10**). Selected Envs represented the range of sensitivities among viruses tested, reflecting the diversity of variants that developed in response to sustained selection for neutralization escape.

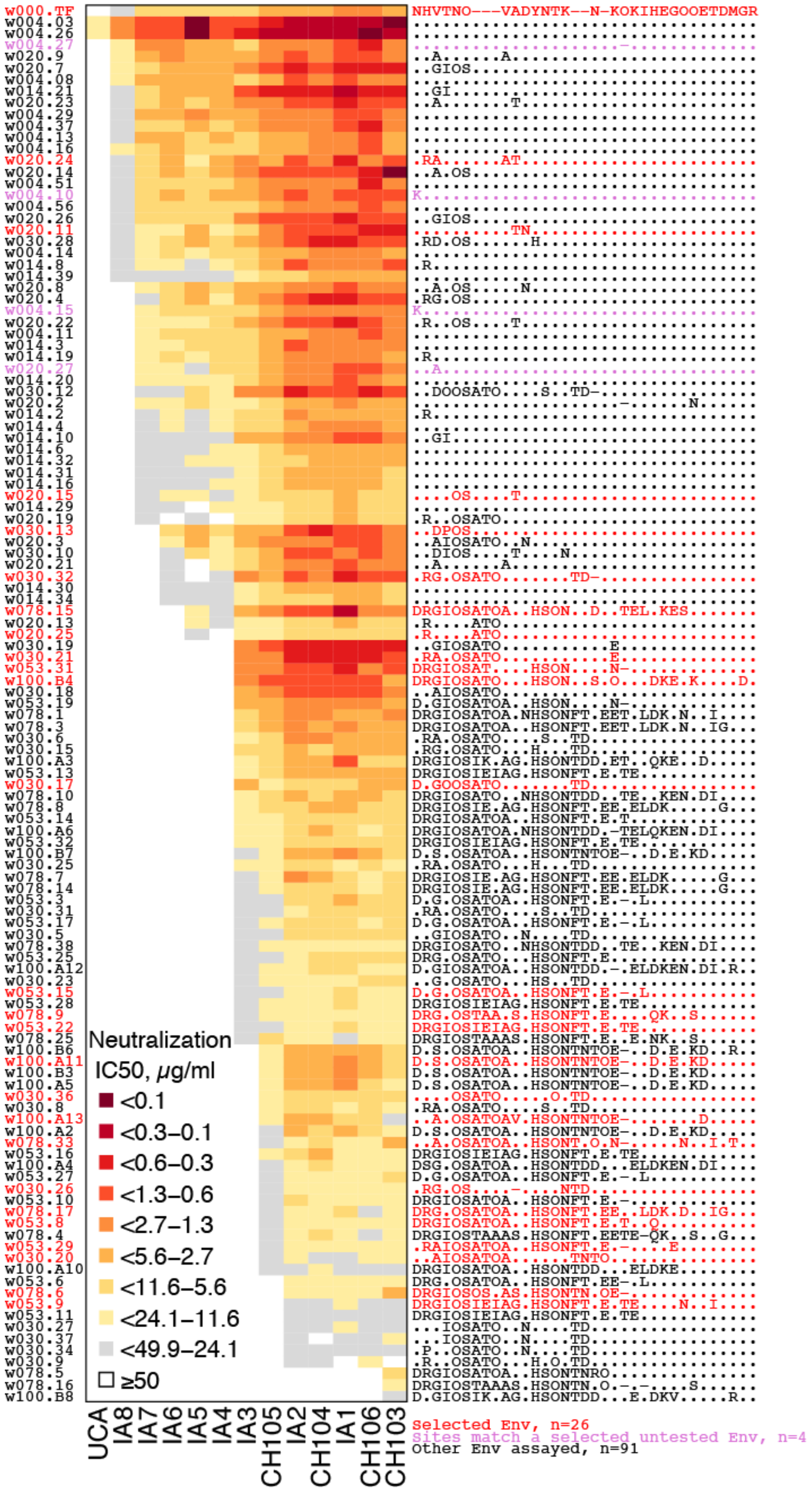

**Figure 10. Selected Envs represent diverse neutralization phenotypes.** Among the swarm of 54 Envs, 26 were cloned into pseudovirus backbones for TZM-bl neutralization assays (red text). Another four of the Env-pseudotyped virus constructs tested contained selected sites that matched with those in selected Envs (purple text). Neutralization IC50s

are represented as colors to indicate sensitivity of each virus to neutralization by each mAb in the CH103 lineage. Selected Env sites correspond to concatamers listed in **Table 2**. The values appear in **Table S4**.

In **Figure S5** (in the Supplement), neutralization titers from all previously hand-selected viruses clearly show the development of neutralization breadth in phylogenetic context. Envs at the top of the tree are broadly susceptible to many antibodies in the CH103 lineage. Envs that evolved later appear lower in the tree. Neutralization breadth was acquired later in bnAb ontogeny, which is clear as a gradient of increasing potency from the unmutated ancestor (left) to the mature CH103 bnAb (right). By selecting Envs that represent genetic diversity sampled during bnAb development, the method selects Envs that represent relevant antigenicity over time.

### *3.3. Swarm Size Adjustments*

We have used the LASSIE approach to analyze samples from two individuals with larger data sets that are currently in preparation. To reduce the number of Envs for inclusion in reagent design below 100 from over 1000 initially sampled sequences (as compared to 385 sequences sampled from CH505) required that we increase the TF loss cutoff above 90%. Selecting an antigenic swarm of fewer than 100 sequences also required increasing the frequency of rare amino acids for representation in the swarm. Both of these adjustments focused outcomes on sites that ranked highest in terms of selective pressure (TF loss threshold) and the mutations within these sites that are most successful under host immune pressure (minimum variant count).

Another way to reduce the number of sites and sequences that result from LASSIE is to exclude sites that were insertions relative to the TF sequence, which occurs most frequently in the hypervariable loops. Due to the evolutionary processes that yield length polymorphisms, rather than point mutations, and the resulting difficulty of consistently aligning homologous sites in these regions, we naturally found disproportionately high diversity of V1, V2, V4, and V5 sites among selected Envs. Resources to produce reagents that capture each recurring point mutation in the hypervariable loops might be better allocated elsewhere. Excluding such sites from the list of selected sites used to choose sequences reduced the number of sequences, while still representing diversity of envelope regions with well-defined structure. However, because these length polymorphisms in variable regions are known to be involved with immune escape (e.g. [73]), representation of the most common forms found in the hypervariable regions should be still included in swarm sets. When exploring the option of excluding TF insertions from the selection criteria, we found that different forms of hypervariable loops were indeed still included in the antigenic swarm, but their representation was of course limited to evolutionary contexts in which they occurred together with sites that have high TF loss. When considering this alternative, one should ensure that the common forms of hypervariable-region variants, such as we have listed in **Tables S1** and **S2** as having the greatest numbers of times observed, are included among the antigenic swarm identified by LASSIE. If not, they might be added as specific additional sites or sequences to the reagent set (see Methods).

### *3.4. Chronic Infection*

These methods were developed initially to select sequences from longitudinal studies beginning early in infection, where the TF virus is reliably inferred, and the progression of escape mutations is readily apparent. This is obviously not true for chronic infection. Still, it is often necessary to select a subset of sequences that represent diversity in serial samples taken during chronic infection. To evaluate the algorithm's ability to select an antigenic swarm from a chronic infection, we applied it to sequences from a study participant enrolled during chronic infection, designated CH0457 [56]. We analyzed 205 plasma SGA Envs from ten sample time-points (median was 20 sequences per time-point; the distribution ranged from 12 to 35). In the chronic enrollment sample, the first available, five of twenty Envs exactly matched the within-time-point consensus. We used one of these as the reference to compute variant frequencies. No variation was detected in 582 of 888 aligned sites, and an 85% cutoff identified 35 sites that were candidates for strong positive selection (**Figure S6**). Nine of the 35 sites are located in gp41.

With singleton variants excluded, the algorithm selected a swarm of 44 Envs (**Figure S7**). The progressive accumulation of mutations among concatamers of selected sites is less clear in this chronically infected subject than in acute infection (cf. **Figure 6b**). Furthermore, sites that appear to be under selection in the interval sampled are not clearly associated with two epitope regions, as was the case of CH505, where there was a strong imprint of CD4bs and V3 antibodies selection, and indeed antibodies with these specificities were isolated from the subject. In the case of CH0457, most of the selected sites were not identified as relevant to known antibodies, although two sites were in the MPER region of gp41 (HXB2 positions 667 and 671) and two sites were predicted signatures of the 2F5 MPER antibody (positions 640 and 351). In addition, one site was in contact with some CD4bs antibodies: a changing glycosylation pattern at 461, which contacts CD4 and the CD4bs bnAbs VRC01 and NIH45-46. Two of the selected sites (651 and 640) have been noted to be CD4bs antibody signatures [67]. A potent CD4bs bnAb CH27 was isolated from subject CH0457, but the virus isolated from CH0457 plasma had escaped from this antibody by the time of enrollment [56]. However, archived provirus from CH0457 cell-associated DNA remained sensitive to neutralization by bnAb CH27 [56]. CH13, a weaker CD4bs nAb capable of neutralizing only heterologous tier 1 viruses, was isolated, and may have been exerting selective pressure in the last weeks sampled [56].

The phylogeny indicated a persistent, divergent secondary clade, represented by 24 of 205 plasma Envs (**Figure S8**). This clade was not introduced by misalignment nor by simple recombination, and was also represented by cellular provirus sequences [56]. Though the divergent clade was undetected among sequences from the enrollment sample, it was represented by 14 of the 44 Envs selected (**Figure S7**). Thus, the algorithm can be applied to both acute and chronic Env sequential sequence analysis and swarm design.

## 4. Discussion

Vertical stripes of mutations in an aligned set of serially sampled sequences, as shown in **Figures 2** and **8**, indicate sites where the transmitted-founder amino acids are lost over time, a useful criterion to identify candidate selected sites. We have presented a systematic approach to identify these sites, track their dynamics graphically, and then identify protein sequences that capture mutations in the selected sites as they first emerge in the quasispecies. The task of selecting representative variants from a larg-er set for follow-up studies from longitudinal samples can be complex when choosing from hundreds to thousands of sequences. LASSIE divides the task into two main parts, automatically identifies and tracks selected sites within a subject, and identifies sequences that represent antigenic diversification in that subject.

First, we use transmitted-founder loss in a longitudinal study as a simple way to identify sites under positive selection pressure. Despite the existence of a variety of methods to test for positive selection by comparing rates of synonymous versus non-synonymous substitution, their utility to identify sites under positive selection in the limited diversity context of within-subject viral evolution is limited by statistical power [44,78]. For example, the strong selective pressure may be exhibited by a single change in a phylogenic reconstruction, as illustrated in **Figure S1**. This could happen with a population bottleneck, or an advantageous mutation or mutations that are strongly favored being carried forward through recombination. Also, allele frequencies can be biased by removing identical sequences in order to obtain dichotomously branching trees. Identical seuquences are unlikely to occur among samples from different hosts, but commonplace when sampling from acute through early infection within a host [5,15,16,46,47].

In contrast, loss of the TF form at any time-point is a simple and inclusive measure. In CH505, sites selected by this criterion were focused in regions highly relevant to the known adaptive immune responses previously identified in the subject [15,16]. This suggested that in future studies, structural localization of selected sites could be used to raise hypotheses about specificities of bnAbs in plasma. Furthermore, the timing of TF loss identifies these important mutational events and could help determine when antibodies are exerting the most selective pressure, indicative of which samples in a longitudinal series are likely to yield antibodies with particular specificities, and when the earliest members of a clonal lineage that target a particular epitope first appear. Such information could help efforts to isolate monoclonal antibodies in subjects with potent nAbs, by focusing on antibody specificities that recognize the epitopes under selection, and by identifying which samples might best be used to isolate new bnAbs from subjects sampled over several years of follow-up.

Admittedly, the TF loss criterion alone cannot distinguish between sites that represent a random sweep, where a mutation is established in a lineage by chance, and selected sites. This is why we refer to the set of sites LASSIE identifies as *candidates* for sites under positive selection. For example, LASSIE could identify a neutral mutation genetically linked to another mutation under positive selection. We aim to be inclusive and evaluate immunological phenotypes of such patterns in our swarm sets, to ascertain which mutations are indeed escape mutations. In this way, LASSIE is designed to minimize type II errors, even with a small cost of increased type I error. Also, even if a mutation were neutral when it first arose, and subsequently maintained in the population, it may represent a pre-adaptive state that allows novel escape mutations to accrue that depend on the presence of the initally neutral mutation. LASSIE "swarms" bring each new mutation into the swarm in the Env genetic context in which it was first sampled, in the simplest form possible, relative to the evolving quasispecies. Thus, subsequent immunological experiments can identify immune-sensitive phenotypic profiles between Envs with minimal amino-acid changes between them. This helps to isolate the candidate sites under pressure from immune selection, to the extent possible in the context of appropriate Envs.

Second, we provide a rational, objective method to guide the selection of Env sets for experimental study from large sequence sets sampled over time. LASSIE can select sets of sequences that represent gradual antigenic diversification induced during bnAb development, ensuring that all variants in sites identified by TF loss are represented in an Env reagent set, while minimizing redundancy, and selecting only as many variants as are necessary to represent diversity in sites selected by TF loss. For a set of aligned sequences sampled from an acute, single-founder infection, the algorithm starts with sequences most like the form that established the infection, and gradually increases diversity in a manner that parallels natural infection. Though a smaller set of sequences could be defined that minimally cover all variants at selected sites, the gradual accumulation of mutations would not be captured by such minimal set, and more subtle transitions may be critical for selection of bnAb breadth, i.e. evolving antibodies may adapt more readily to serial introductions of single mutations in an epitope, rather than to those same changes introduced simultaneously.

We used LASSIE to identify selected sites and representative sequence subsets in longitudinal samples from one acutely infected subject and one subject sampled only during chronic infection. We analyzed SGA sequences, which provide intact *env* gp160 genes with no recombination artifacts and minimal error [5,14,21,45-47]. While such data provide ideal conditions, the approach could also be used in other longitudinal study designs and sequencing strategies (e.g. [41]).

In related research, sequence selection has been represented as a set-coverage problem [79], to identify networks of covarying sites in a population-level alignment, which represents a particular clade [80], rather than a serially sampled, within-subject alignment as in this study. A limitation of our approach, which we intend to address in the future, is that sites are treated independently, while covariation between sites may influence variant suitability and TF loss. Considering covariation may potentially facilitate identification of smaller representative swarm sets, and may better accommodate rampant recombination between divergent lineages within a host.

By progressively adopting mutations in the context of variant sequences where they first arise in our sequence sets, our swarm sets – by definition – allow the study of mutations in the context of the natural combinations of mutations as they occurred *in vivo*. This strategy could complement, or in some instances provide sufficient information to replace, traditionally used site-specific mutagenesis, which necessarily studies mutations in isolation. As mentioned above, a mutation observed in a later timepoint and introduced into the TF, for example, may not be viable or may not have the same phenotypic consequences as it does in the background of the Env in which it arose, so the ability to study related natural variants isolated serially may ultimately be more informative.

A working hypothesis to explain the observation that bnAbs tend to arise late infection, after antigenic diversification has arisen in the subject, is that serial immune escape *in vivo* drives antibody lineages to adapt to the emerging viral variants, and eventually enable recognition of diverse forms of the targeted epitope from the circulating population [16,18]. Exposure of a neutralizing antibody lineage during affinity maturation to increasing antigen diversity could result in selection of antibodies with increased breadth [15,18,31,37,38]. Mimicking *in vivo* diversification has therefore been proposed as a possible vaccination strategy for bnAb induction [15,18,81-83]. Related work has suggested that Env variants sampled during development of heterologous neutralization breadth could be administered as immunogens [84-86].

With recent technological advances, it is becoming feasible to test vaccine designs that not only include 5-10 antigens, but potentially between 50-100 antigens, administered as DNA either in series or in combination [87,88]. Because LASSIE uses an efficient algorithm to identify candidate sets of antigens with progressively increasing diversity at important sites in polymorphic viral proteins, it could be used to help with the design of such antigen-swarm vaccines. This method could be applied to other large, longitudinally sampled sets of sequences, such as from hepatitis C virus [89-93]. Use of this method to analyze antibody sequences, which complement the evolving viral sequences, could identify selected sites and select a representative subset of sequences from antibody clonal lineages.

## 5. Conclusions

In summary, we have developed computational methods to identify and track selected sites in longitudinal sequence data, and to use these selected sites to aid in down-selecting sequence sets for reagent design, or for testing the "antigenic swarm" vaccine concept. When applied to longitudinal HIV samples, a retrospective evaluation of viral sequences from the intensively studied subject CH505 showed that the LASSIE provided meaningful results. High TF loss highlights mutations in selected sites that were indeed under immune selective pressure. An efficient algorithm builds a non-redundant collection of sequences tailored to characterize the phenotypic consequences of the mutations in those sites. LASSIE may be useful in many contexts, for reagent selection, to assist with bnAb isolation, potentially for vaccine design, as well studies of other viral infections, and studies of antibody evolution.

## Supplementary Materials

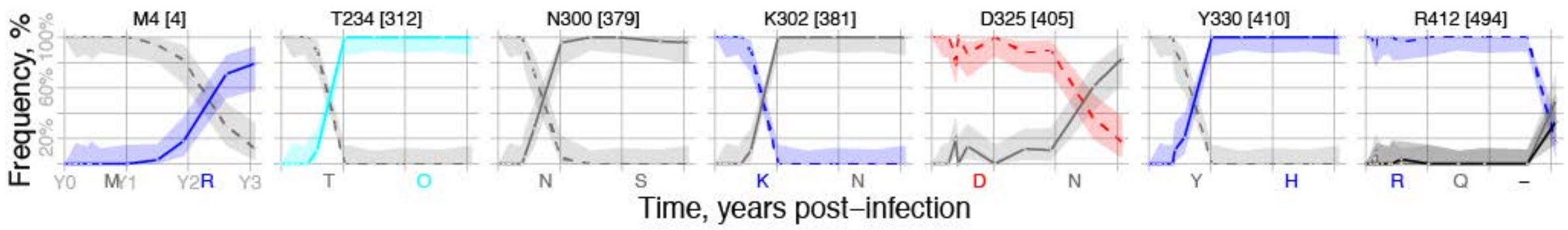

**Figure S1.** Sites with high TF loss not detected by MEME. The 7 sites selected with over 80% peak TF loss that were not selected by MEME (q-values above 0.2 and p-values above 0.1) show abrupt transitions from the TF to a mutated amino acid, described as dynamic category *i* in the Section 3.1.1 of the main text. For clarity, mutations that never attain 15% frequency in any sample are not shown.

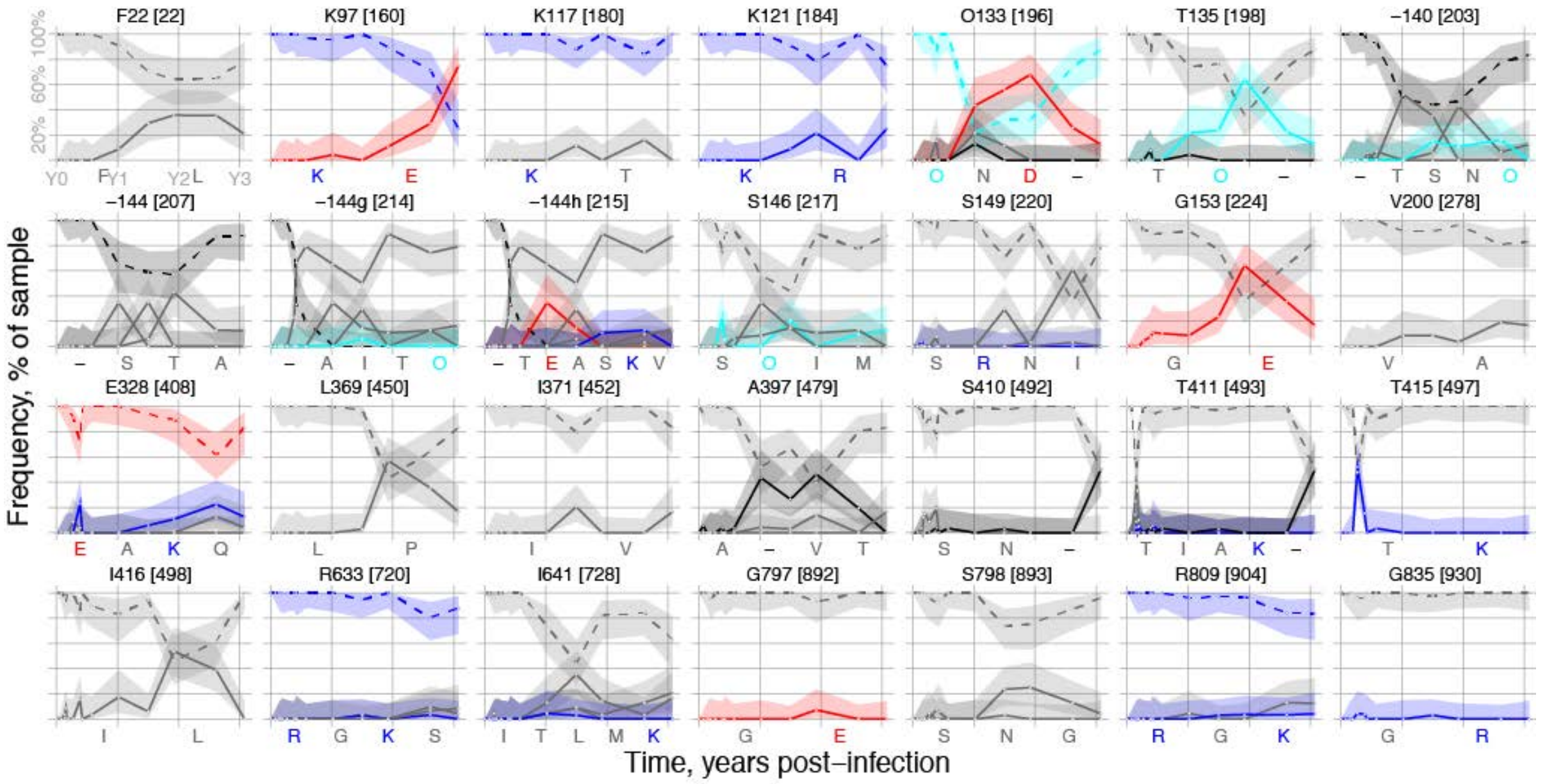

**Figure S2.** Sites without high TF loss detected with positive selection by MEME. The 28 sites selected by MEME with q-values below 0.2 and p-values above 0.1, which were not selected by the 80% peak TF loss criterion, included dynamic categories *i*, *iii,* and i*v* described in Section 3.1.1 of the main text, and included at least 4 sites without dynamics.

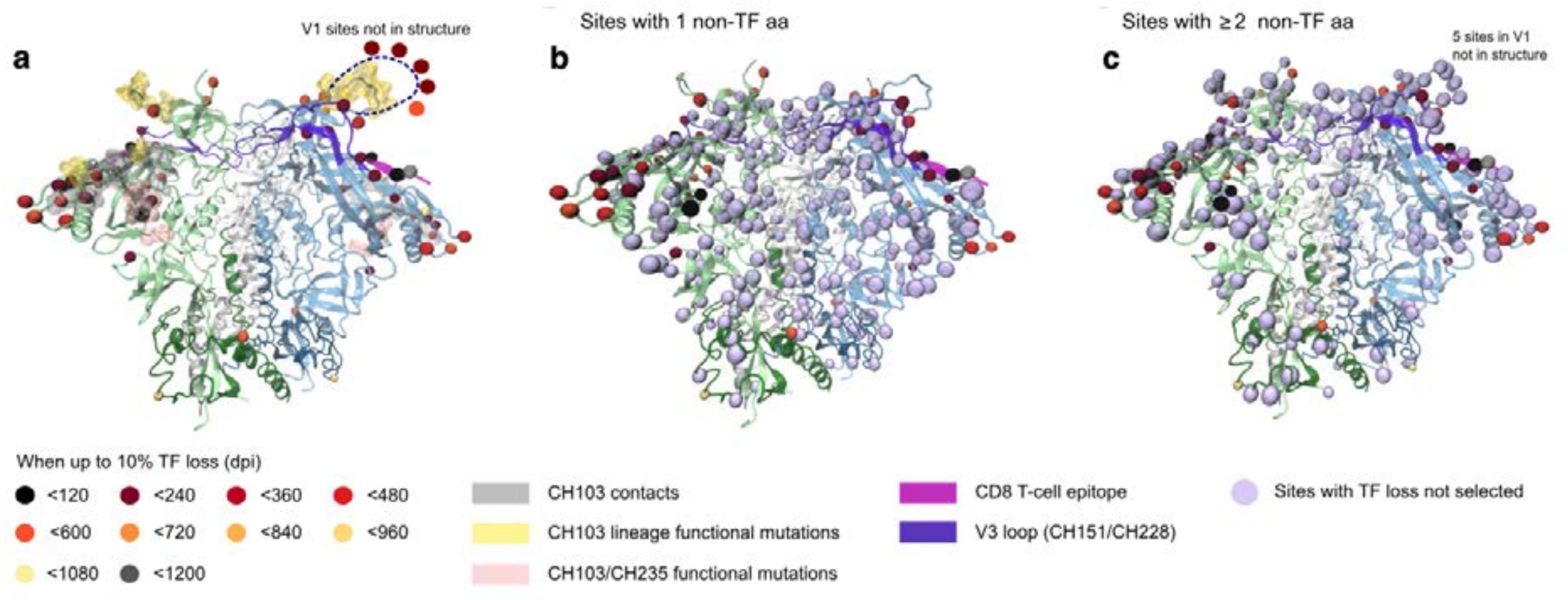

**Figure S3. Locations of selected and non-selected Env sites in CH505.** (**a**) Sites selected by high TF loss are depicted by beads, whose colors indicate when each site exceeded 10% TF loss, as listed in Table 1. For structural context, the immunologically relevant mutations and regions described in Fig 4 are also shown. V1 sites missing from the structure are illustrated schematically (top left). (**b**) Sites that mutated only once among 385 CH505 Env sequences (0.25% TF loss over all time-points), and present in the structure, are identified by lilac beads. These sites were not selected, due to low TF loss. (**c**) Sites with two or more mutations among 385 CH505 Env sequences (at least 0.5% TF loss over all time-points), but less than 80% peak TF loss, are marked by lilac beads.

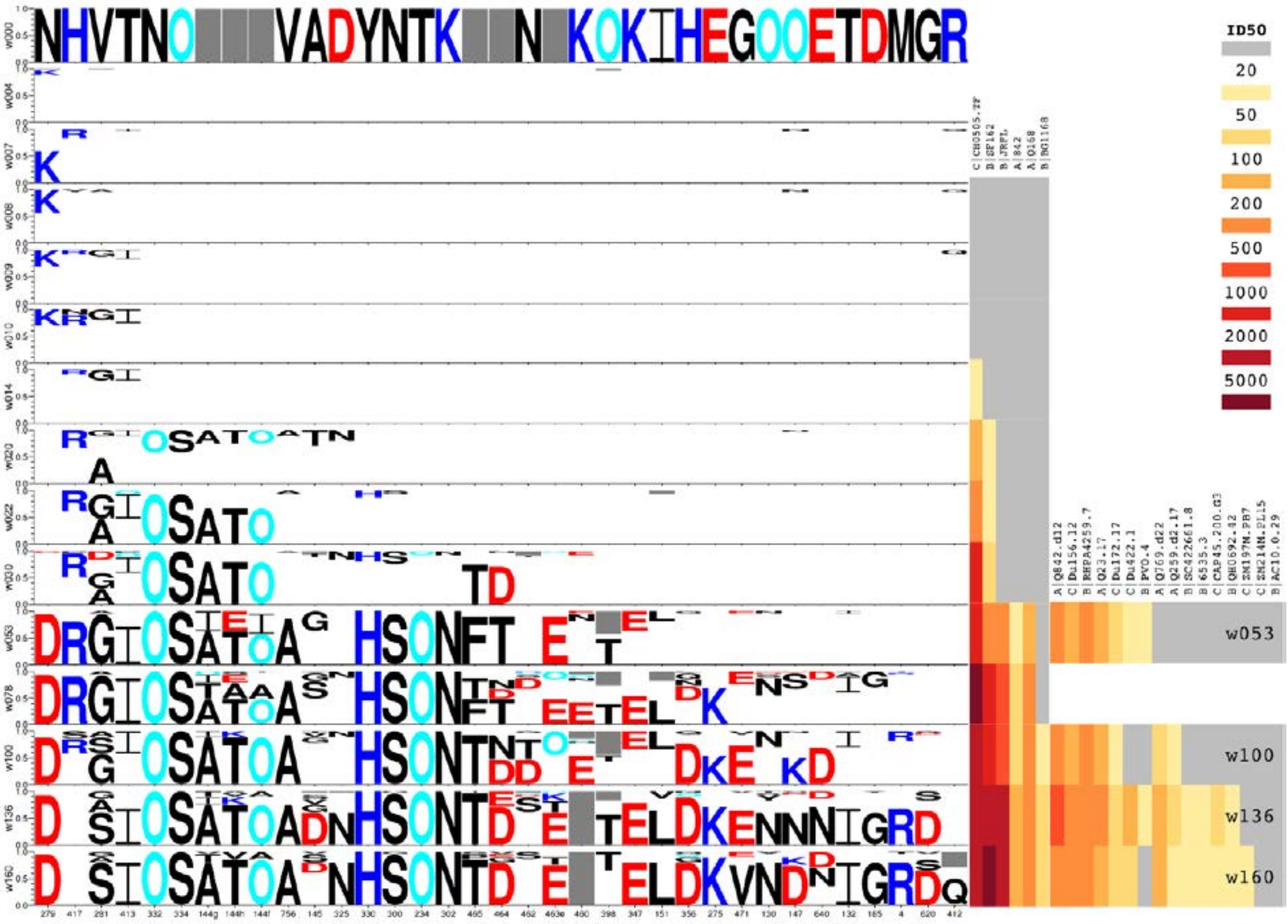

**Figure S4. Comparative timing of TF loss and neutralization breadth.** Sample times (weeks post-infection) increase from top to bottom, and align sequence logos (left) from **Figure 5b** with neutralization titers from contemporaneous plasmas tested against autologous virus (CH0505.TF), a panel of five Tier 1 envelope pseudotyped viruses (SF162 through BG1168), and sixteen Tier 2 viruses (Q842.d12 through AC10.0.29). ID50 titers below the sensitivity limit of 20 μg/ml are indicated by grey, and darker red colors indicate greater neutralization potency. The clade of each virus (A, B, or C) appears before its name.

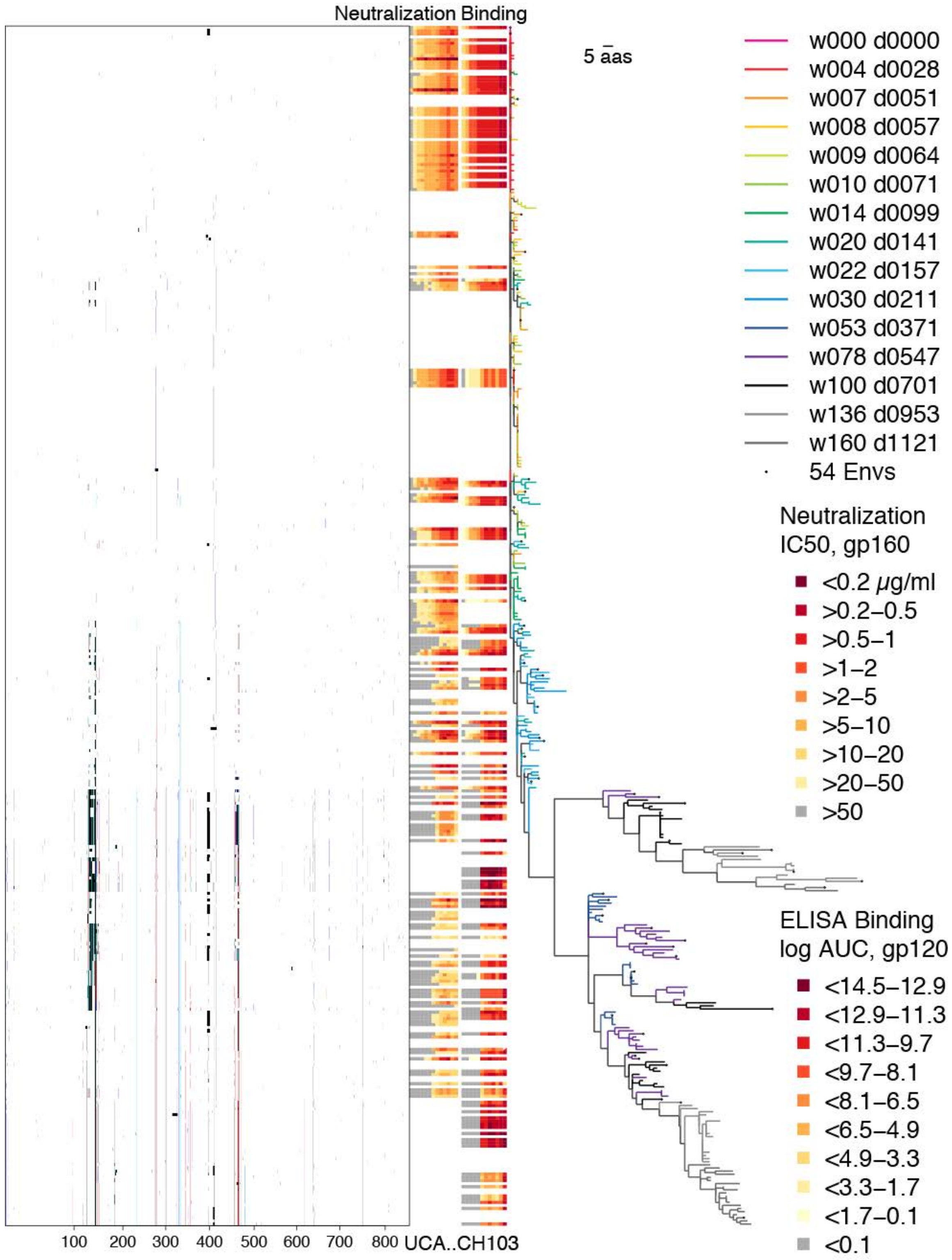

**Figure S5. CH505 Env variants and antibody development.** Neutralization IC50 titers and ELISA log-AUC binding affinities of hand-picked envelopes from assays against the CH103 bnAb lineage are juxtaposed with the Env polymorphism and phylogeny shown in **Figure 8**. Column order for bnAb titers follows that shown in **Figures 9** and **10** (from left to right: UCA, IA8, IA7, IA6, IA5, IA4, IA3, CH105, IA2, CH104, IA1, CH106, CH103).

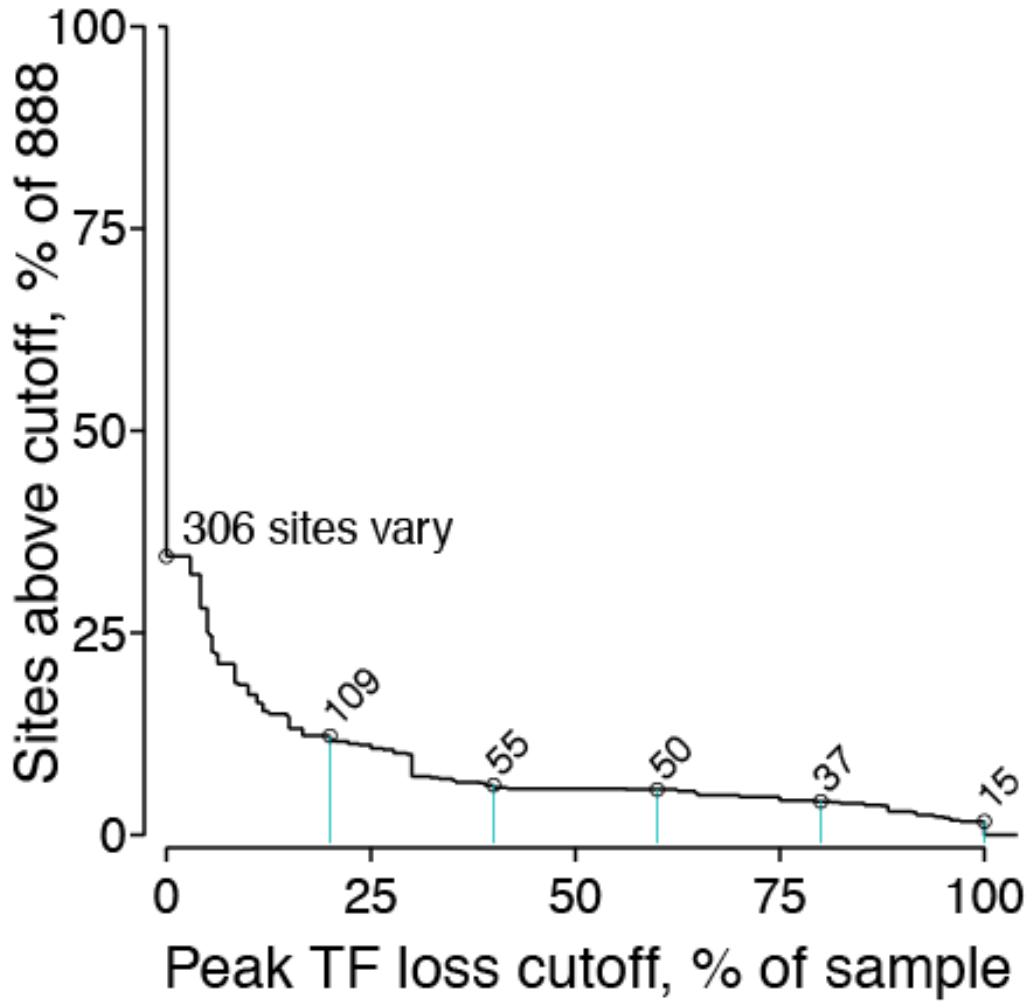

**Figure S6. Number of sites varied with cutoff in chronically infected donor CH0457.** Increasing the cutoff decreased number of sites.

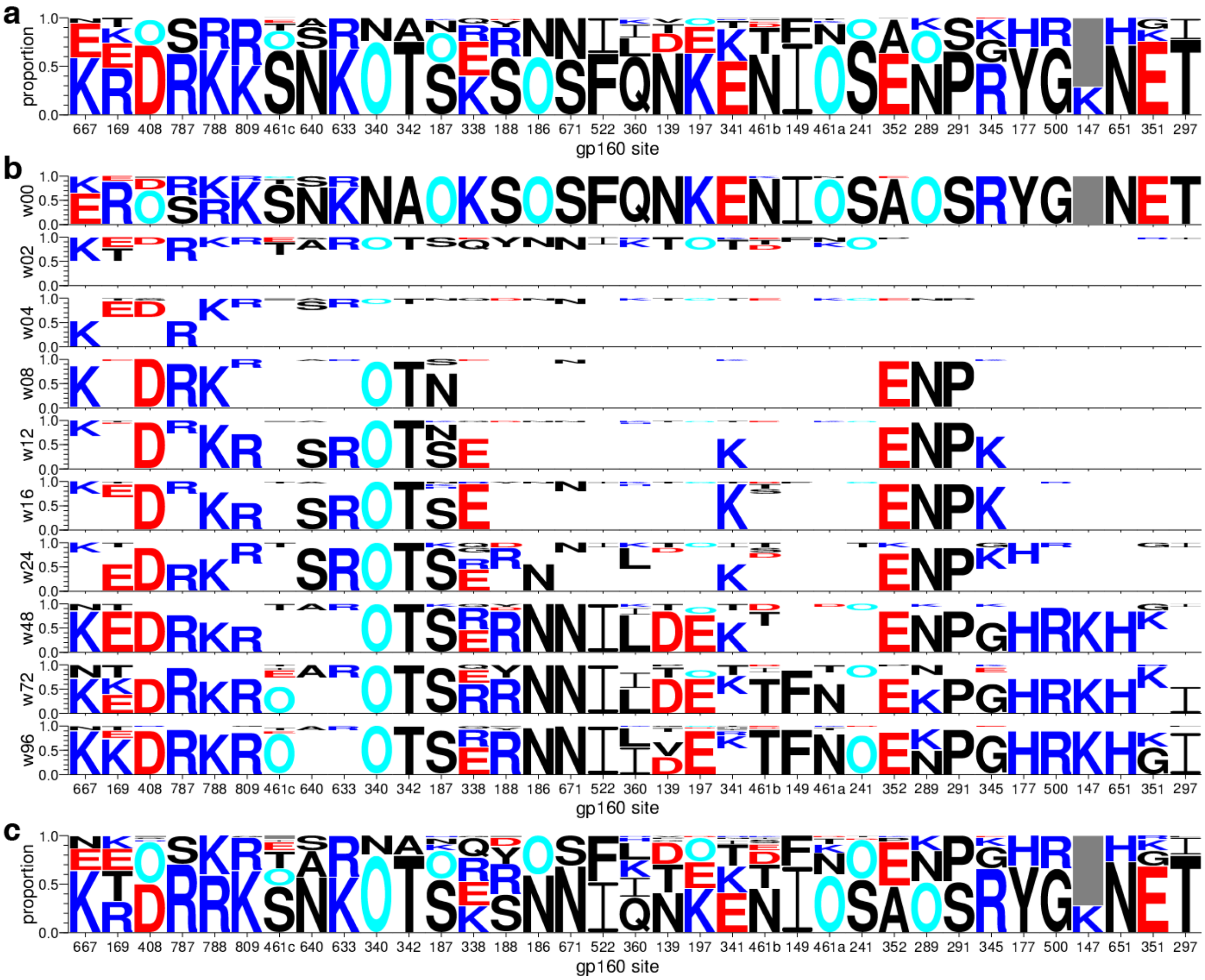

**Figure S7. Variant frequencies among selected sites in chronic infection.** Frequencies from CH0457, computed among (A) all sequences, pooled; (B) sequences stratified by time; and (C) 44 selected Envs. Colors indicate positive (blue) and negative (red) charges and "O" indicates a potentially gylcosylated asparagine (cyan). Indels appear as grey boxes.

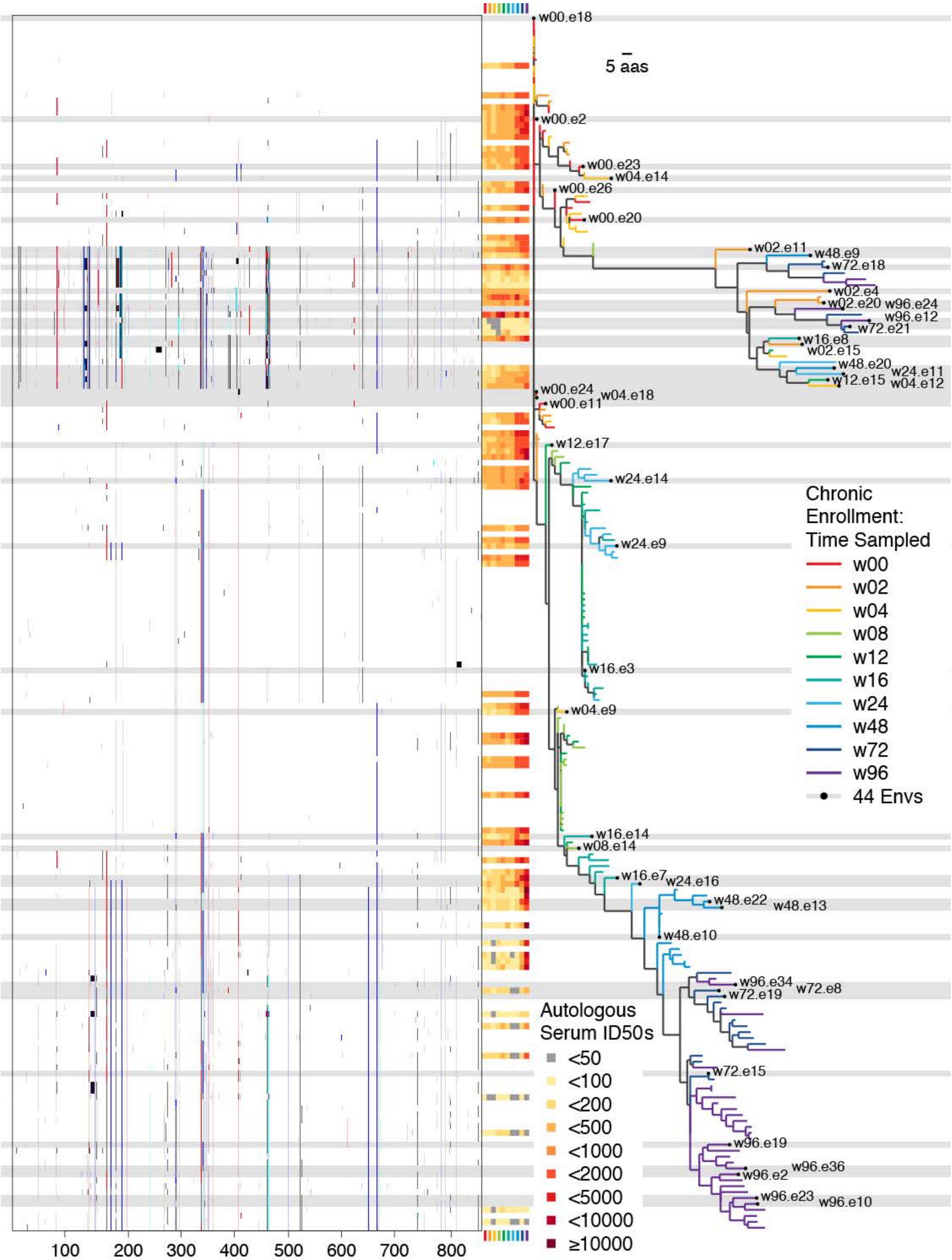

**Figure S8. Diversity in chronic infection.** CH0457 Env mutations (left), neutralization ID50 titers against autologous contemporaneous plasmas (center), and maximum-likelihood Env phylogeny (right). The 44 selected Envs are emphasized among all Envs

sampled. The divergent clade appears above the "Time Sampled" legend. This representation follows **Figure 8**, with neutralization titers, one column per time-point. Neutralization responses were profiled for 84 Env-pseudotyped viruses chosen before the swarm-selection algorithm existed, and tested against autologous sera from each time-point sampled.

**Table S1. V1 hypervariable region.** This table shows variants in HXB2 positions 136 through 145, codons 199-215 in the CH505 alignment, a region not present in the transmitted-founder virus. This region accepted many distinct insertions, up to 17 amino acids long, often followed by deletions within the insertion, or subsequent mutations, throughout the period of the infection sampled. We found 29 distinct forms in this region. An example of each unique sequence selected from the first time-point in which each pattern was observed, with a representative sequence named, which includes the time of isolation, in weeks post-infection. The V1 loop noted in the Los Alamos database spans positions HXB2 132-153. In subject CH505 indels in this broader region are almost certainly also impacting evolutionary patterns. Dashes maintain alignment consistency.

| Example sequence | V1 Insertion | Number of times observed |
|---|---|---|
| w000.TF | ----------------- | 211 |
| w020.19 | AS------------NAT | 19 |
| w020.28 | ASN----------TNAT | 3 |
| w022.10 | --------------NAT | 74 |
| w022.17 | -------------TNAT | 12 |
| w030.18 | A-------------NAT | 11 |
| w030.33 | TSNSS------------ | 1 |
| w030.19 | AR---------NCTNAT | 1 |
| w030.32 | ASNSSI-----NCTNAT | 1 |
| w053.2 | ASNATAS-------NAT | 5 |
| w053.22 | A---TASNS----SIIE | 11 |
| w053.13 | ASNATASNS----SIIE | 3 |
| w078.42 | ASNATASNATAS---NS | 2 |
| w078.34 | ASNATASNATAS-NTTA | 1 |
| w078.16 | T-NATASNATAS-NATA | 2 |
| w078.36 | ASNATASNATAS-NATA | 7 |
| w078.5 | ATASNTNATAS-NINAT | 12 |
| w078.33 | ATAS----------NAT | 1 |
| w100.A3 | ATASNSSI-------IK | 3 |
| w100.T2 | ATAS--------NINAT | 3 |
| w100.A13 | ATASNANATAS-NTNAT | 2 |
| w136.B7 | D----ANATASNTNATV | 1 |
| w136.B2 | ATASNTNATVSNIKATV | 1 |
| w136.B3 | ATASNANATASNTNATV | 1 |
| w136.B24 | ATASNA--TASNTNATA | 1 |
| w136.B23 | DTASNSSI-------IK | 3 |
| w160.C1 | ANATASNI------NVT | 1 |
| w160.C12 | ATASNANATVSNTNATV | 3 |
| w160.D6 | --------------NTT | 1 |

**Table S2. V5 hypervariable region.** This table summarizes variants in HXB2 positions 463 through 466, spanning codons 546-552 in the CH505 alignment, a region that was not present in the transmitted-founder virus. This region accepted many distinct insertions, up to 7 amino acids long, often followed by deletions within the insertion, or subsequent mutations, throughout the period of the infection sampled. We found 21 distinct forms of indels in this region. We selected an example of each unique sequence from the time-point in which each pattern was first observed, and a representative sequence named, which includes the week of isolation. The V5 loop region noted in the Los Alamos database spans HXB2 positions 460-467. In subject CH505, indels in this broader region are almost certainly also impacting the evolutionary patterns. Dashes maintain the alignment.

| Example sequence | V5 Insertion | Number of times observed |
|---|---|---|
| w000.TF | ------- | 241 |
| w030.11 | -----DT | 40 |
| w030.20 | DGGKNNT | 1 |
| w053.15 | ----ETF | 40 |
| w078.42 | -GGKNNT | 5 |
| w078.16 | --GKNNT | 1 |
| w078.29 | -----NT | 1 |
| w078.33 | ------T | 1 |
| w078.5 | -RGKNNT | 1 |
| w100.B7 | DGGNNNT | 11 |
| w136.B1 | ----EDT | 38 |
| w136.B23 | ----TET | 5 |
| w136.B5 | ----KET | 4 |
| w136.B26 | ----NDT | 1 |
| w136.B28 | ---DTDT | 1 |
| w136.B2 | -----DI | 1 |
| w136.B7 | -----DM | 1 |
| w160.C3 | -----DP | 2 |
| w160.C10 | ETSETVS | 1 |
| w160.C1 | ----TGT | 1 |
| w160.C6 | ----EGT | 1 |

**Table S3. ELISA binding assay log-AUC of CH103 lineage Abs against autologous Env gp120s.**

| gp120 | UCA | IA8 | IA7 | IA6 | IA4 | IA3 | CH105 | IA2 | CH104 | IA1 | CH106 | CH103 |
|---|---|---|---|---|---|---|---|---|---|---|---|---|
| **w000.TF** | 3.5 | 5.5 | 9.2 | 9.1 | 10.1 | 11 | 11.2 | 10.8 | 10.4 | 10.4 | 11.3 | 12.6 |
| **w004.26** | 3.2 | 5.5 | 9.1 | 9 | 9.5 | 10.4 | 10.4 | 10.6 | 11.1 | 10.3 | 11 | 12.4 |
| **w004.54** | <0.1 | 0.5 | 2.3 | 2.9 | 2.8 | 5.1 | 8.3 | 6.8 | 8.1 | 6.2 | 8.1 | 9.2 |
| **w014.10** | 0.3 | 2.2 | 5.1 | 6.3 | 7.1 | 9 | 10.8 | 9.9 | 10.7 | 9.4 | 10.4 | 11.9 |
| **w014.2** | 0.8 | 3.2 | 7.2 | 6.9 | 8.5 | 9.7 | 9.6 | 9.7 | 9.5 | 9.1 | 9.6 | 10.9 |
| **w014.21** | 0.4 | 2.8 | 6.2 | 7.8 | 8 | 9.4 | 11.4 | 10.2 | 11.4 | 9.7 | 11.1 | 13.1 |
| **w014.3** | 1 | 4.4 | 7.7 | 7.6 | 9.2 | 10.3 | 10.8 | 10.6 | 10.6 | 9.6 | 10.6 | 11.9 |
| **w014.32** | 1.6 | 4.2 | 7.9 | 7.8 | 8.5 | 9.9 | 10.7 | 10.2 | 10.9 | 9.5 | 10.5 | 11.9 |
| **w014.8** | <0.1 | 0.9 | 0.6 | 2.4 | 2.6 | 1.8 | 6 | 3.6 | 3.9 | 3.4 | 7.9 | 11.2 |
| **w020.11** | <0.1 | 0.9 | 0.1 | 0.8 | 0.8 | 0.8 | 3.6 | 2.6 | 2.2 | 1.8 | 5.4 | 9.6 |
| **w020.14** | 0.3 | 3.4 | 7.2 | 7.9 | 8.6 | 9.5 | 10.4 | 11 | 10.4 | 10.3 | 11.2 | 12.6 |
| **w020.15** | 1.6 | 4.2 | 8.2 | 7.8 | 9.1 | 10.2 | 10.8 | 10.5 | 10.5 | 9.9 | 10.5 | 11.8 |
| **w020.19** | 0.8 | 3.5 | 7.2 | 6.5 | 8.1 | 9.4 | 9.5 | 9.3 | 9.7 | 9 | 9.7 | 11.3 |

| gp120 | UCA | IA8 | IA7 | IA6 | IA4 | IA3 | CH105 | IA2 | CH104 | IA1 | CH106 | CH103 |
|---|---|---|---|---|---|---|---|---|---|---|---|---|
| w020.22 | 1.2 | 4 | 8.3 | 8.4 | 8.8 | 10.2 | 10.5 | 10.6 | 10.7 | 10.1 | 10.7 | 11.5 |
| w020.23 | 0.9 | 3.6 | 7.2 | 7.9 | 8.6 | 9.7 | 11.2 | 10.7 | 11.6 | 9.8 | 11.3 | 12.8 |
| w020.25 | 0.8 | 2.4 | 6.4 | 6 | 7.3 | 8.6 | 8.2 | 9 | 8.3 | 8.6 | 9.4 | 10.3 |
| w020.26 | <0.1 | 0.7 | 2.7 | 4 | 5.1 | 7.1 | 9 | 9 | 8.7 | 8.9 | 9.6 | 11.3 |
| w020.3 | 0.8 | 2.6 | 5.9 | 6 | 7 | 7.7 | 10.1 | 9.3 | 9.8 | 8.9 | 10.3 | 12.4 |
| w020.30 | 0.8 | 4.2 | 7.3 | 7.9 | 8.5 | 9 | 10.3 | 10.5 | 10.1 | 9.8 | 10.8 | 11.7 |
| w020.4 | 0.9 | 3.6 | 8.2 | 8.7 | 9.4 | 10.6 | 11.1 | 11.6 | 11.6 | 11.3 | 11.5 | 12.5 |
| w020.7 | 0.6 | 2.7 | 7.6 | 8.1 | 8.7 | 10.4 | 10.6 | 11.3 | 11.1 | 10.8 | 11.3 | 12.5 |
| w020.8 | 0.4 | 2.6 | 5.9 | 6.7 | 7.4 | 8.8 | 10.6 | 10.5 | 10.5 | 9.9 | 11 | 12.4 |
| w020.9 | 0.4 | 2.8 | 4.9 | 6.8 | 7.4 | 8.1 | 9.3 | 9.7 | 8.9 | 9 | 10 | 11.9 |
| w030.10 | 0.2 | 1.3 | 3.2 | 5.1 | 5.8 | 7.3 | 9.3 | 11 | 10.3 | 10.1 | 11.1 | 12.5 |
| w030.11 | <0.1 | <0.1 | 0.7 | 0.5 | 0.5 | 2.4 | 9 | 9.1 | 9.5 | 8.4 | 9.4 | 10.9 |
| w030.12 | <0.1 | 0.4 | 1 | 1.2 | 1.9 | 4.1 | 10.7 | 11.6 | 11.6 | 10.6 | 11.8 | 13.3 |
| w030.13 | 0.3 | 2 | 4.7 | 6.5 | 7.4 | 9 | 10.5 | 11.4 | 11.3 | 10.5 | 11.9 | 12.9 |
| w030.15 | <0.1 | <0.1 | <0.1 | <0.1 | <0.1 | <0.1 | 9.3 | 8.4 | 9.7 | 8 | 9.3 | 10.1 |
| w030.17 | <0.1 | <0.1 | <0.1 | <0.1 | <0.1 | <0.1 | 6.7 | 7 | 7 | 5.8 | 8.2 | 9.9 |
| w030.18 | <0.1 | <0.1 | <0.1 | <0.1 | <0.1 | <0.1 | 11.1 | 9.4 | 10.9 | 9.9 | 10.5 | 13 |
| w030.19 | <0.1 | <0.1 | <0.1 | <0.1 | <0.1 | <0.1 | 9.5 | 8.2 | 8.9 | 8.2 | 7.4 | 12.9 |
| w030.20 | <0.1 | <0.1 | <0.1 | <0.1 | <0.1 | <0.1 | 7.1 | 6.3 | 7.5 | 5.7 | 7.3 | 9.5 |
| w030.21 | <0.1 | <0.1 | <0.1 | <0.1 | <0.1 | <0.1 | 10.6 | 11.5 | 11.3 | 11.8 | 10.9 | 12.2 |
| w030.23 | <0.1 | <0.1 | <0.1 | <0.1 | <0.1 | 0.7 | 9.2 | 8.9 | 9.6 | 8.4 | 10 | 11.1 |
| w030.25 | <0.1 | <0.1 | 0.4 | 0.5 | 0.4 | 1.4 | 10 | 9.9 | 10.3 | 9.1 | 10.2 | 12.5 |
| w030.27 | <0.1 | <0.1 | <0.1 | <0.1 | <0.1 | 0.4 | 7.9 | 7.9 | 8.2 | 7.3 | 8.2 | 9.1 |
| w030.28 | <0.1 | 1.6 | 3.5 | 6.3 | 6.5 | 7.7 | 9.1 | 11.1 | 10.7 | 10.1 | 11.7 | 12.8 |
| w030.36 | <0.1 | <0.1 | <0.1 | <0.1 | <0.1 | <0.1 | 7.5 | 7.3 | 7.8 | 6.7 | 8.5 | 9.1 |
| w030.5 | <0.1 | <0.1 | <0.1 | <0.1 | <0.1 | 0.3 | 9 | 8.1 | 9.4 | 7.8 | 9.4 | 11.4 |
| w030.6 | <0.1 | <0.1 | <0.1 | <0.1 | <0.1 | <0.1 | 6.5 | 5.4 | 6.1 | 4.5 | 6.6 | 8 |
| w030.9 | <0.1 | <0.1 | <0.1 | <0.1 | <0.1 | 0.2 | 8.7 | 8.6 | 8.9 | 7.8 | 9.6 | 10.2 |
| w053.13 | <0.1 | <0.1 | <0.1 | <0.1 | <0.1 | <0.1 | 9.4 | 10.5 | 10 | 9.9 | 10.2 | 11.2 |
| w053.15 | <0.1 | <0.1 | <0.1 | <0.1 | <0.1 | <0.1 | 9.4 | 9.5 | 10.4 | 9.2 | 10.1 | 11.2 |
| w053.19 | <0.1 | <0.1 | <0.1 | <0.1 | <0.1 | <0.1 | 13.8 | 12.6 | 13.8 | 13.6 | 13.2 | 13.5 |
| w053.22 | <0.1 | <0.1 | <0.1 | <0.1 | 0.2 | 1.1 | 9 | 9.3 | 9.9 | 8.8 | 9.8 | 11.6 |
| w053.25 | <0.1 | <0.1 | <0.1 | <0.1 | <0.1 | <0.1 | 10 | 10.2 | 10.6 | 10 | 10.9 | 11.6 |
| w053.29 | <0.1 | <0.1 | <0.1 | <0.1 | <0.1 | <0.1 | 8.5 | 9.2 | 9.8 | 8.8 | 9.7 | 10.2 |
| w053.31 | <0.1 | <0.1 | <0.1 | <0.1 | <0.1 | <0.1 | 13.5 | 13.3 | 13.7 | 13.4 | 13.4 | 13.6 |
| w053.6 | <0.1 | <0.1 | <0.1 | <0.1 | <0.1 | <0.1 | 11.7 | 11.9 | 12 | 11.4 | 10.9 | 12.7 |
| w078.1 | <0.1 | <0.1 | <0.1 | <0.1 | <0.1 | <0.1 | 7.5 | 7.7 | 8.3 | 7.5 | 7.3 | 11 |
| w078.10 | <0.1 | <0.1 | <0.1 | <0.1 | <0.1 | <0.1 | 7.6 | 6.6 | 8.1 | 6.6 | 7.8 | 11.1 |
| w078.15 | <0.1 | <0.1 | 0.7 | 1 | 1.3 | 3 | 10.1 | 11.5 | 10.8 | 10.9 | 11 | 10.7 |
| w078.17 | <0.1 | <0.1 | <0.1 | <0.1 | <0.1 | <0.1 | 9.5 | 9.9 | 10 | 9 | 8.7 | 11.3 |
| w078.25 | <0.1 | <0.1 | <0.1 | <0.1 | <0.1 | <0.1 | 3.3 | 3.2 | 3.5 | 2.4 | 4.6 | 8.7 |
| w078.33 | <0.1 | <0.1 | <0.1 | <0.1 | <0.1 | <0.1 | 8.9 | 9 | 9 | 8.2 | 9.5 | 11.1 |
| w078.38 | <0.1 | <0.1 | <0.1 | <0.1 | <0.1 | <0.1 | 4 | 4.7 | 4.2 | 4.4 | 5.3 | 7.7 |
| w078.4 | <0.1 | <0.1 | <0.1 | <0.1 | <0.1 | <0.1 | 9.5 | 10.2 | 10.1 | 9.8 | 9.2 | 11.7 |
| w078.5 | <0.1 | <0.1 | <0.1 | <0.1 | <0.1 | <0.1 | 6.6 | 6.6 | 7.7 | 5.6 | 7.9 | 10.2 |
| w078.6 | <0.1 | <0.1 | <0.1 | <0.1 | <0.1 | <0.1 | 3.8 | 4.6 | 4.5 | 3 | 6.7 | 9.9 |
| w078.7 | <0.1 | <0.1 | <0.1 | <0.1 | <0.1 | <0.1 | 9.4 | 9.2 | 9.5 | 8.9 | 8.7 | 11.4 |
| w078.9 | <0.1 | <0.1 | <0.1 | <0.1 | <0.1 | <0.1 | 6.6 | 7.2 | 6.9 | 6.3 | 8.3 | 8.9 |
| w100.A10 | <0.1 | <0.1 | <0.1 | <0.1 | <0.1 | 0.1 | 9.1 | 10.5 | 10.1 | 9.9 | 10.4 | 11.7 |
| w100.A12 | <0.1 | <0.1 | <0.1 | <0.1 | <0.1 | <0.1 | 9.6 | 9.4 | 10.4 | 8.6 | 9.7 | 4.7 |
| w100.A13 | <0.1 | <0.1 | <0.1 | <0.1 | <0.1 | <0.1 | 11.4 | 12.5 | 12.9 | 12.6 | 12 | 12.9 |
| w100.A3 | <0.1 | <0.1 | <0.1 | <0.1 | <0.1 | <0.1 | 7.2 | 9.3 | 10 | 8.5 | 8.6 | 11.4 |
| w100.A4 | <0.1 | <0.1 | <0.1 | <0.1 | <0.1 | <0.1 | 8.1 | 9 | 9.1 | 8.6 | 9 | 10.7 |
| w100.A6 | <0.1 | <0.1 | <0.1 | <0.1 | <0.1 | <0.1 | 10.1 | 11.4 | 11.6 | 11 | 11.1 | 12.1 |
| w100.B2 | <0.1 | <0.1 | <0.1 | <0.1 | <0.1 | <0.1 | 7.5 | 7.5 | 6.1 | 7.3 | 7.8 | 3.2 |

| gp120 | UCA | IA8 | IA7 | IA6 | IA4 | IA3 | CH105 | IA2 | CH104 | IA1 | CH106 | CH103 |
|---|---|---|---|---|---|---|---|---|---|---|---|---|
| **w100.B4** | <0.1 | <0.1 | <0.1 | <0.1 | <0.1 | <0.1 | 11.9 | 13.4 | 13.1 | 13.7 | 12.6 | 9.7 |
| **w100.B6** | <0.1 | <0.1 | <0.1 | <0.1 | <0.1 | <0.1 | 11 | 12.1 | 11 | 12.2 | 11.8 | 7.1 |
| **w100.B7** | <0.1 | <0.1 | <0.1 | <0.1 | <0.1 | <0.1 | 11.6 | 12.4 | 11.6 | 11.9 | 11.5 | 7.8 |
| **w100.C7** | <0.1 | <0.1 | <0.1 | <0.1 | <0.1 | <0.1 | 9 | 9.1 | 9.2 | 8.9 | 9.3 | 4.8 |
| **w136.B10** | <0.1 | <0.1 | <0.1 | <0.1 | <0.1 | <0.1 | 8.5 | 9.7 | 8.9 | 9.2 | 9 | 11.5 |
| **w136.B12** | <0.1 | <0.1 | <0.1 | <0.1 | <0.1 | <0.1 | 9.7 | 12.4 | 10.7 | 12.2 | 11.9 | 12.7 |
| **w136.B2** | <0.1 | <0.1 | <0.1 | <0.1 | <0.1 | <0.1 | 12.4 | 13.1 | 13.2 | 13.2 | 12.7 | 10.8 |
| **w136.B20** | <0.1 | <0.1 | <0.1 | <0.1 | <0.1 | <0.1 | 10.7 | 12.7 | 11.7 | 12.4 | 12.1 | 12.4 |
| **w136.B23** | <0.1 | <0.1 | <0.1 | <0.1 | <0.1 | <0.1 | 13.7 | 14.3 | 14.2 | 14.4 | 13.3 | 11.8 |
| **w136.B27** | <0.1 | <0.1 | <0.1 | <0.1 | <0.1 | <0.1 | 10.4 | 11.9 | 11.5 | 11.2 | 11.2 | 12.6 |
| **w136.B29** | <0.1 | <0.1 | <0.1 | <0.1 | <0.1 | <0.1 | 10.2 | 12.5 | 11.4 | 12.2 | 12 | 12.7 |
| **w136.B3** | <0.1 | <0.1 | <0.1 | <0.1 | <0.1 | 0.1 | 12 | 13.1 | 13.1 | 13.1 | 12.6 | 9.9 |
| **w136.B36** | <0.1 | <0.1 | <0.1 | <0.1 | <0.1 | <0.1 | 12.8 | 13.6 | 13.8 | 13.8 | 13.1 | 13.7 |
| **w136.B4** | <0.1 | <0.1 | <0.1 | <0.1 | <0.1 | <0.1 | 10.7 | 12.2 | 12.2 | 11.8 | 11.8 | 13.4 |
| **w136.B5** | <0.1 | <0.1 | <0.1 | <0.1 | <0.1 | <0.1 | 10.5 | 9.9 | 10.6 | 9.5 | 10.9 | 4.3 |
| **w136.B8** | <0.1 | <0.1 | <0.1 | <0.1 | <0.1 | <0.1 | 9.9 | 13.2 | 11.7 | 12.9 | 11.7 | 11.9 |
| **w160.A1** | <0.1 | <0.1 | <0.1 | <0.1 | <0.1 | <0.1 | 6.7 | 7.6 | 7.3 | 6.4 | 7.6 | 10.7 |
| **w160.C11** | <0.1 | <0.1 | <0.1 | <0.1 | <0.1 | <0.1 | 7.7 | 8.5 | 7.4 | 7.7 | 8.2 | 11.9 |
| **w160.C12** | <0.1 | <0.1 | <0.1 | <0.1 | <0.1 | <0.1 | 11.1 | 12.9 | 11.3 | 12.9 | 12.2 | 7.8 |
| **w160.C14** | <0.1 | <0.1 | <0.1 | <0.1 | <0.1 | <0.1 | 6.6 | 7 | 6.8 | 5.8 | 6.9 | 12.1 |
| **w160.C2** | <0.1 | <0.1 | <0.1 | <0.1 | <0.1 | 1.3 | 8.6 | 10.1 | 9.2 | 9.6 | 9.3 | 12.2 |
| **w160.C4** | <0.1 | <0.1 | <0.1 | <0.1 | <0.1 | <0.1 | 6.6 | 6.2 | 7.2 | 5.3 | 6.8 | 12.2 |
| **w160.D1** | <0.1 | <0.1 | <0.1 | <0.1 | <0.1 | <0.1 | 6 | 7.3 | 5.7 | 6.4 | 7.1 | 11.5 |
| **w160.D5** | <0.1 | <0.1 | <0.1 | <0.1 | <0.1 | <0.1 | 4.6 | 5 | 4.5 | 3.8 | 5.6 | 10.9 |
| **w160.T2** | <0.1 | <0.1 | <0.1 | <0.1 | <0.1 | <0.1 | 4.9 | 5.6 | 5.4 | 4.7 | 6.2 | 11.2 |
| **w160.T4** | <0.1 | <0.1 | <0.1 | <0.1 | <0.1 | <0.1 | 12.2 | 13.4 | 12.8 | 13.6 | 12.3 | 9 |

## Table S4. TZM-bl neutralization IC50 titers of CH103 lineage Abs with autologous *env* gp160s.

| gp160 | UCA | IA8 | IA7 | IA6 | IA5 | IA4 | IA3 | CH105 | IA2 | CH104 | IA1 | CH106 | CH103 |
|---|---|---|---|---|---|---|---|---|---|---|---|---|---|
| **w000.TF** | >50 | 44.67 | 10.17 | 10.17 | 6.82 | 9.72 | 6.62 | 5.1 | 2.73 | 3.08 | 1.42 | 2.29 | 4.14 |
| **w004.03** | 15.12 | 2.05 | 0.71 | 0.99 | 0.13 | 0.67 | 0.85 | 0.38 | 0.29 | 0.29 | 0.16 | 0.16 | 0.13 |
| **w004.08** | >50 | 16.75 | 4.2 | 4.11 | 3.38 | 3.24 | 6.09 | 2.92 | 1.29 | 1.62 | 0.95 | 1.39 | 1.89 |
| **w004.10** | >50 | 41.36 | 8.66 | 5.32 | 5.66 | 3.8 | 3.68 | 2.44 | 1.17 | 1.31 | 0.74 | 0.71 | 1.1 |
| **w004.11** | >50 | >50 | 14.35 | 10.32 | 8.17 | 8.76 | 7.71 | 3.62 | 2.79 | 2.32 | 1.71 | 1.2 | 2.37 |
| **w004.13** | >50 | 25.24 | 3.08 | 3.5 | 9.78 | 4.08 | 7.88 | 3.4 | 3.46 | 2.09 | 1.45 | 1.08 | 3.15 |
| **w004.14** | >50 | 47.61 | 11.92 | 9.71 | 13.78 | 11.02 | 10.62 | 5.22 | 3.53 | 2.34 | 2.14 | 1.32 | 4 |
| **w004.15** | >50 | >50 | 16.45 | 9.82 | 6.43 | 8.2 | 6.23 | 4.03 | 1.92 | 1.49 | 1.34 | 1.41 | 1.37 |
| **w004.16** | >50 | 21.26 | 8.43 | 4.33 | 7.32 | 6.95 | 6.4 | 4.33 | 2.78 | 2.66 | 1.39 | 1.23 | 2.83 |
| **w004.26** | 19.5 | 1.5 | 0.77 | 0.88 | 0.07 | 0.67 | 0.41 | 0.24 | 0.15 | 0.27 | 0.25 | 0.09 | 0.24 |
| **w004.27** | >50 | 12.39 | 2.32 | 2.57 | 4.69 | 2.7 | 5.7 | 2.4 | 1.89 | 1.44 | 1.17 | 0.53 | 1.94 |
| **w004.29** | >50 | 31.68 | 5.57 | 4.72 | 2.48 | 4.7 | 3.4 | 2.24 | 1.3 | 1.36 | 1.1 | 0.79 | 1.9 |
| **w004.37** | >50 | 25.63 | 8.66 | 7.97 | 4.76 | 5.61 | 4.15 | 2.19 | 1.68 | 1.65 | 1 | 0.43 | 1.87 |
| **w004.51** | >50 | 34.15 | 6.83 | 5.77 | 8.62 | 6.54 | 5.45 | 2.11 | 1.54 | 1.59 | 1.36 | 0.56 | 2.04 |
| **w004.56** | >50 | 34.78 | 8.46 | 8.51 | 6.42 | 6.35 | 5.83 | 3.36 | 2 | 1.59 | 1.39 | 0.8 | 2.4 |
| **w014.10** | >50 | >50 | 28.28 | 29.42 | 25.98 | 28.93 | 4.98 | 2.82 | 1.72 | 1.9 | 1.27 | 0.98 | 1.6 |
| **w014.16** | >50 | >50 | 42.21 | 27.23 | 32.29 | 18.85 | 11.75 | 8.42 | 4.3 | 4.48 | 4.08 | 4.45 | 6.27 |
| **w014.19** | >50 | >50 | 15.49 | 10.14 | 20.09 | 9.2 | 6.06 | 4.39 | 1.82 | 2.47 | 1.36 | 2.47 | 4.66 |
| **w014.2** | >50 | >50 | 26.66 | 14.92 | 25.33 | 13.63 | 10.3 | 6.84 | 2.28 | 3.24 | 2.87 | 2.86 | 4.01 |
| **w014.20** | >50 | >50 | 23.56 | 12.68 | 20.3 | 15.94 | 6.27 | 3.02 | 1.96 | 1.88 | 1.07 | 1.43 | 2.92 |
| **w014.21** | >50 | 24.97 | 11.01 | 5.22 | 4.07 | 3.36 | 0.94 | 0.6 | 0.39 | 0.37 | 0.24 | 0.41 | 0.53 |
| **w014.29** | >50 | >50 | 37.65 | 39.45 | >50 | 22.36 | 14.88 | 12.54 | 5.66 | 6.14 | 4.02 | 6.33 | 6.91 |
| **w014.3** | >50 | >50 | 23.5 | 13.51 | 8.13 | 10.07 | 5.83 | 3.64 | 1.18 | 1.69 | 1.84 | 1.45 | 2.57 |
| **w014.30** | >50 | >50 | >50 | 46.59 | 39.45 | 34.46 | 15.52 | 8.11 | 4.94 | 4.2 | 3.01 | 3.41 | 6.17 |
| **w014.31** | >50 | >50 | 36.38 | 25.99 | 31.67 | 28.03 | 13.04 | 7.22 | 4.05 | 3.69 | 4.28 | 3.19 | 5.84 |
| **w014.32** | >50 | >50 | 41.85 | 25.02 | 21.56 | 22.67 | 16 | 8.26 | 5.66 | 3.96 | 3.66 | 4.41 | 5.51 |
| **w014.34** | >50 | >50 | >50 | 43.94 | 48.04 | 49.79 | 22.64 | 11.24 | 7.05 | 5.72 | 5.09 | 4.66 | 7.32 |
| **w014.39** | >50 | 43.27 | 40.99 | 24.91 | 34.07 | 32.87 | 15.15 | 8.48 | 5.77 | 4.73 | 3.09 | 4.52 | 5.44 |
| **w014.4** | >50 | >50 | 24.72 | 20.23 | 28.95 | 15.59 | 10.41 | 6.04 | 2.77 | 2.35 | 2.25 | 2.55 | 3.37 |
| **w014.6** | >50 | >50 | 26.84 | 25.13 | 32.85 | 20.88 | 12.23 | 6.99 | 3.37 | 3.14 | 3.07 | 3.02 | 4.14 |
| **w014.8** | >50 | 28.47 | 23.12 | 17.3 | 34.15 | 12.93 | 8.81 | 4.59 | 1.06 | 1.41 | 1.54 | 1.6 | 1.06 |

| gp160 | UCA | IA8 | IA7 | IA6 | IA5 | IA4 | IA3 | CH105 | IA2 | CH104 | IA1 | CH106 | CH103 |
|---|---|---|---|---|---|---|---|---|---|---|---|---|---|
| **w020.11** | >50 | 32.68 | 21.45 | 13.04 | 4.41 | 12.53 | 5.97 | 1.82 | 1.14 | 0.85 | 1.17 | 0.61 | 0.43 |
| **w020.13** | >50 | >50 | >50 | >50 | 19.84 | 31.77 | 20.17 | 14.16 | 6.7 | 5.6 | 3.62 | 6.15 | 5.82 |
| **w020.14** | >50 | 42.47 | 10.13 | 5.45 | 6.33 | 4.88 | 1.99 | 1.04 | 1.22 | 0.67 | 0.72 | 0.46 | 0.08 |
| **w020.15** | >50 | >50 | 42.09 | 21 | 19.46 | 29.54 | 16.25 | 11.42 | 6.41 | 5.9 | 3.12 | 6.16 | 6.91 |
| **w020.19** | >50 | >50 | 43.58 | >50 | 43.43 | >50 | 21.65 | 13.42 | 7.58 | 6.12 | 5.05 | 7.69 | 8.34 |
| **w020.2** | >50 | >50 | 20.17 | 18.4 | 7.09 | 19.49 | 11.8 | 6.2 | 4.08 | 3.29 | 2.12 | 2.78 | 3.5 |
| **w020.21** | >50 | >50 | >50 | 29.58 | >50 | 21.02 | 6.36 | 3.64 | 1.88 | 1.81 | 1.23 | 1.56 | 2.54 |
| **w020.22** | >50 | >50 | 20.77 | 12.38 | 14.1 | 8.22 | 3.48 | 2.21 | 1.25 | 0.88 | 0.53 | 1.01 | 1.45 |
| **w020.23** | >50 | 26.79 | 8.07 | 4.74 | 10.6 | 3.45 | 2.65 | 1.48 | 1.12 | 0.87 | 0.54 | 0.68 | 0.87 |
| **w020.24** | >50 | 28.92 | 9.44 | 5.57 | 12.97 | 4.73 | 2.51 | 3.15 | 1.12 | 1.52 | 0.57 | 1.54 | 1.11 |
| **w020.25** | >50 | >50 | >50 | >50 | 37.72 | >50 | 23.37 | 14 | 11.15 | 9.01 | 6.41 | 7.96 | 7.95 |
| **w020.26** | >50 | 48.25 | 10.45 | 7.31 | 8.65 | 6.79 | 2.06 | 1.23 | 0.9 | 1.29 | 0.51 | 0.77 | 0.88 |
| **w020.27** | >50 | >50 | 20.82 | 14.21 | 25.37 | 9.51 | 5.7 | 3.89 | 2.24 | 1.85 | 1.2 | 1.22 | 1.81 |
| **w020.3** | >50 | >50 | >50 | 24.06 | 4.17 | 22.41 | 4.93 | 2.35 | 1.63 | 1.46 | 0.77 | 1.04 | 1.41 |
| **w020.4** | >50 | >50 | 25.99 | 9.46 | 3.98 | 11.83 | 2.77 | 1.36 | 0.88 | 0.6 | 0.55 | 0.67 | 0.67 |
| **w020.7** | >50 | 23.68 | 6.85 | 2.85 | 2.92 | 3.83 | 1.39 | 0.86 | 0.62 | 0.95 | 0.41 | 0.6 | 0.52 |
| **w020.8** | >50 | >50 | 14.67 | 8.41 | 3.78 | 7.53 | 3.34 | 3.54 | 1.37 | 1.63 | 1.14 | 1.7 | 2.02 |
| **w020.9** | >50 | 14.87 | 5.46 | 2.07 | 4.3 | 3.06 | 3.62 | 1.92 | 1.16 | 1.33 | 1.06 | 1.13 | 1.37 |
| **w030.10** | >50 | >50 | >50 | 35 | 6.93 | 23.97 | 8.07 | 3.88 | 1.11 | 1.08 | 1.61 | 1.14 | 1.56 |
| **w030.12** | >50 | >50 | 36.65 | 25.87 | 6.44 | 18.39 | 1.92 | 1.88 | 0.52 | 0.95 | 0.65 | 0.62 | 1.11 |
| **w030.13** | >50 | >50 | >50 | 8.43 | 2.77 | 9.64 | 5.19 | 3.86 | 1 | 0.43 | 1.06 | 1.08 | 1.43 |
| **w030.15** | >50 | >50 | >50 | >50 | >50 | >50 | 11.44 | 7.43 | 3.5 | 5.79 | 4.6 | 2.8 | 3.98 |
| **w030.17** | >50 | >50 | >50 | >50 | >50 | >50 | 3.86 | 14.24 | 7.28 | 6.78 | 6.86 | 3.9 | 4.28 |
| **w030.18** | >50 | >50 | >50 | >50 | >50 | >50 | 2.92 | 2.1 | 1.12 | 1.16 | 1.08 | 0.88 | 1.36 |
| **w030.19** | >50 | >50 | >50 | >50 | >50 | >50 | 1.45 | 0.81 | 0.47 | 0.51 | 0.39 | 0.54 | 0.58 |
| **w030.20** | >50 | >50 | >50 | >50 | >50 | >50 | >50 | 44.19 | 20.42 | 28.33 | 31.48 | 17.56 | 13.2 |
| **w030.21** | >50 | >50 | >50 | >50 | >50 | >50 | 2.35 | 1.49 | 0.43 | 0.53 | 0.62 | 0.56 | 1.05 |
| **w030.23** | >50 | >50 | >50 | >50 | >50 | >50 | 33.12 | 23.24 | 18.56 | 12.81 | 17.21 | 9.39 | 9.51 |
| **w030.25** | >50 | >50 | >50 | >50 | >50 | >50 | 20.19 | 12.84 | 11.02 | 8.69 | 12.9 | 7.7 | 11.99 |
| **w030.26** | >50 | >50 | >50 | >50 | >50 | >50 | >50 | 32.38 | 17.23 | 18.9 | 12.5 | 12.48 | 12.11 |
| **w030.27** | >50 | >50 | >50 | >50 | >50 | >50 | >50 | >50 | 43.69 | 38.02 | 18.28 | 31.87 | 30.02 |
| **w030.28** | >50 | 44 | 23.07 | 7.18 | 3.78 | 6.92 | 4.35 | 2.13 | 0.77 | 0.59 | 0.49 | 0.77 | 0.77 |
| **w030.31** | >50 | >50 | >50 | >50 | >50 | >50 | 43.08 | 27.73 | 11.73 | 9.79 | 6.71 | 10 | 8.79 |
| **w030.32** | >50 | >50 | >50 | 33.9 | >50 | 40.93 | 2.56 | 2.89 | 1.09 | 1.52 | 0.54 | 1.25 | 1.03 |
| **w030.34** | >50 | >50 | >50 | >50 | >50 | >50 | >50 | >50 | 44.85 | 47.07 | 49.72 | 39.83 | 26.37 |
| **w030.36** | >50 | >50 | >50 | >50 | >50 | >50 | >50 | 21.15 | 10.96 | 11.07 | 8.12 | 8.13 | 9.24 |
| **w030.37** | >50 | >50 | >50 | >50 | >50 | >50 | >50 | >50 | 29.75 | >50 | 25.19 | 40.42 | 22.3 |
| **w030.5** | >50 | >50 | >50 | >50 | >50 | >50 | 45.16 | 34.12 | 6.76 | 7.45 | 9.67 | 10.04 | 8.05 |
| **w030.6** | >50 | >50 | >50 | >50 | >50 | >50 | 15.27 | 9.23 | 1.3 | 1.3 | 4.99 | 4.51 | 4.15 |
| **w030.8** | >50 | >50 | >50 | >50 | >50 | >50 | >50 | 14.11 | 10.35 | 7.07 | 12.83 | 9.91 | 12.56 |
| **w030.9** | >50 | >50 | >50 | >50 | >50 | >50 | >50 | >50 | 44.49 | 42.98 | >50 | 19.82 | 49.94 |
| **w053.10** | >50 | >50 | >50 | >50 | >50 | >50 | >50 | 31.68 | 10.59 | 20.07 | 13.21 | 12.73 | 18.87 |
| **w053.11** | >50 | >50 | >50 | >50 | >50 | >50 | >50 | >50 | 26.22 | 32.63 | 32.31 | 37.41 | 30.37 |
| **w053.13** | >50 | >50 | >50 | >50 | >50 | >50 | 12.38 | 6.64 | 4.78 | 4.47 | 3.6 | 4.66 | 5.3 |
| **w053.14** | >50 | >50 | >50 | >50 | >50 | >50 | 20.93 | 15.52 | 6 | 6.46 | 5.76 | 8.28 | 8.01 |
| **w053.15** | >50 | >50 | >50 | >50 | >50 | >50 | 31.54 | 29.48 | 12.11 | 12.65 | 8.19 | 17.72 | 13.82 |
| **w053.16** | >50 | >50 | >50 | >50 | >50 | >50 | >50 | 21.73 | 10.68 | 3.44 | 17.37 | 18.39 | 12.31 |
| **w053.17** | >50 | >50 | >50 | >50 | >50 | >50 | 43.24 | 20.89 | 10.62 | 7.25 | 10.32 | 13.8 | 10.61 |
| **w053.19** | >50 | >50 | >50 | >50 | >50 | >50 | 4.58 | 2.84 | 2.59 | 2.66 | 1.91 | 2.55 | 3.16 |
| **w053.22** | >50 | >50 | >50 | >50 | >50 | >50 | 41.2 | 29.51 | 15.81 | 12.31 | 11.58 | 19.36 | 12.19 |
| **w053.25** | >50 | >50 | >50 | >50 | >50 | >50 | 43.14 | 28.05 | 13.07 | 10.1 | 7.56 | 10.47 | 9.26 |
| **w053.27** | >50 | >50 | >50 | >50 | >50 | >50 | >50 | 27.02 | 17.47 | 13.61 | 14.56 | 14.99 | 10.47 |
| **w053.28** | >50 | >50 | >50 | >50 | >50 | >50 | 46.24 | 21.84 | 14.19 | 14.45 | 7.67 | 15.86 | 12.65 |
| **w053.29** | >50 | >50 | >50 | >50 | >50 | >50 | >50 | 42.24 | 21.79 | 17.38 | 14.58 | 22.52 | 21.95 |
| **w053.3** | >50 | >50 | >50 | >50 | >50 | >50 | 49.31 | 27.54 | 10.51 | 8.82 | 4.42 | 9.46 | 8.37 |
| **w053.31** | >50 | >50 | >50 | >50 | >50 | >50 | 2.38 | 2.06 | 1.06 | 1.09 | 0.43 | 1.68 | 1.24 |
| **w053.32** | >50 | >50 | >50 | >50 | >50 | >50 | 22.8 | 13.89 | 7.53 | 6.29 | 6.09 | 10.78 | 6.76 |
| **w053.6** | >50 | >50 | >50 | >50 | >50 | >50 | >50 | >50 | 19.43 | 19 | 16.67 | 12.57 | 11.88 |
| **w053.8** | >50 | >50 | >50 | >50 | >50 | >50 | >50 | 34.75 | 21.53 | 17.15 | 16.78 | 20.92 | 16.01 |
| **w053.9** | >50 | >50 | >50 | >50 | >50 | >50 | >50 | >50 | 31.44 | 32.21 | 22.7 | 34.57 | 28.81 |
| **w078.1** | >50 | >50 | >50 | >50 | >50 | >50 | 7.06 | 4.07 | 1.3 | 1.68 | 3.31 | 4.63 | 2.69 |
| **w078.10** | >50 | >50 | >50 | >50 | >50 | >50 | 15.43 | 8.67 | 4.42 | 6.1 | 3.04 | 5.49 | 7.08 |
| **w078.14** | >50 | >50 | >50 | >50 | >50 | >50 | 28.34 | 17.7 | 10.25 | 9.25 | 6.58 | 17.81 | 11.33 |
| **w078.15** | >50 | >50 | >50 | >50 | 21.97 | 49.04 | 2.97 | 2.55 | 0.76 | 0.96 | 0.25 | 2.08 | 1.51 |
| **w078.16** | >50 | >50 | >50 | >50 | >50 | >50 | >50 | >50 | >50 | >50 | >50 | >50 | 12.73 |
| **w078.17** | >50 | >50 | >50 | >50 | >50 | >50 | >50 | 30.44 | 16.62 | 14.81 | 9.56 | 26.32 | 15.23 |
| **w078.25** | >50 | >50 | >50 | >50 | >50 | >50 | 30.45 | 10.5 | 21.4 | 22.67 | 29.85 | 15.19 | 17.72 |
| **w078.3** | >50 | >50 | >50 | >50 | >50 | >50 | 7.14 | 5.34 | 2.28 | 3.11 | 1.41 | 5.17 | 3.74 |

| gp160 | UCA | IA8 | IA7 | IA6 | IA5 | IA4 | IA3 | CH105 | IA2 | CH104 | IA1 | CH106 | CH103 |
|---|---|---|---|---|---|---|---|---|---|---|---|---|---|
| **w078.33** | >50 | >50 | >50 | >50 | >50 | >50 | >50 | 28.41 | 12.23 | 11.06 | 15.79 | 12.2 | 4.13 |
| **w078.38** | >50 | >50 | >50 | >50 | >50 | >50 | 33.25 | 15.49 | 13.81 | 12.51 | 13.26 | 12.85 | 12.32 |
| **w078.4** | >50 | >50 | >50 | >50 | >50 | >50 | >50 | 24.76 | 19.03 | 10.56 | 23.86 | 29.17 | 22.48 |
| **w078.5** | >50 | >50 | >50 | >50 | >50 | >50 | >50 | >50 | >50 | >50 | >50 | >50 | 10.9 |
| **w078.6** | >50 | >50 | >50 | >50 | >50 | >50 | >50 | >50 | 22.34 | 23.56 | 20.68 | 17.31 | 3.8 |
| **w078.7** | >50 | >50 | >50 | >50 | >50 | >50 | 43.42 | 18.53 | 2.58 | 4.4 | 6.67 | 15.23 | 6.79 |
| **w078.8** | >50 | >50 | >50 | >50 | >50 | >50 | 17.56 | 9.7 | 6.07 | 4.85 | 5.27 | 9.33 | 6.41 |
| **w078.9** | >50 | >50 | >50 | >50 | >50 | >50 | 47.83 | 15.44 | 13.28 | 17.02 | 16.17 | 14 | 17.32 |
| **w100.A10** | >50 | >50 | >50 | >50 | >50 | >50 | >50 | 48.74 | 25.12 | 21.8 | 27.54 | 37.29 | 36.75 |
| **w100.A11** | >50 | >50 | >50 | >50 | >50 | >50 | >50 | 12.24 | 4.13 | 3.13 | 2.11 | 3.99 | 10.74 |
| **w100.A12** | >50 | >50 | >50 | >50 | >50 | >50 | 33.48 | 20.53 | 13.17 | 10.77 | 5.98 | 17.72 | 18.02 |
| **w100.A13** | >50 | >50 | >50 | >50 | >50 | >50 | >50 | 17.24 | 3.96 | 3.49 | 10.36 | 6.63 | 26.34 |
| **w100.A2** | >50 | >50 | >50 | >50 | >50 | >50 | >50 | 25.34 | 4.85 | 7.63 | 3.42 | 9.21 | 21.36 |
| **w100.A3** | >50 | >50 | >50 | >50 | >50 | >50 | 13.27 | 7.42 | 2.87 | 3.81 | 1.02 | 6.1 | 5.76 |
| **w100.A4** | >50 | >50 | >50 | >50 | >50 | >50 | >50 | 25.45 | 12.83 | 10.51 | 12.93 | 14.06 | 13.65 |
| **w100.A5** | >50 | >50 | >50 | >50 | >50 | >50 | >50 | 20.69 | 3.7 | 3.34 | 1.85 | 5.64 | 15.63 |
| **w100.A6** | >50 | >50 | >50 | >50 | >50 | >50 | 16.02 | 12.64 | 5.79 | 5.27 | 6.34 | 12.01 | 11.51 |
| **w100.B3** | >50 | >50 | >50 | >50 | >50 | >50 | >50 | 12.56 | 3.18 | 3.42 | 2.06 | 4.64 | 10.72 |
| **w100.B4** | >50 | >50 | >50 | >50 | >50 | >50 | 1.64 | 0.75 | 1.13 | 0.64 | 0.91 | 0.74 | 3.46 |
| **w100.B6** | >50 | >50 | >50 | >50 | >50 | >50 | >50 | 13.02 | 2.77 | 3 | 3.15 | 3.41 | 10.97 |
| **w100.B7** | >50 | >50 | >50 | >50 | >50 | >50 | 41.77 | 14.31 | 4.89 | 3.66 | 2.38 | 5.38 | 11.33 |
| **w100.B8** | >50 | >50 | >50 | >50 | >50 | >50 | >50 | >50 | >50 | >50 | >50 | >50 | 44.9 |

## Acknowledgments

We thank James Theiler, Will Fischer, Peter Kwong, and the LANL Interactive Intelligence team. This research was funded by the Division of AIDS of the National Institute of Allergy and Infectious Disease via the Center for HIV/AIDS Vaccine Immunology-Immunogen Discovery grant (CHAVI-ID; grant UM1 AI100645).

## Author Contributions

P.H., B.K., T.B., and B.F.H. conceived and designed the computational experiments; P.H., B.K., K.W., E.E.G., and A.S.L. performed the computational experiments; P.H., B.K., K.W., T.B., E.E.G., S.G., A.S.L. G.H.L., E.K., G.M.S., and B.F.H. analyzed the data; F.G., H.X.L., D.M., S.M.A., M.B., M.A.M., E.K., and Y.L., contributed experimental data; P.H., B.K., K.W., T.B., and B.F.H. wrote the paper; E.E.G., E.F.K., B.H.H, G.M.S, and M.A.M. edited the paper.

## Conflicts of Interest

Some coauthors have filed a provisional patent for combinations of sequences identified using methods described in this manuscript for use in a vaccine. The funding sponsors had no role in the design of the study; in the collection, analyses, or interpretation of data; in the writing of the manuscript, or in the decision to publish the results.

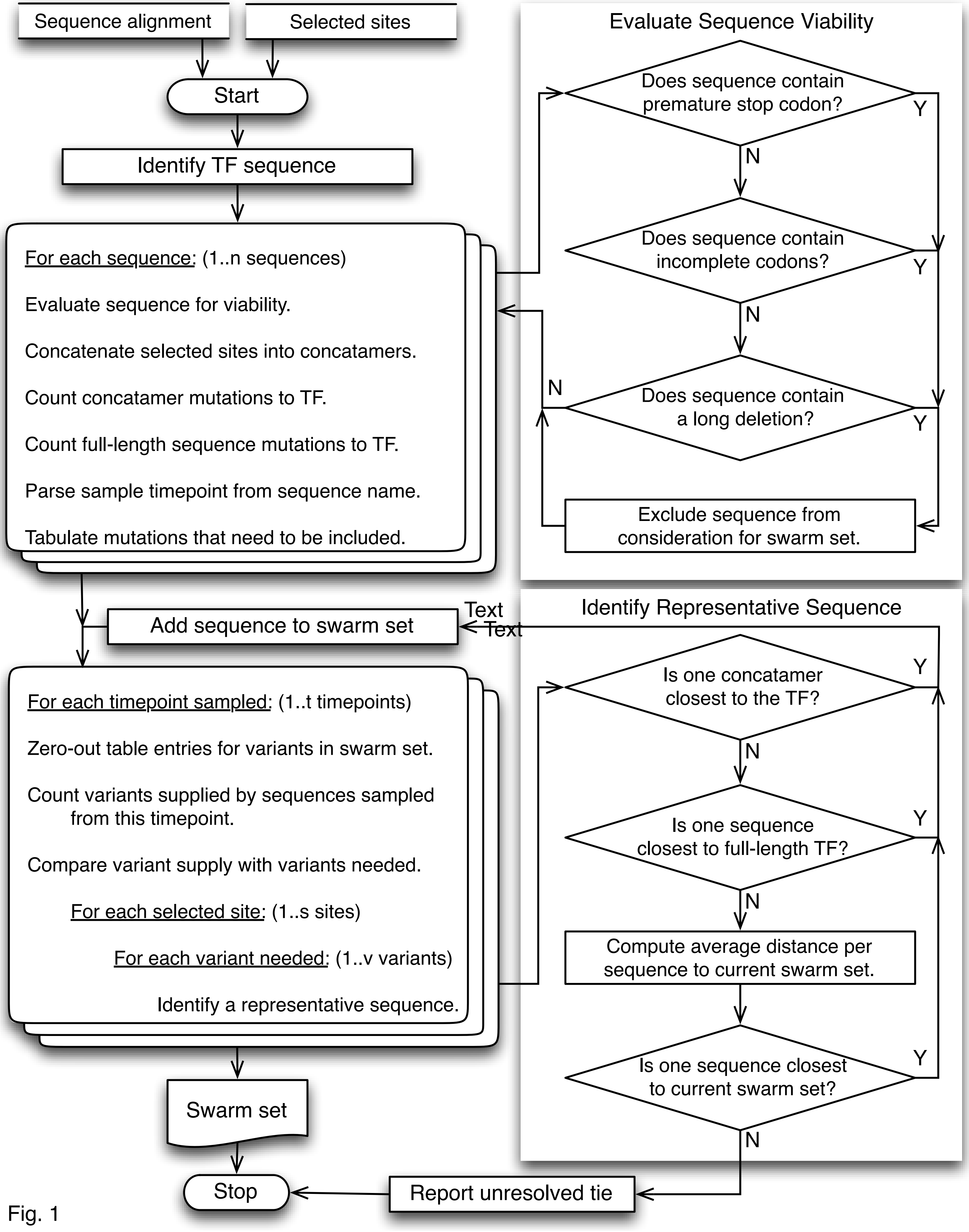
Sequence alignment
Selected sites
Start
Identify TF sequence
For each sequence: (1..n sequences)
Evaluate sequence for viability.
Concatenate selected sites into concatamers.
Count concatamer mutations to TF.
Count full-length sequence mutations to TF.
Parse sample timepoint from sequence name.
Tabulate mutations that need to be included.
Evaluate Sequence Viability
Does sequence contain premature stop codon?
Does sequence contain incomplete codons?
Does sequence contain a long deletion?
Exclude sequence from consideration for swarm set.
Y
N
Add sequence to swarm set
Text
Text
Identify Representative Sequence
For each timepoint sampled: (1..t timepoints)
Zero-out table entries for variants in swarm set.
Count variants supplied by sequences sampled from this timepoint.
Compare variant supply with variants needed.
For each selected site: (1..s sites)
For each variant needed: (1..v variants)
Identify a representative sequence.
Is one concatamer closest to the TF?
Is one sequence closest to full-length TF?
Compute average distance per sequence to current swarm set.
Is one sequence closest to current swarm set?
Swarm set
Stop
Report unresolved tie

Fig. 1

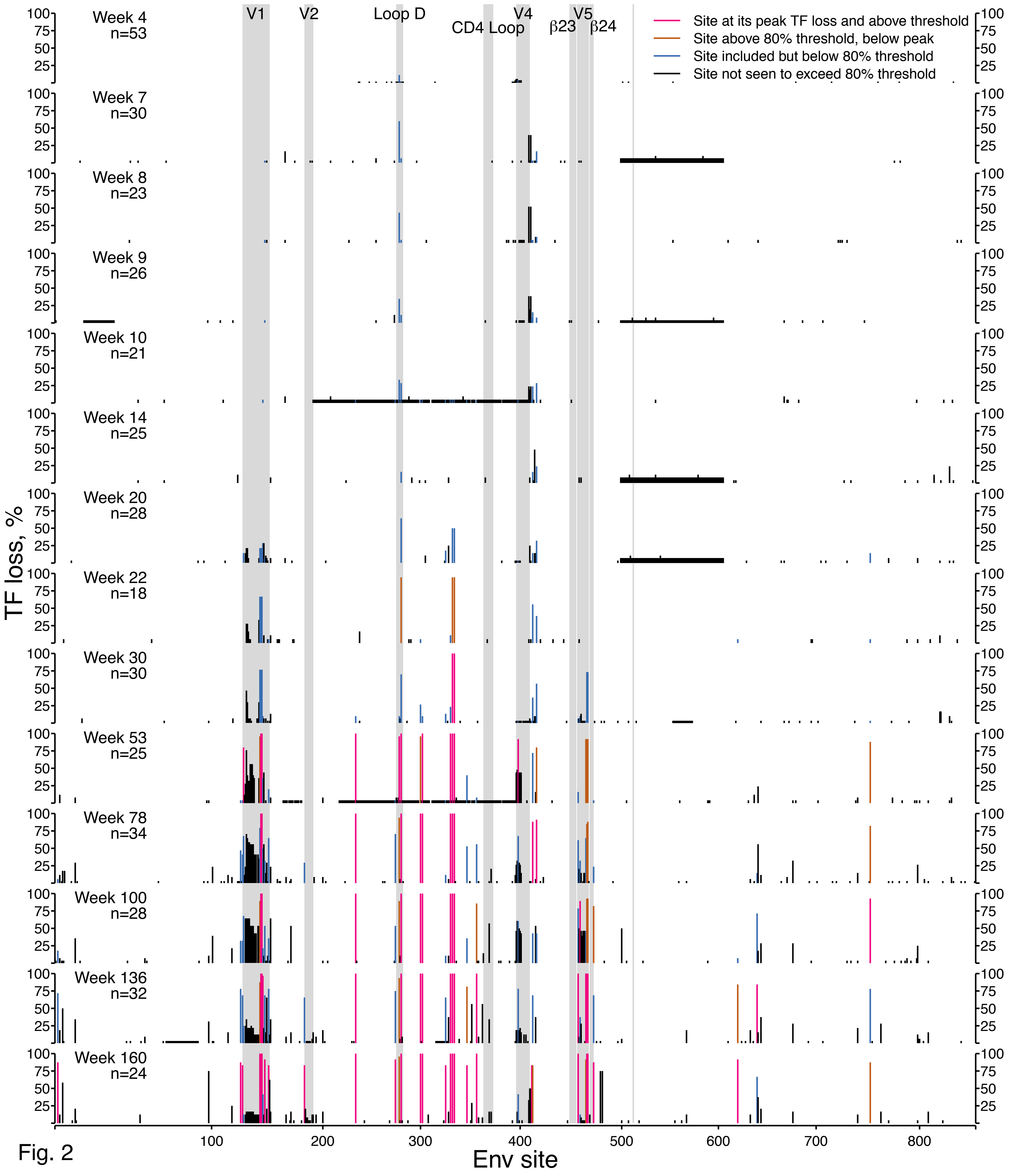

V1
V2
Loop D
CD4 Loop
V4
β23
V5
β24
Site at its peak TF loss and above threshold
Site above 80% threshold, below peak
Site included but below 80% threshold
Site not seen to exceed 80% threshold
Week 4 n=53
Week 7 n=30
Week 8 n=23
Week 9 n=26
Week 10 n=21
Week 14 n=25
Week 20 n=28
Week 22 n=18
Week 30 n=30
Week 53 n=25
Week 78 n=34
Week 100 n=28
Week 136 n=32
Week 160 n=24
TF loss, %
Env site
100
200
300
400
500
600
700
800

Fig. 2

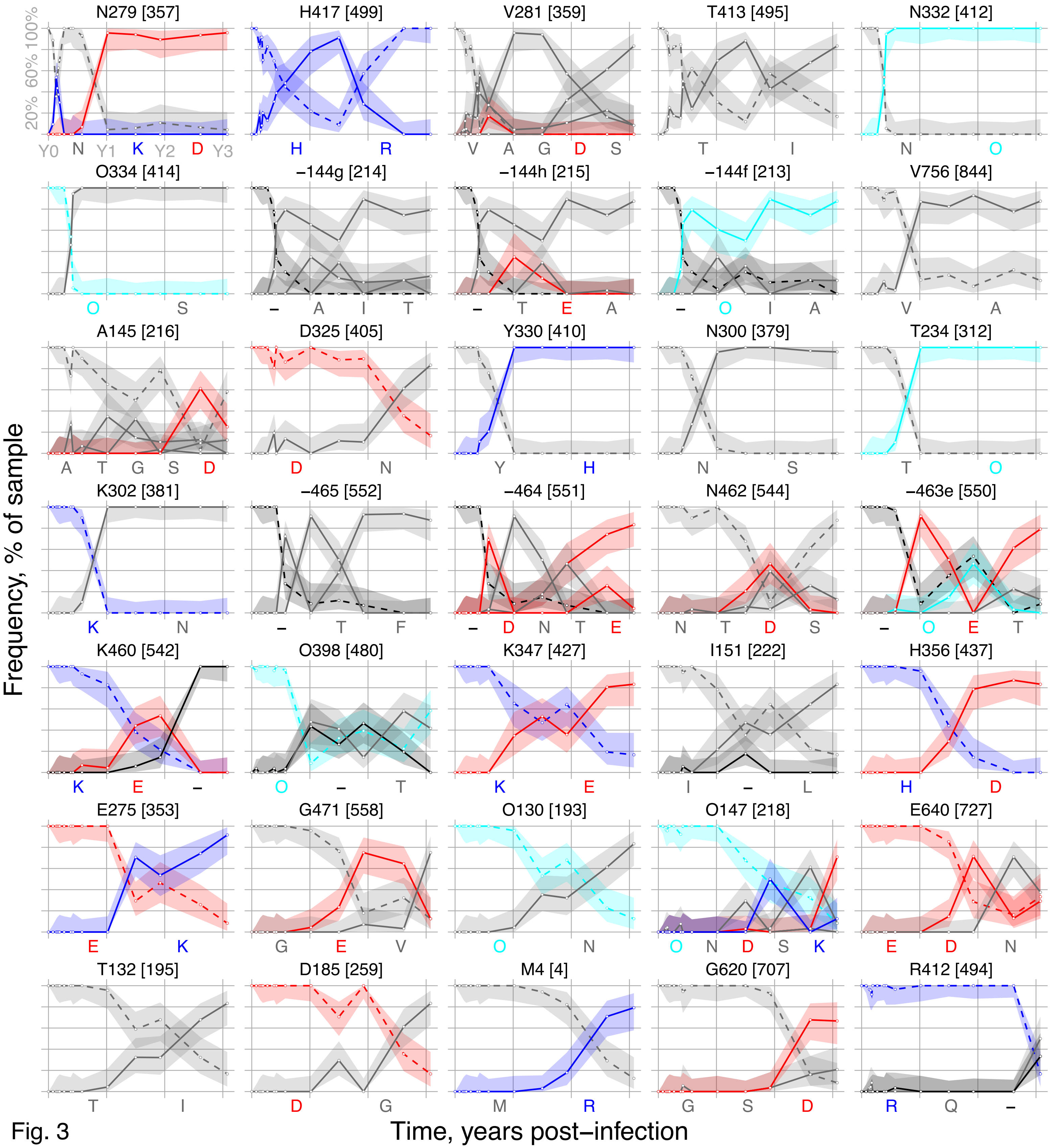

N279 [357]
H417 [499]
V281 [359]
T413 [495]
N332 [412]
20% 60% 100%
Y0 N Y1 K Y2 D Y3
H R
V A G D S
T I
N O
O334 [414]
−144g [214]
−144h [215]
−144f [213]
V756 [844]
O S
– A I T
– T E A
– O I A
V A
A145 [216]
D325 [405]
Y330 [410]
N300 [379]
T234 [312]
A T G S D
D N
Y H
N S
T O
K302 [381]
−465 [552]
−464 [551]
N462 [544]
−463e [550]
K N
– T F
– D N T E
N T D S
– O E T
K460 [542]
O398 [480]
K347 [427]
I151 [222]
H356 [437]
K E –
O – T
K E
I – L
H D
E275 [353]
G471 [558]
O130 [193]
O147 [218]
E640 [727]
E K
G E V
O N
O N D S K
E D N
T132 [195]
D185 [259]
M4 [4]
G620 [707]
R412 [494]
T I
D G
M R
G S D
R Q –
Frequency, % of sample
Time, years post−infection

Fig. 3

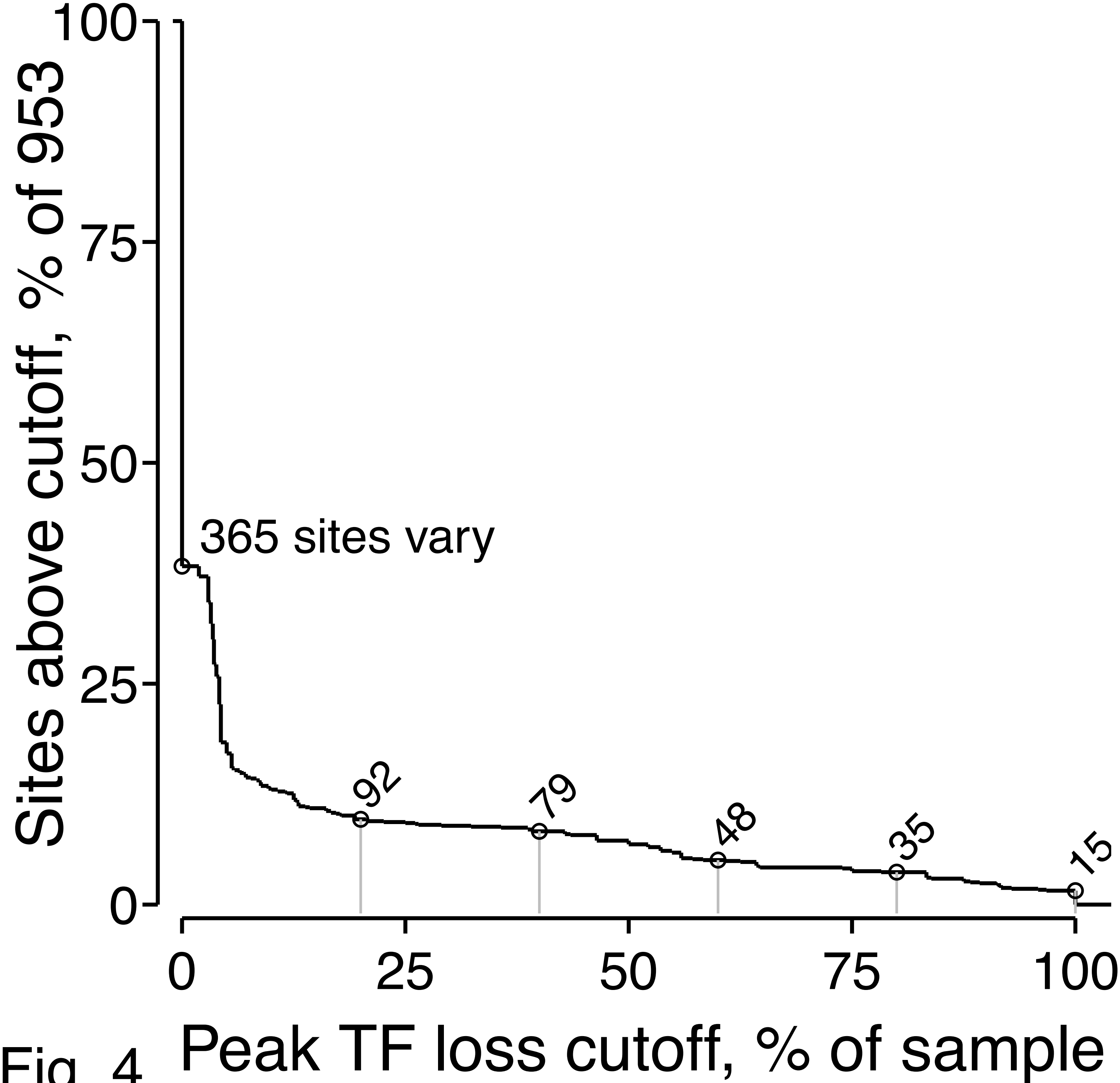

Fig. 4

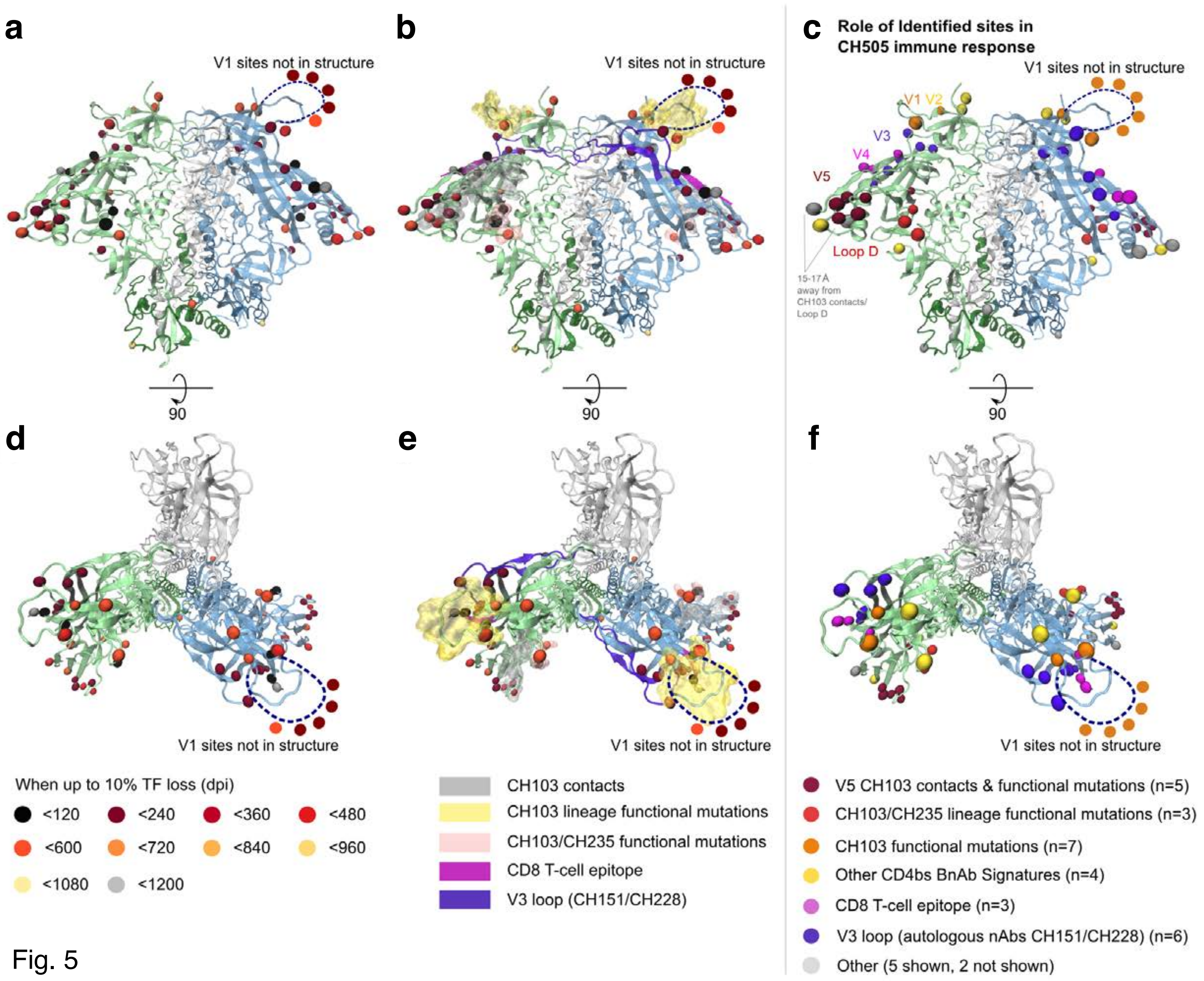
a
b
c
Role of Identified sites in CH505 immune response
V1 sites not in structure
V1 sites not in structure
V1 sites not in structure
V1
V2
V3
V4
V5
Loop D
15-17 Å away from CH103 contacts/ Loop D
90
90
90
d
e
f
V1 sites not in structure
V1 sites not in structure
V1 sites not in structure
When up to 10% TF loss (dpi)
<120
<240
<360
<480
<600
<720
<840
<960
<1080
<1200
CH103 contacts
CH103 lineage functional mutations
CH103/CH235 functional mutations
CD8 T-cell epitope
V3 loop (CH151/CH228)
V5 CH103 contacts & functional mutations (n=5)
CH103/CH235 lineage functional mutations (n=3)
CH103 functional mutations (n=7)
Other CD4bs BnAb Signatures (n=4)
CD8 T-cell epitope (n=3)
V3 loop (autologous nAbs CH151/CH228) (n=6)
Other (5 shown, 2 not shown)

Fig. 5

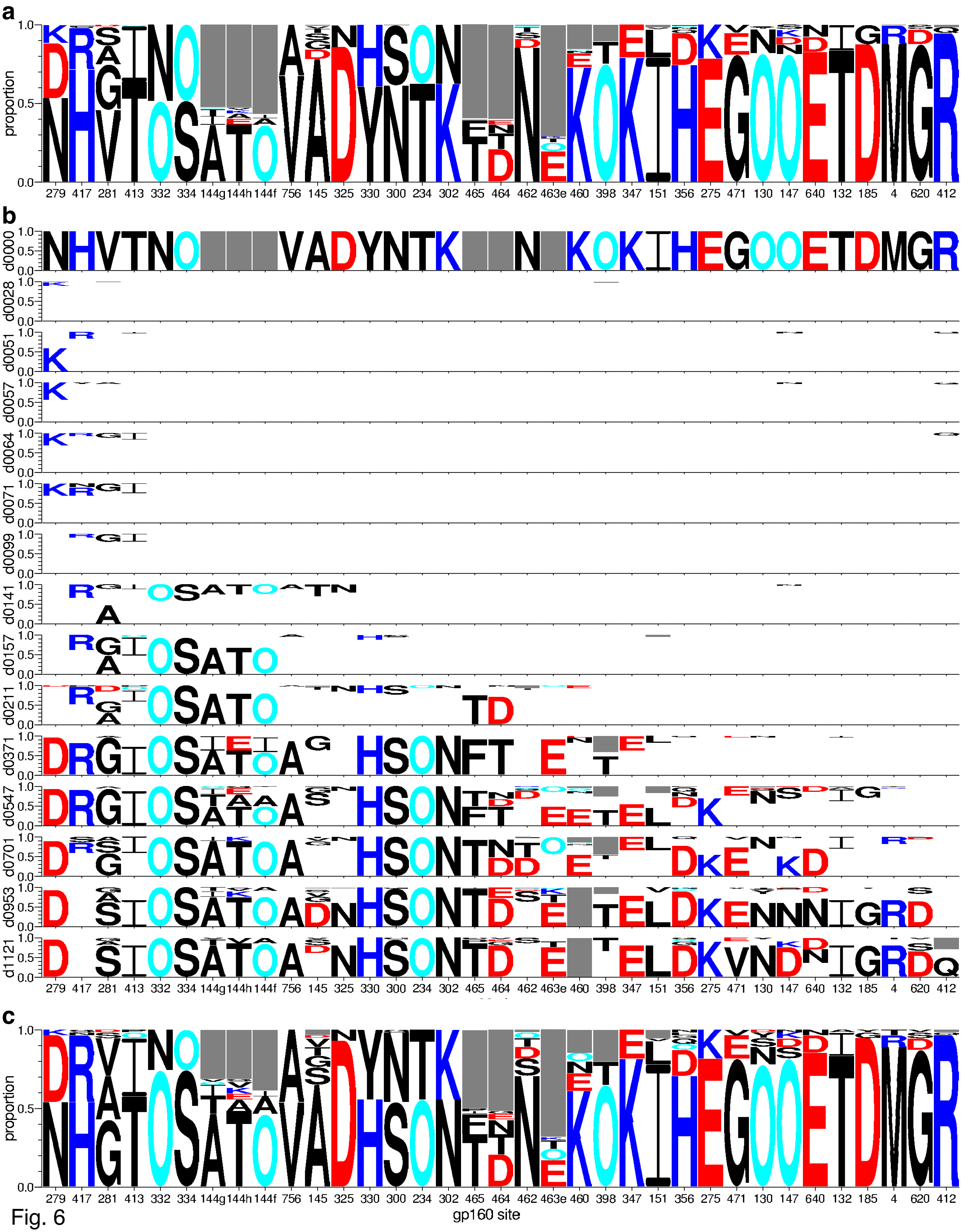
a
proportion
1.0
0.5
0.0
279 417 281 413 332 334 144g 144h 144f 756 145 325 330 300 234 302 465 464 462 463e 460 398 347 151 356 275 471 130 147 640 132 185 4 620 412
b
d0000
d0028
d0051
d0057
d0064
d0071
d0099
d0141
d0157
d0211
d0371
d0547
d0701
d0953
d1121
279 417 281 413 332 334 144g 144h 144f 756 145 325 330 300 234 302 465 464 462 463e 460 398 347 151 356 275 471 130 147 640 132 185 4 620 412
c
proportion
1.0
0.5
0.0
279 417 281 413 332 334 144g 144h 144f 756 145 325 330 300 234 302 465 464 462 463e 460 398 347 151 356 275 471 130 147 640 132 185 4 620 412
gp160 site

Fig. 6

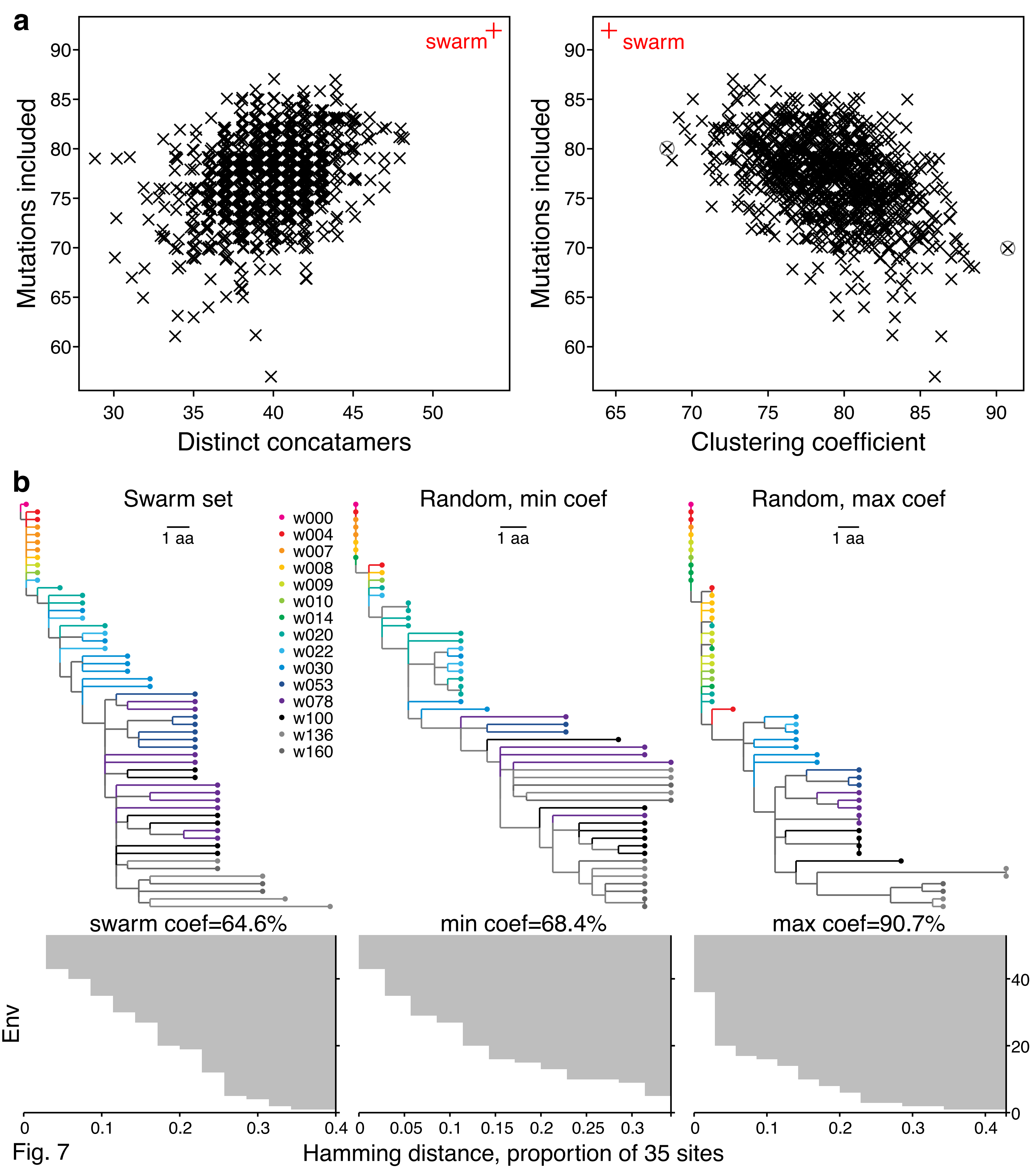

Fig. 7

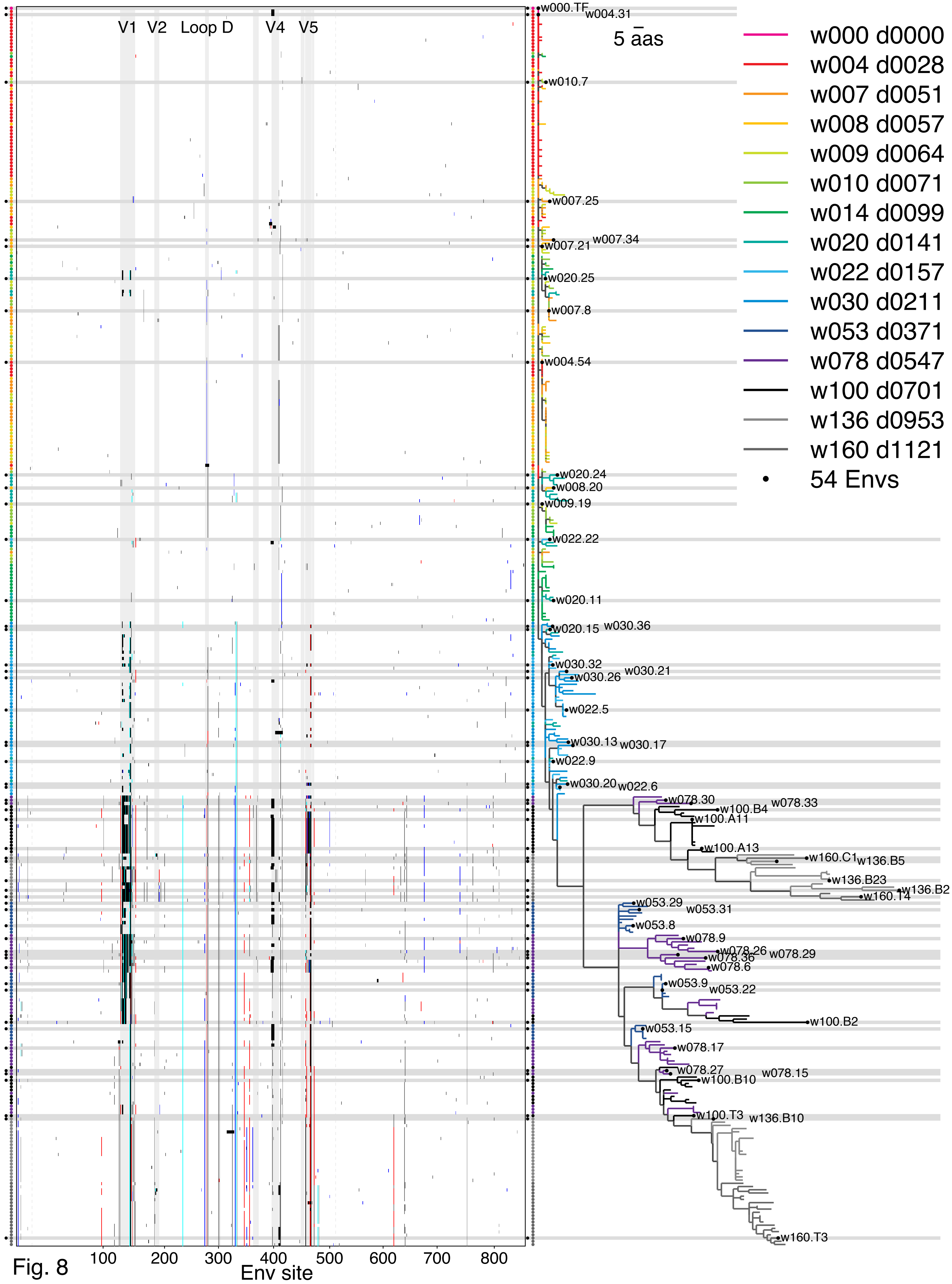
V1
V2
Loop D
V4
V5
100
200
300
400
500
600
700
800
Env site
5 aas
w000 d0000
w004 d0028
w007 d0051
w008 d0057
w009 d0064
w010 d0071
w014 d0099
w020 d0141
w022 d0157
w030 d0211
w053 d0371
w078 d0547
w100 d0701
w136 d0953
w160 d1121
54 Envs
w000.TF
w004.31
w010.7
w007.25
w007.21
w007.34
w020.25
w007.8
w004.54
w020.24
w008.20
w009.19
w022.22
w020.11
w020.15
w030.36
w030.32
w030.26
w030.21
w022.5
w030.13
w030.17
w022.9
w030.20
w022.6
w078.30
w100.B4
w078.33
w100.A11
w100.A13
w160.C1
w136.B5
w136.B23
w136.B2
w160.T4
w053.29
w053.31
w053.8
w078.9
w078.26
w078.29
w078.36
w078.6
w053.9
w053.22
w100.B2
w053.15
w078.17
w078.27
w100.B10
w078.15
w100.T3
w136.B10
w160.T3

Fig. 8

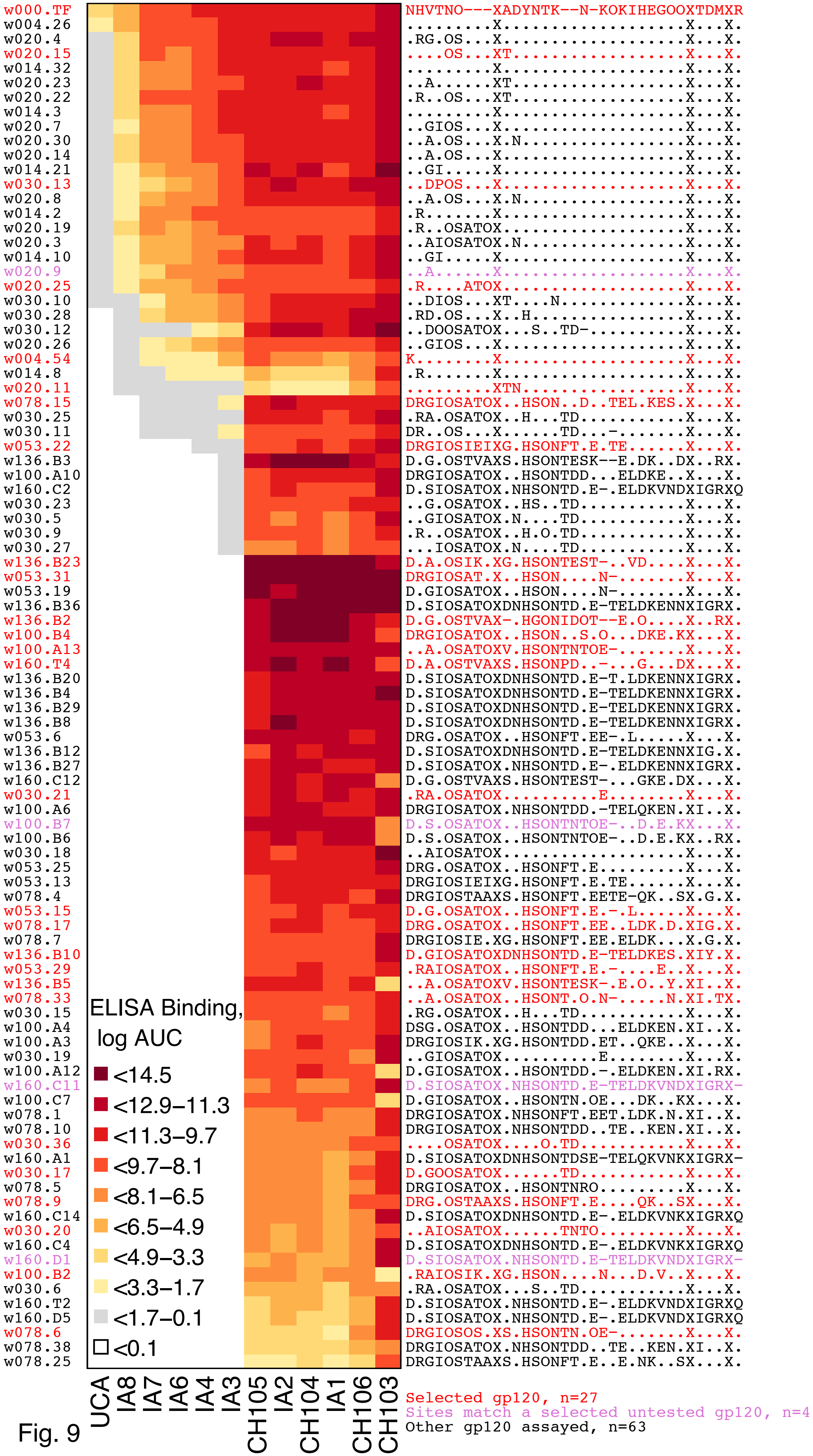

w000.TF NHVTNO---XADYNTK--N-KOKIHEGOOXTDMXR
w004.26 .........X...................X...X.
w020.4 .RG.OS...X...................X...X.
w020.15 ....OS...XT..................X...X.
w014.32 .........X...................X...X.
w020.23 ..A......XT..................X...X.
w020.22 .R..OS...XT..................X...X.
w014.3 .........X...................X...X.
w020.7 ..GIOS...X...................X...X.
w020.30 ..A.OS...X.N.................X...X.
w020.14 ..A.OS...X...................X...X.
w014.21 ..GI.....X...................X...X.
w030.13 ..DPOS...X...................X...X.
w020.8 ..A.OS...X.N.................X...X.
w014.2 .R.......X...................X...X.
w020.19 .R..OSATOX...................X...X.
w020.3 ..AIOSATOX.N.................X...X.
w014.10 ..GI.....X...................X...X.
w020.9 ..A......X...................X...X.
w020.25 .R....ATOX...................X...X.
w030.10 ..DIOS...XT....N.............X...X.
w030.28 .RD.OS...X..H................X...X.
w030.12 ..DOOSATOX...S..TD-..........X...X.
w020.26 ..GIOS...X...................X...X.
w004.54 K........X...................X...X.
w014.8 .R.......X...................X...X.
w020.11 .........XTN.................X...X.
w078.15 DRGIOSATOX..HSON..D..TEL.KES.X...X.
w030.25 .RA.OSATOX..H...TD...........X...X.
w030.11 DR..OS...X......TD...-.......X...X.
w053.22 DRGIOSIEIXG.HSONFT.E.TE......X...X.
w136.B3 D.G.OSTVAXS.HSONTESK--E.DK..DX..RX.
w100.A10 DRGIOSATOX..HSONTDD...ELDKE..X...X.
w160.C2 D.SIOSATOX.NHSONTD.E-.ELDKVNDXIGRXQ
w030.23 ..G.OSATOX..HS..TD...........X...X.
w030.5 ..GIOSATOX.N....TD...........X...X.
w030.9 .R..OSATOX..H.O.TD...........X...X.
w030.27 ...IOSATOX.N....TD...........X...X.
w136.B23 D.A.OSIK.XG.HSONTEST-..VD....X...X.
w053.31 DRGIOSAT.X..HSON....N-.......X...X.
w053.19 D.GIOSATOX..HSON....N-.......X...X.
w136.B36 D.SIOSATOXDNHSONTD.E-TELDKENNXIGRX.
w136.B2 D.G.OSTVAX-.HGONIDOT--E.O....X..RX.
w100.B4 DRGIOSATOX..HSON..S.O...DKE.KX...X.
w100.A13 ..A.OSATOXV.HSONTNTOE-.......X...X.
w160.T4 D.A.OSTVAXS.HSONPD..-...G...DX...X.
w136.B20 D.SIOSATOXDNHSONTD.E-T.LDKENNXIGRX.
w136.B4 D.SIOSATOXDNHSONTD.E-TELDKENNXIGRX.
w136.B29 D.SIOSATOXDNHSONTD.E-TELDKENNXIGRX.
w136.B8 D.SIOSATOXDNHSONTD.E-TELDKENNXIGRX.
w053.6 DRG.OSATOX..HSONFT.EE-.L.....X...X.
w136.B12 D.SIOSATOXDNHSONTD.E-TELDKENNXIG.X.
w136.B27 D.SIOSATOX.NHSONTD.E-.ELDKENNXIGRX.
w160.C12 D.G.OSTVAXS.HSONTEST-...GKE.DX...X.
w030.21 .RA.OSATOX.........E.........X...X.
w100.A6 DRGIOSATOX.NHSONTDD.-TELQKEN.XI..X.
w100.B7 D.S.OSATOX..HSONTNTOE-..D.E.KX...X.
w100.B6 D.S.OSATOX..HSONTNTOE-..D.E.KX..RX.
w030.18 ..AIOSATOX...................X...X.
w053.25 DRG.OSATOX..HSONFT.E.........X...X.
w053.13 DRGIOSIEIXG.HSONFT.E.TE......X...X.
w078.4 DRGIOSTAAXS.HSONFT.EETE-QK..SX.G.X.
w053.15 D.G.OSATOX..HSONFT.E.-.L.....X...X.
w078.17 DRG.OSATOX..HSONFT.EE..LDK.D.XIG.X.
w078.7 DRGIOSIE.XG.HSONFT.EE.ELDK...X.G.X.
w136.B10 D.GIOSATOXDNHSONTD.E-TELDKES.XIY.X.
w053.29 .RAIOSATOX..HSONFT.E.-....E..X...X.
w136.B5 ..A.OSATOXV.HSONTESK-.E.O..Y.XI..X.
w078.33 ..A.OSATOX..HSONT.O.N-.....N.XI.TX.
w030.15 .RG.OSATOX..H...TD...........X...X.
w100.A4 DSG.OSATOX..HSONTDD...ELDKEN.XI..X.
w100.A3 DRGIOSIK.XG.HSONTDD.ET..QKE..X...X.
w030.19 ..GIOSATOX.........E.........X...X.
w100.A12 D.GIOSATOX..HSONTDD.-.ELDKEN.XI.RX.
w160.C11 D.SIOSATOX.NHSONTD.E-TELDKVNDXIGRX-
w100.C7 D.GIOSATOX..HSONTN.OE...DK..KX...X.
w078.1 DRGIOSATOX.NHSONFT.EET.LDK.N.XI..X.
w078.10 DRGIOSATOX.NHSONTDD..TE..KEN.XI..X.
w030.36 ....OSATOX....O.TD...........X...X.
w160.A1 D.SIOSATOXDNHSONTDSE-TELQKVNKXIGRX-
w030.17 D.GOOSATOX......TD...........X...X.
w078.5 DRGIOSATOX..HSONTNRO.........X...X.
w078.9 DRG.OSTAAXS.HSONFT.E....QK..SX...X.
w160.C14 D.SIOSATOXDNHSONTD.E-.ELDKVNKXIGRXQ
w030.20 ..AIOSATOX......TNTO.........X...X.
w160.C4 D.SIOSATOXDNHSONTD.E-.ELDKVNKXIGRXQ
w160.D1 D.SIOSATOX.NHSONTD.E-TELDKVNDXIGRX-
w100.B2 .RAIOSIK.XG.HSON....N...D.V..X...X.
w030.6 .RA.OSATOX...S..TD...........X...X.
w160.T2 D.SIOSATOX.NHSONTD.E-.ELDKVNDXIGRXQ
w160.D5 D.SIOSATOX.NHSONTD.E-.ELDKVNDXIGRXQ
w078.6 DRGIOSOS.XS.HSONTN.OE-.......X...X.
w078.38 DRGIOSATOX.NHSONTDD..TE..KEN.XI..X.
w078.25 DRGIOSTAAXS.HSONFT.E..E.NK..SX...X.
ELISA Binding, log AUC
<14.5
<12.9–11.3
<11.3–9.7
<9.7–8.1
<8.1–6.5
<6.5–4.9
<4.9–3.3
<3.3–1.7
<1.7–0.1
<0.1
UCA
IA8
IA7
IA6
IA4
IA3
CH105
IA2
CH104
IA1
CH106
CH103
Selected gp120, n=27
Sites match a selected untested gp120, n=4
Other gp120 assayed, n=63

Fig. 9

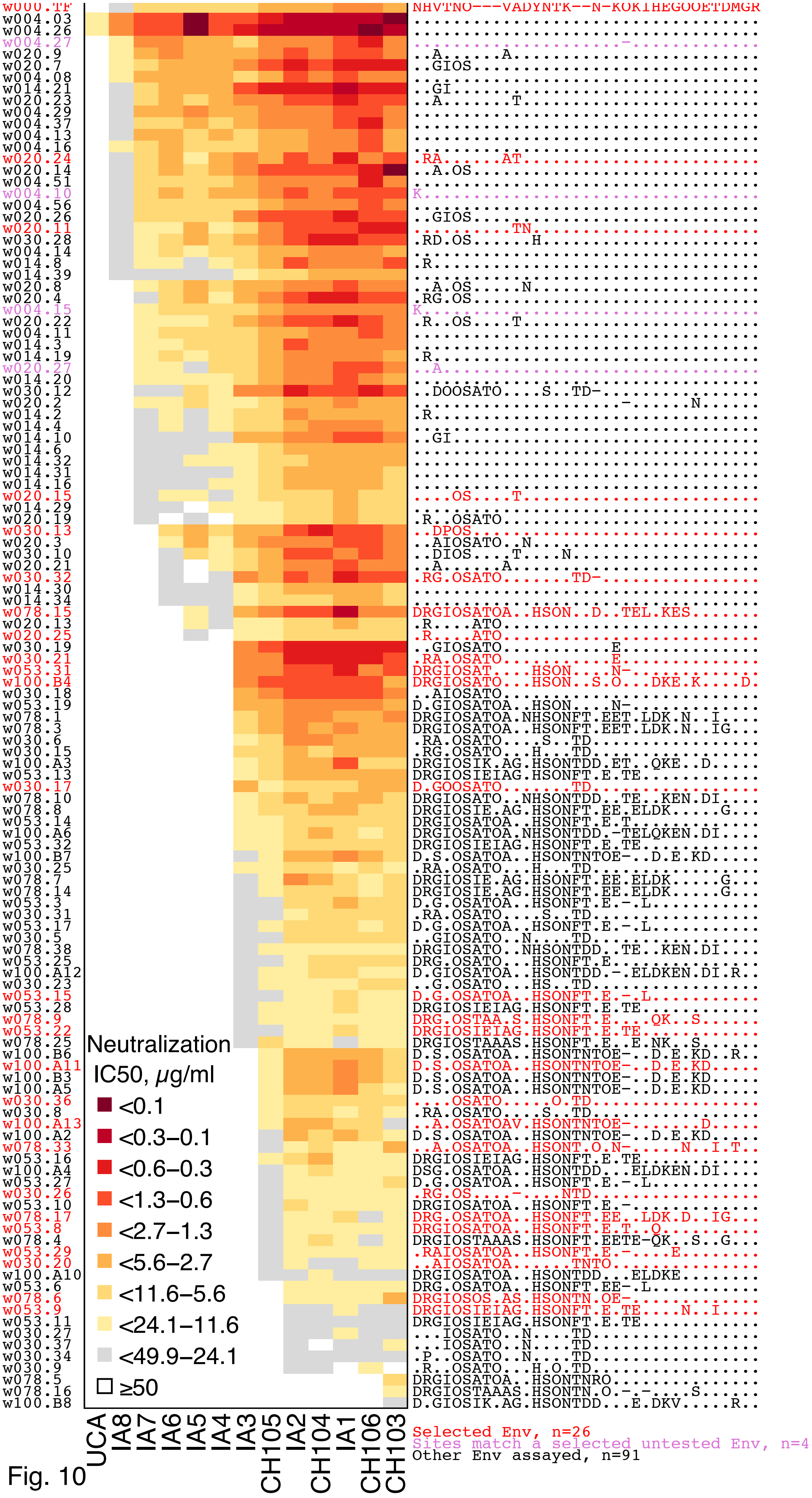
Neutralization
IC50, μg/ml
<0.1
<0.3–0.1
<0.6–0.3
<1.3–0.6
<2.7–1.3
<5.6–2.7
<11.6–5.6
<24.1–11.6
<49.9–24.1
≥50
UCA
IA8
IA7
IA6
IA5
IA4
IA3
CH105
IA2
CH104
IA1
CH106
CH103
Selected Env, n=26
Sites match a selected untested Env, n=4
Other Env assayed, n=91

Fig. 10

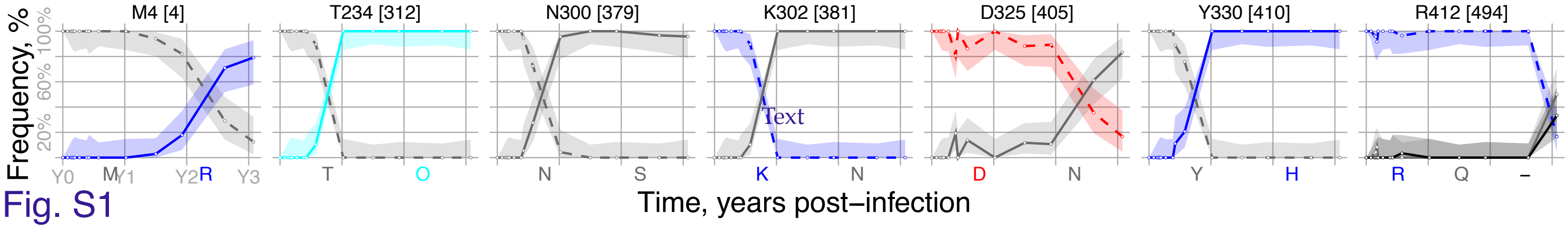
M4 [4]
T234 [312]
N300 [379]
K302 [381]
D325 [405]
Y330 [410]
R412 [494]
Frequency, %
100%
60%
20%
Y0
Y1
Y2
Y3
M
R
T
O
N
S
K
N
Text
D
N
Y
H
R
Q
–
Time, years post–infection
Fig. S1

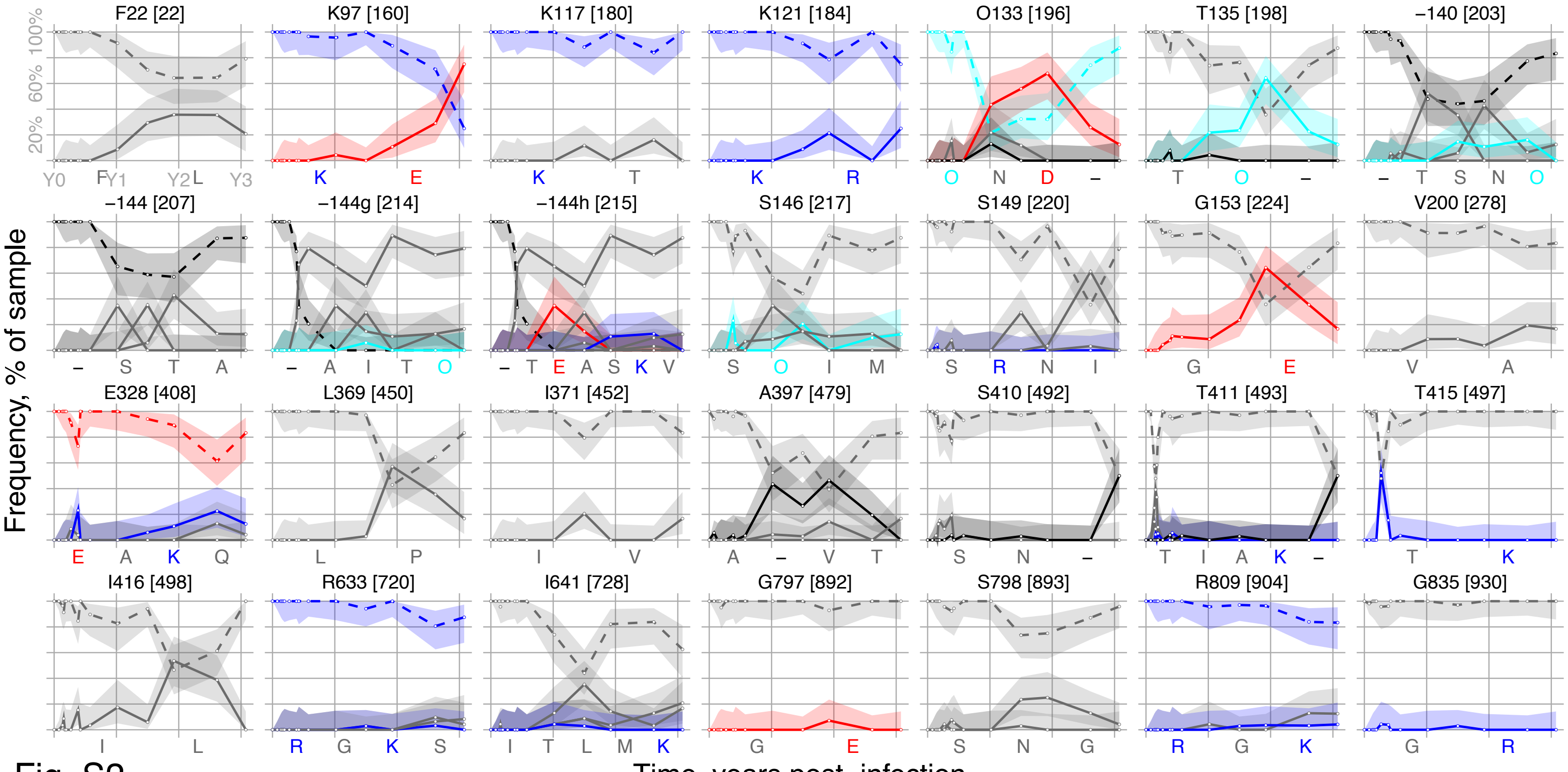

F22 [22]
K97 [160]
K117 [180]
K121 [184]
O133 [196]
T135 [198]
–140 [203]
–144 [207]
–144g [214]
–144h [215]
S146 [217]
S149 [220]
G153 [224]
V200 [278]
E328 [408]
L369 [450]
I371 [452]
A397 [479]
S410 [492]
T411 [493]
T415 [497]
I416 [498]
R633 [720]
I641 [728]
G797 [892]
S798 [893]
R809 [904]
G835 [930]
20%
60%
100%
Y0
Y1
Y2
Y3
Frequency, % of sample
Time, years post–infection

Fig. S2

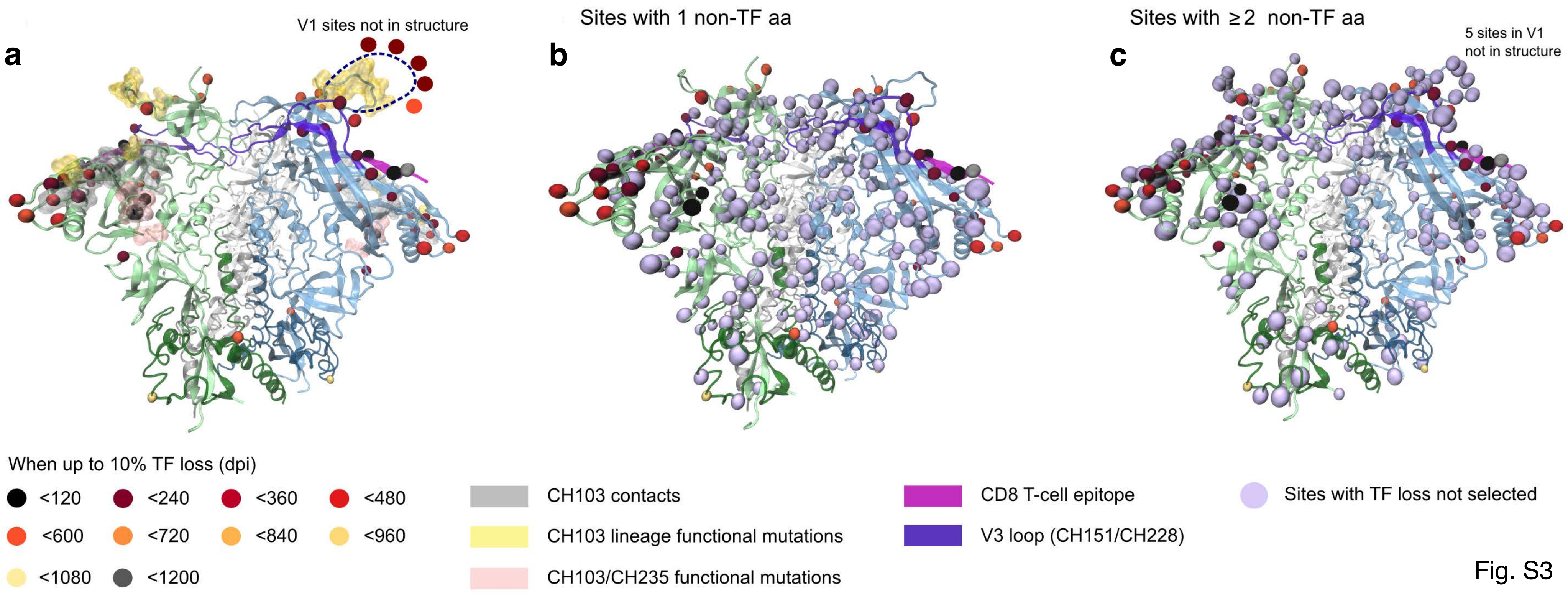

a
V1 sites not in structure
b
Sites with 1 non-TF aa
c
Sites with ≥2 non-TF aa
5 sites in V1
not in structure
When up to 10% TF loss (dpi)
<120
<240
<360
<480
<600
<720
<840
<960
<1080
<1200
CH103 contacts
CH103 lineage functional mutations
CH103/CH235 functional mutations
CD8 T-cell epitope
V3 loop (CH151/CH228)
Sites with TF loss not selected

Fig. S3

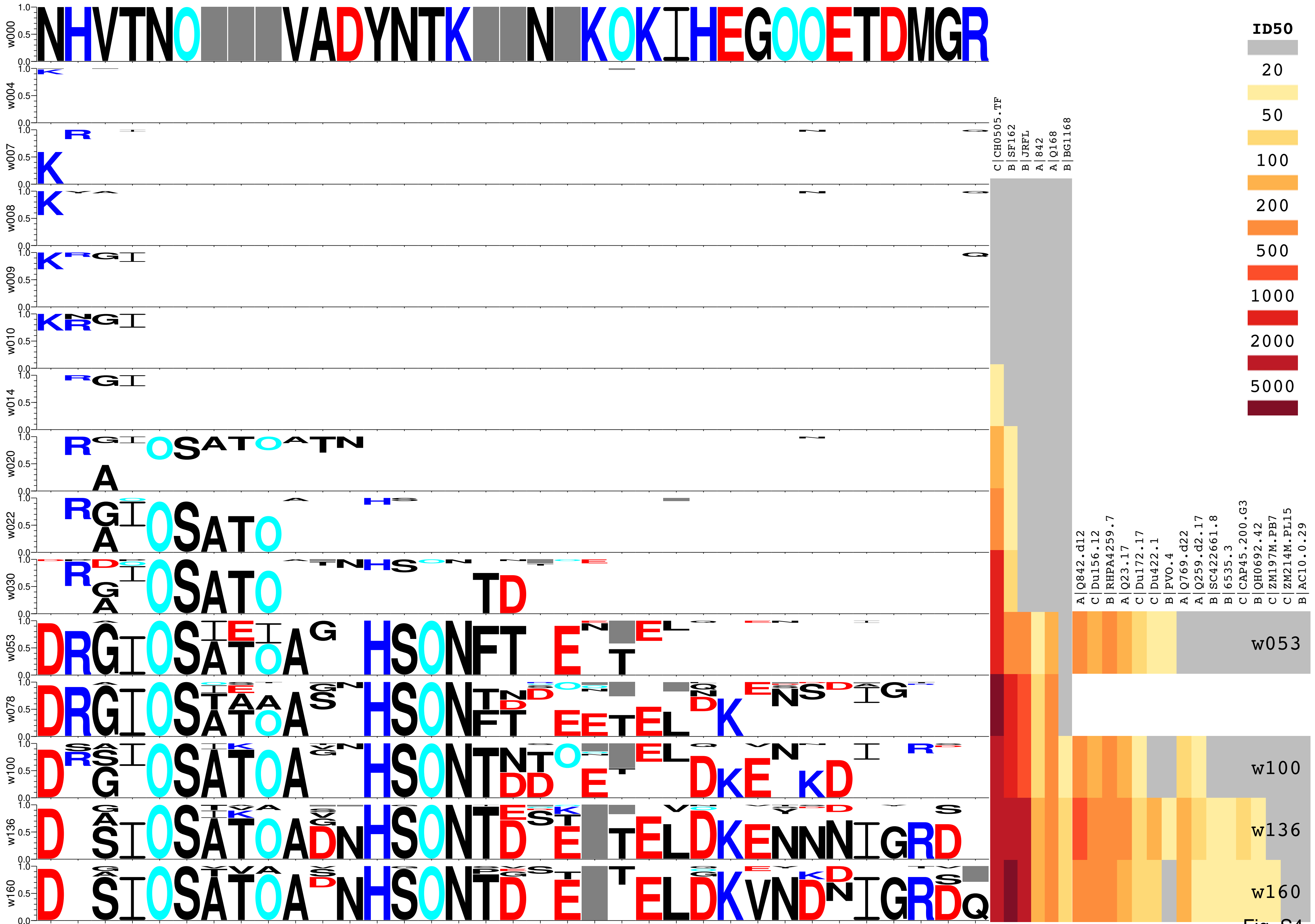

w000
w004
w007
w008
w009
w010
w014
w020
w022
w030
w053
w078
w100
w136
w160
279 417 281 413 332 334 144g 144h 144f 756 145 325 330 300 234 302 465 464 462 463e 460 398 347 151 356 275 471 130 147 640 132 185 4 620 412
C|CH0505.TF
B|SF162
B|JRFL
A|842
A|Q168
B|BG1168
A|Q842.d12
C|Du156.12
B|RHPA4259.7
A|Q23.17
C|Du172.17
C|Du422.1
B|PVO.4
A|Q769.d22
A|Q259.d2.17
B|SC422661.8
B|6535.3
C|CAP45.200.G3
B|QH0692.42
C|ZM197M.PB7
C|ZM214M.PL15
B|AC10.0.29
w053
w100
w136
w160
ID50
20
50
100
200
500
1000
2000
5000

Fig. S4

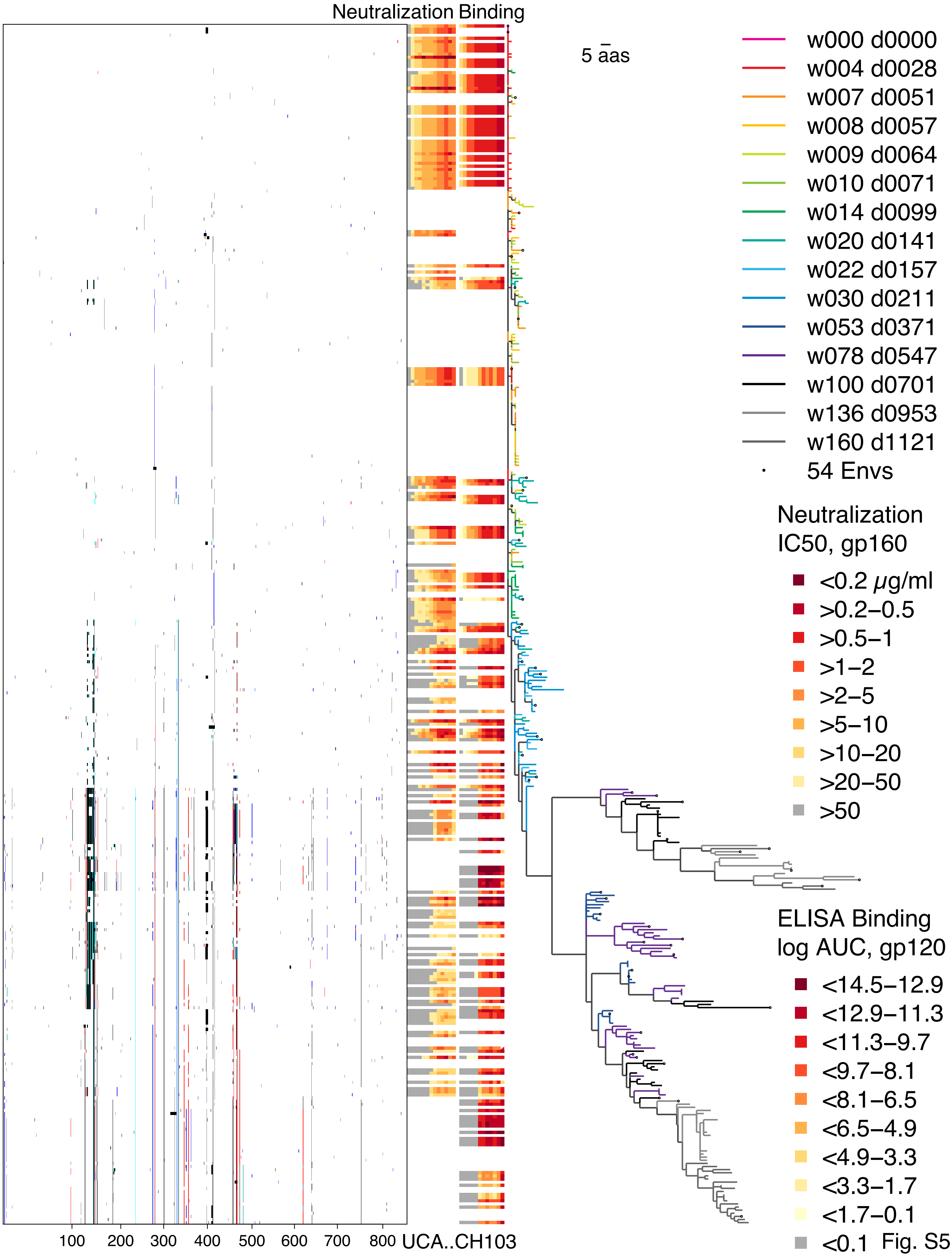

Neutralization Binding
5 aas
w000 d0000
w004 d0028
w007 d0051
w008 d0057
w009 d0064
w010 d0071
w014 d0099
w020 d0141
w022 d0157
w030 d0211
w053 d0371
w078 d0547
w100 d0701
w136 d0953
w160 d1121
54 Envs
Neutralization
IC50, gp160
<0.2 μg/ml
>0.2–0.5
>0.5–1
>1–2
>2–5
>5–10
>10–20
>20–50
>50
ELISA Binding
log AUC, gp120
<14.5–12.9
<12.9–11.3
<11.3–9.7
<9.7–8.1
<8.1–6.5
<6.5–4.9
<4.9–3.3
<3.3–1.7
<1.7–0.1
<0.1
100 200 300 400 500 600 700 800 UCA..CH103

Fig. S5

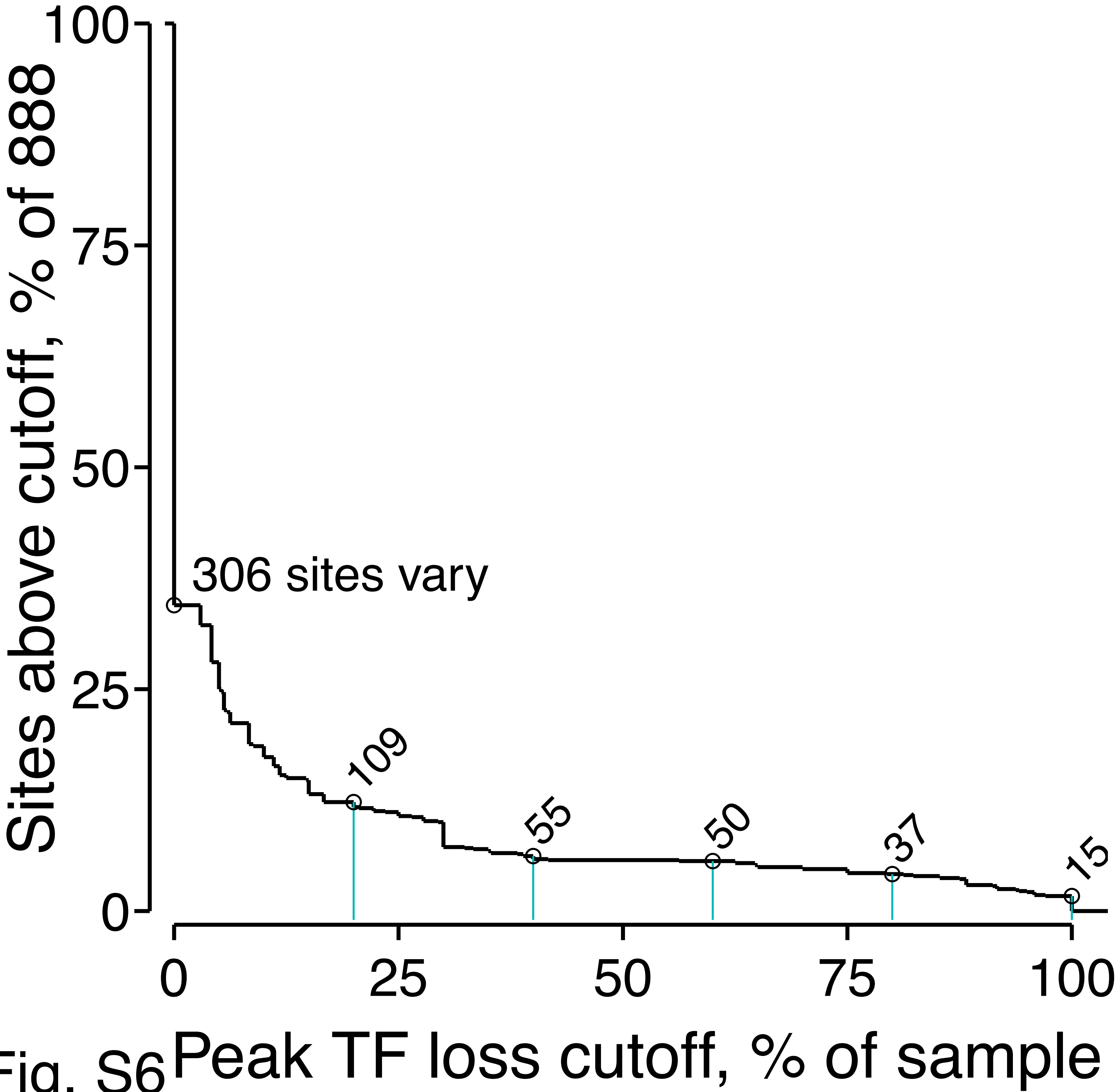

Fig. S6

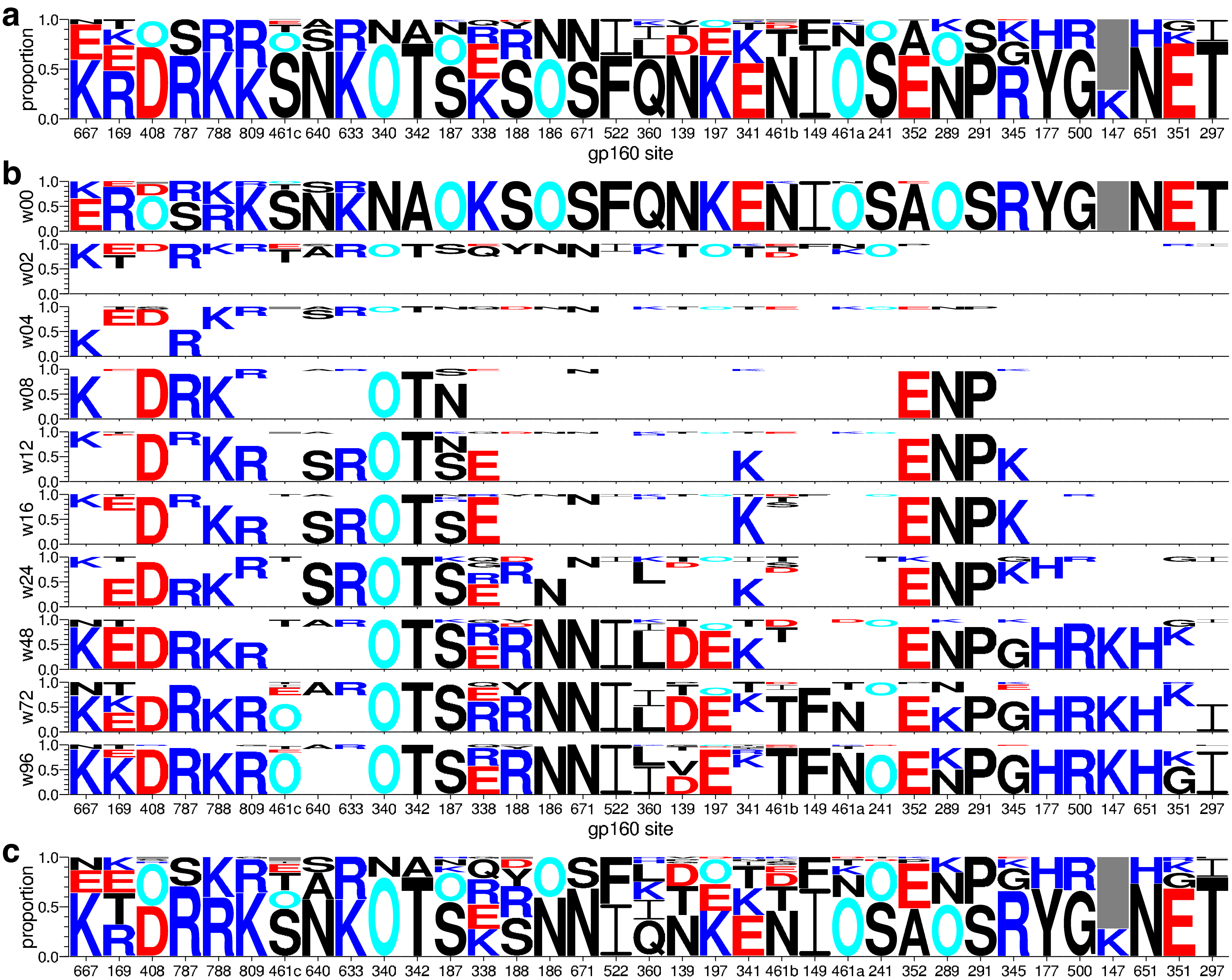
a
b
c
proportion
1.0
0.5
0.0
w00
w02
w04
w08
w12
w16
w24
w48
w72
w96
667 169 408 787 788 809 461c 640 633 340 342 187 338 188 186 671 522 360 139 197 341 461b 149 461a 241 352 289 291 345 177 500 147 651 351 297
gp160 site

Fig. S7

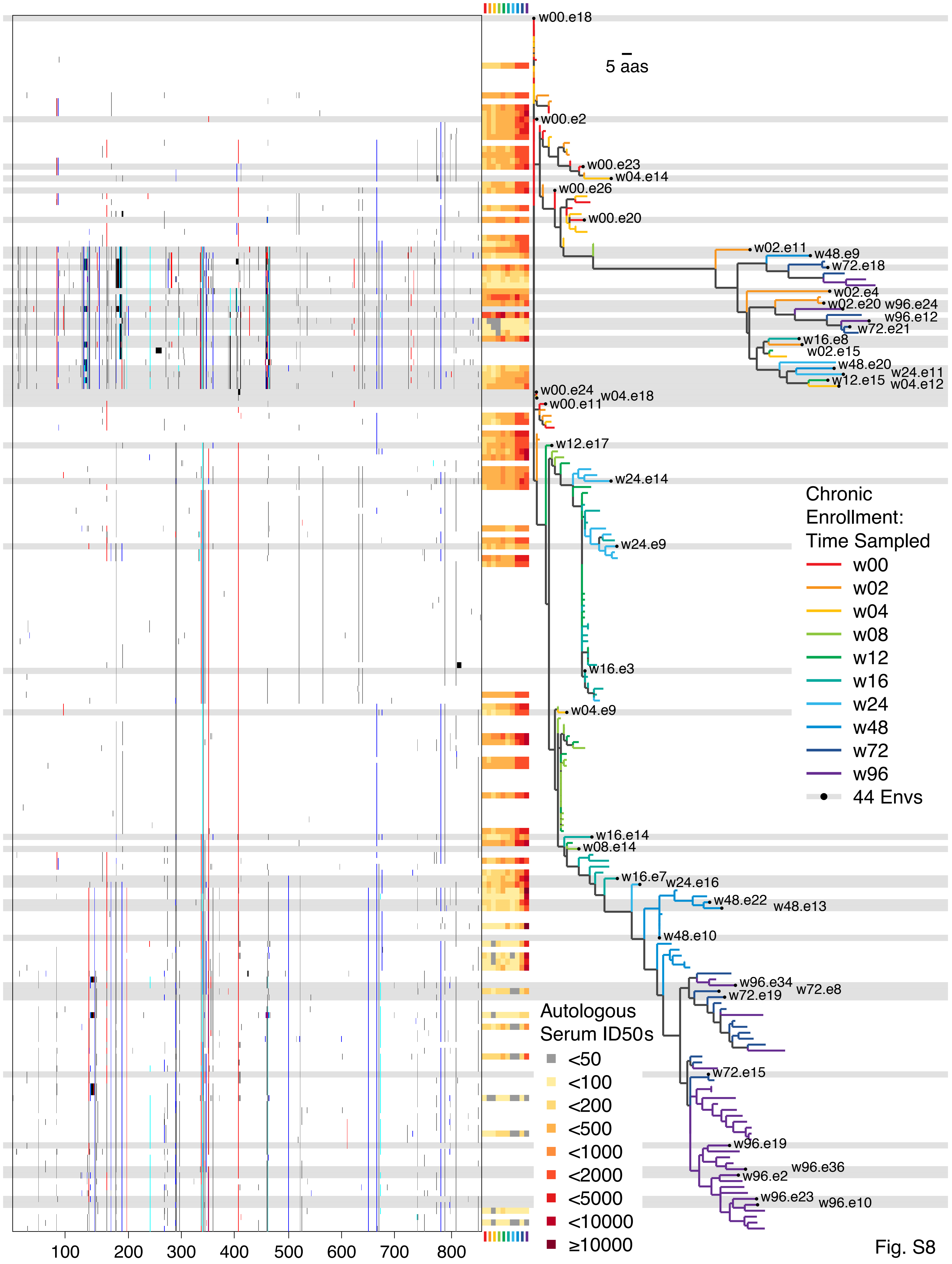

Fig. S8